\documentclass[mathpazo]{cicp}

\usepackage{amsmath,amssymb,newlfont}
\usepackage{graphicx}
\usepackage{verbatim}

\usepackage{dsfont}
\usepackage{xspace}
\usepackage{amsgen}
\usepackage{wasysym}
\usepackage{upgreek}
\usepackage{amsfonts}
\usepackage{stmaryrd}
\usepackage{mathtools}
\usepackage{relsize}
\usepackage[mathcal, mathscr]{euscript}
\usepackage{mathrsfs}

\usepackage{amsbsy}
\usepackage{amsthm}
\usepackage{amssymb}

\usepackage{cite}                                   

\usepackage{xcolor}
\usepackage{hyperref}
\definecolor{linkcolor}{rgb}{0.9,0,0}
\definecolor{citecolor}{rgb}{0,0.6,0}
\definecolor{urlcolor}{rgb}{0,0,1}
\hypersetup{
    colorlinks, linkcolor={linkcolor},
    citecolor={citecolor}, urlcolor={urlcolor}
}

\usepackage{color}

\usepackage{lineno}

\newcommand{\Pnh}{\power}
\newcommand{\power}{\raisebox{.15\baselineskip}{\large\ensuremath{\wp}}}

\newcommand{\nm}[1]{\textsc{#1}}
\newcommand{\cf}{\emph{c.f.}\xspace}
\newcommand{\ie}{\emph{i.e.}\xspace}

\newcommand{\etc}{\emph{etc.}\xspace}

\newcommand{\NS}{\boldsymbol{\mathsf{NS}}}
\newcommand{\SW}{\boldsymbol{\mathsf{SV}}}
\newcommand{\Eu}{\boldsymbol{\mathsf{Eul}}}
\newcommand{\Po}{\boldsymbol{\mathsf{Pot}}}
\newcommand{\SGN}{\boldsymbol{\mathsf{SGN}}}
\newcommand{\Bouss}{\boldsymbol{\mathsf{Bouss}}}

\renewcommand{\H}{\mathcal{H}}

\newcommand{\No}{$\mathrm{N}^\circ$}

\newcommand{\ub}{\bar{\boldsymbol{u}}}
\newcommand{\ut}{\under{\boldsymbol{u}}}

\newcommand{\under}[1]{\mkern 1.5mu\underline{\mkern-1.5mu#1\mkern-1.5mu}\mkern 1.5mu}

\usepackage[acronym, toc]{glossaries}

\makeglossaries

\newacronym{1d}{1D}{one-dimensional}
\newacronym{2d}{2D}{two-dimensional}
\newacronym{3d}{3D}{three-dimensional}
\newacronym{vof}{VoF}{Volume-of-Fluid}
\newacronym{npp}{NPP}{Nuclear Power Plant}
\newacronym{lng}{LNG}{Liquefied Natural Gas}
\newacronym{cpu}{CPU}{Central Processing Unit}
\newacronym{fee}{FEE}{full \nm{Euler} equations}
\newacronym{nswe}{NSWE}{Nonlinear Shallow Water Equations}
\newacronym{sgn}{SGN equations}{\nm{Serre}--\nm{Green}--\nm{Naghdi} Equations}

\usepackage[explicit]{titlesec}

\newcommand\invisiblesection[1]{%
  \addcontentsline{toc}{section}{#1}%
  \sectionmark{#1}}

\newcommand{\vect}[1]{\boldsymbol{#1}}

\begin{document}
\title[short title]{Wave/partially immersed body interaction. Part II }
\title{Long wave interaction with a partially immersed body. Part II: Numerical results}

 \author[Khakimzyanov et~al.]{Gayaz Khakimzyanov\affil{1},
       Denys Dutykh\affil{2}\comma\corrauth, and Oleg Gusev\affil{1}}
\address{\affilnum{1}\ Federal Research Center for Information and Computational Technologies,
          Academician M.A. Lavrentiev avenue, 6, 630090, Novosibirsk, Russia \\
           \affilnum{2}\ Univ. Grenoble Alpes, Univ. Savoie Mont Blanc, CNRS, LAMA, 73000 Chamb\'ery, France}
 \emails{{\tt gayaz.khakimzyanov@gmail.com} (G.~Khakimzyanov), {\tt Denys.Dutykh@univ-smb.fr} (D.~Dutykh),
          {\tt gusev\_oleg\_igor@mail.ru} (O.~Gusev)}

\begin{abstract}
In this manuscript we perform an extensive numerical study of the long wave
interaction problem with a fixed partially immersed body into a fluid layer. The
incident wave is assumed to be an isolated solitary wave. The body in this study is
assumed to be fixed with a rectangular section which is not touching the bottom
of the channel. The mathematical modelling of this problem is based on Part I
\cite{Khakimzyanov2018a} of this series and considered models include the Nonlinear Shallow Water
Equations (NSWE), fully nonlinear weakly dispersive Serre–Green–Naghdi
Equations (SGN equations) (completed with appropriate compatibility conditions on solid/fluid boundaries) and the free surface irrotational full Euler
equations (FEE). We study the influence of the floating body elongation, immersion depth
and incident wave amplitude on the wave field before and after the obstacle.
The comparison of all three models predictions and the data of small-scale laboratory experiments is performed. Moreover, in the
framework of the FEE model we investigate the anomalous wave run-up behind
the floating body in the close presence of a vertical wall. We demonstrate the
cases where the vertical wall creates extreme wave amplitudes behind the body,
but also we show the cases where the wall attenuates wave amplitudes comparing
to the wave field without a wall.
\end{abstract}

\pac{[2010] 47.35.Bb (primary), 47.35.Fg, 02.60.Lj (secondary)}

\maketitle

\section{Introduction}
\label{sec1} 
In the first part \cite{Khakimzyanov2018a} of this study, we considered the following hierarchy of mathematical models describing the interaction of water waves with an immersed floating body:
\begin{align}\label{eq:hie}
  & \xLeftarrow{\text{Simplified}} \nonumber \\
  \SW\ \Longrightarrow\ \Bouss\ \Longrightarrow\ \SGN\ &\Longrightarrow\ \ldots\ \Longrightarrow\ \Po\ \Longrightarrow\ \Eu\ \Longrightarrow\ \NS \\
  & \xRightarrow{\text{More complete}} \nonumber
\end{align}
The arrows show the direction of increasing model complexity and the models are
\begin{itemize}
  \item Rotational incompressible ideal fluid flow model (\acrshort{fee}) \cite{Lamb1932} ($\Eu$),
  \item Potential flow model (irrotational \acrshort{fee}) \cite{Stoker1958b} ($\Po$),
  \item Fully nonlinear weakly dispersive wave model (\acrshort{sgn}) \cite{Serre1953, Serre1953a, Green1976, Green1974} ($\SGN$),
  \item \nm{Boussinesq}-type weakly nonlinear and weakly dispersive model ($\Bouss$) \cite{Boussinesq1877, DMII, Brocchini2013}
  \item Nonlinear shallow water (nonlinear non-hydrostatic or \nm{Saint-Venant} or \nm{Airy}) equations (\acrshort{nswe}) \cite{SV1871} ($\SW$).
\end{itemize}
In the present study we particularly focus on three models: $\Po$, $\SGN$ and $\SW$. Our choice can be explained, firstly, by the fact that we are interested by the propagation of relatively long waves. Secondly, $\SW$ and $\SGN$ equations are widely used in the wave modelling practice. Finally, we had to include the base model $\Po$ to have a reference solution to assess correctly the predictions of various approximate models. Moreover, nowadays, only approximate depth-integrated models can be applied on large scales due to the prohibitive computational cost of complete governing equations (\ie $\Po$, $\Eu$, $\NS$) with free surface. That is why the limitations of various approximations have to be understood in order to apply them only in situations where they are relevant.

When we consider depth-integrated models such as $\SW$ and $\SGN$, the flow domain is divided into the outer\footnote{Outside of the floating body when one makes the vertical projection along the gravity acceleration vector.} and inner\footnote{Under the floating body under the same projection.} parts \cite{Khakimzyanov2018a}. This division comes from the fact that in the outer domain the flow is in the free surface regime while in the inner part it is rather a closed-channel flow. It has the implication on the choice of dynamic variables which describe the flow in various regions. For instance, in the outer domain we describe it with $(\,\H,\,\ub\,)\,$, while in the inner domain it will be some pressure-related quantities together with $\ut\,$. Here, $\ub$ and $\ut$ are depth-averaged horizontal velocities in outer and inner domains correspondingly and $\H$ is the total fluid layer depth (in the outer domain). The fluid layer depth in the inner domain is supposed to be known. Moreover, in \cite{Khakimzyanov2018a} we proposed several conditions which allow to glue the solutions at the boundary between two domains.

The idea behind this study is to consider a hierarchy of models instead of working with a single (favourite) one. The increasing complexity in the hierarchy allows us to determine the applicability limits of various models and, thus, to find the best trade-off between the model complexity/accuracy depending on the situation being modelled. A similar research effort has been undertaken for free water wave propagation in \cite{Khakimzyanov2019} over globally flat and spherical geometries.

In the modelling practice, one wishes to obtain the most accurate predictions by spending the least \acrshort{cpu} time to produce them. However, it is difficult to assess the accuracy of obtained results with lower order models without recomputing the same case with a higher order one and corroborating the results. Of course, it cannot be done all the time. That is why we need to elaborate some general recommendations and rules of thumb to accompany the engineers and modelling practitioners. We are well aware that precise limits of mathematical models applicability are inaccessible and depend on the user error tolerance. One can mention a few general principles. For example, it is well known that the application of linear models should be limited to small amplitude waves, shallow water models are applicable only to the modelling of long waves, \etc In reality, the situation is even more complicated because it is not difficult to give examples where such general principles provide misguidance. Let us consider, for example, the problem of the solitary wave run-up on a vertical wall \cite{Chen2015a, Cooker1997, Byatt-Smith1988}. Here, the simplest mathematical model is given by the analytical formula proposed in \cite{Su1980} and based on the small amplitude assumption. However, it turns out that this approximate formula gives reasonable predictions for solitary waves of moderate and large amplitudes \cite{Cooker1997, Khakimzyanov2018, Khakimzyanov2016}. Another classical hydraulic example is the so-called dam-break problem. It is a very complex phenomenon whose modelling is performed using various models. For example, the wave/wall interaction and run-up problem after a dam-break event was critically investigated in \cite{Dutykh2010c}. The limitations of the $\SW$ model, when it comes to hitting the wall, have been demonstrated against the two-fluid $\NS$ system with the air/water interface resolved by the \acrfull{vof} method \cite{Hirt1981}. A common sense says that the standard shallow water models should not be applicable here since they were derived under the explicit assumption of slow variation of flow parameters in space and in time, while in the dam-break problem we have an abrupt local change in the flow, especially at the initial rupture stages. However, if the goal of the modelling consists in predicting the main front height and propagation speed only, then the classical $\SW$ model, even in the \acrfull{1d} case, turns out to be quite helpful. To make a conclusion, the problem of delimiting a mathematical model applicability domain is extremely complicated and practically important in the same time.

In this manuscript we investigate a very particular instance of this problem. Namely, we take a hierarchy of three models, and we try to determine their applicability limits in the simplest wave/body interaction problem: a solitary wave run-up on a partially immersed fixed body of rectangular cross-section. Despite the geometric simplicity of the considered solid body, this problem remains practically important since the projects of floating highly technological offshore structures are being developed around the world. We can mention the offshore \acrfull{npp} in \nm{Russian} far east regions and floating \acrfull{lng} storage tanks to give a few important applications. The design of such mega-structures has to take into account all possible risks including the risk of tsunamis, as the \nm{Tohoku} 2011 event notoriously demonstrated to us \cite{Mori2011, Sublime2021}. The result of a tsunami wave/body interaction may be catastrophic for the environment when the body is a floating \acrshort{npp}.

Let us review the available scientific results regarding the modelling of wave/fixed body interaction problem. First of all, we would like to mention the seminal historical study by \nm{Mei} \& \nm{Black} (1969) \cite{Mei1969} where this problem was investigated in the framework of the linear potential flow model. The analytical approximations were derived and some practical conclusions were drawn based on these formulas. The theoretical investigation of this problem is much more recent \cite{Lannes2017}. The numerical investigations are slightly more numerous. For example, the transformation of a fixed amplitude wave on a floating body was investigated in \cite{Lin2006a} using the $\sigma-$co\"ordinate method. The incorporation of floating structures into the \texttt{NHWAVE} model was discussed in \cite{Orzech2016}. A spectral element method based on unstructured meshes was proposed in \cite{Engsig-Karup2017} to model the solitary wave run-up on a fixed body of rectangular cross-section. The fluid was modeled using the potential flow equations $\Po$ in the spirit of the earlier study \cite{Kamynin2010}. The numerical simulations using the $\SW$ and $\Po$ models simultaneously was done in \cite{Khakimzyanov2002} where the solitary wave run-up on a partially immersed fixed body of rectangular section was investigated. Of course, the body is supposed not to touch the bottom.

We would like to mention also some significant works on the numerical modelling of the surface wave interaction with (fixed) floating bodies of rectangular cross-section. A detailed study of a solitary wave interaction with a fixed partially immersed floating body using an integrated analytical-numerical method was presented in \cite{Lu2015}. In the outer domain (\cf \cite{Khakimzyanov2018a}), the \acrshort{1d} generalized \nm{Boussinesq} equations were solved using the finite difference method. In this way, the free surface excursion and the depth-averaged velocity potential were found. In the inner domain (\cf \cite{Khakimzyanov2018a}), the \acrshort{2d} \nm{Laplace} equation for the velocity potential is solved using a spectral numerical method along with appropriate impermeability conditions on solid boundaries (on the body and the bottom). On the interfaces between the inner and outer domains, the values of the velocity potential along with some (horizontal) derivatives are required to be continuous. Using this numerical method, the dependence of the reflected and transmitted wave amplitudes on other parameters of the problem is studied. Additionally to numerous numerical experiments, some laboratory measurements were performed as well for several lengths of the floating body and several incident wave amplitudes. The same problem was solved numerically with finite difference methods in \cite{Khakimzyanov2002} without dividing the computational domain in several sub-domains. Namely, the problem was solved in the framework of the \acrshort{2d} $\Po$ formulation using curvilinear grids. Similar results were achieved also in \acrshort{2d} in \cite{Chang2017} and in \acrshort{3d} in \cite{Chang2017a}. The common conclusion of all these studies is that the floating body length and its immersion depth greatly influence the wave field in front and behind the obstacle. It goes without saying that maximal values of the wave run-up on both sides is also sensitive to these parameters.

In the present work we also consider the same problem of the solitary wave interaction with a fixed floating partially immersed body. However, the particularity of our approach consists in considering this problem in the framework of a hierarchy of mathematical models described hereinabove: $\Po$, $\SGN$ and $\SW$. Of course, the considered waves must be in the shallow water regime to make the comparisons meaningful. In various long wave models, the computational domain has to be divided into the inner and outer sub-domains. The communication between these domains and the global solution construction are realized using the so-called compatibility conditions on the common interfaces between sub-domains as it was explained in Part~I of this study \cite{Khakimzyanov2018a}. Another particularity of our work consists in the fact that we provide a detailed description of the numerical methods and algorithms for all the models we consider in the present study. Our goal is to provide the reader with the complete information so that our methods can be used in practice by other researchers as well. As numerical experiments, we study the influence of the obstacle elongation and the immersion depth on the incident wave run-up and the wave field in wave-ward and lee-ward sides. As a particular case, we consider also the configuration where a vertical wall is located just behind the floating obstacle. In this case, we show an unexpected result: under certain conditions, the maximal wave run-up on the vertical wall in the presence of a floating body can be higher than in the free space. Hence, a floating body can be an amplifying factor in producing extreme wave run-up heights.

The purpose of using a hierarchy of models consists in being able to perform the comparisons among various models predictions. Based on these comparisons, we can issue some recommendations regarding the applicability ranges of different approximations. However, we have to say that, strictly speaking, our recommendations are valid only in the situations similar to those studied in our manuscript.

The present manuscript is organized as follows. The mathematical problem is formulated in \acrshort{2d} for the $\Po$, $\SGN$ and $\SW$ models in Section~\ref{sec2}. The developed numerical algorithms for these models are presented in Section~\ref{sec3}. The calculation results are discussed in Section~\ref{sec4}.  Finally, in Section~\ref{sec:concl} we outline the main conclusions and perspectives of the present study.

\section{Problem formulation}
\label{sec2}

\begin{figure}[h!]
\centering
\includegraphics[width=0.7\textwidth]{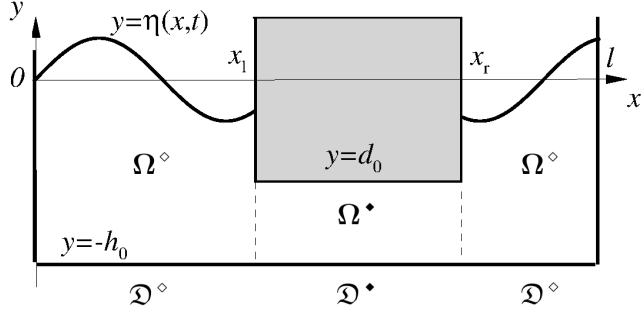}
\\
{\caption{Flow domain diagram in the problem of interaction of surface waves with a semi-submerged stationary object located in a basin with a horizontal bottom and vertical impermeable walls}
\label{scheme_of_task}}
\end{figure}

In contrast to the paper \cite{Khakimzyanov2018a}, which considers a three-dimensional mathematical formulation of the problem in the Cartesian coordinate system $Ox_1x_2y$, in this study we assume that the flow parameters and the geometry of the region do not depend on one of the horizontal coordinates, for certainty from $x_2$. We will use the notation $x$ for the first horizontal coordinate $x_1$, $u(x,t)$ for the first velocity component of shallow water models, $U(x,y,t)$ for models of Euler equations, while the second velocity components are assumed to be zero: $u_2\equiv 0$, $U_2\equiv 0$. Moreover, we assume that both the bottom of the basin and the bottom of the body are horizontal and stationary and are defined by the equations $y=-h_0={\textrm {const}}$ and $y=d_0={\textrm {const}}$ ($-h_0<d_0<0$), respectively. Thus, we consider a stationary semi-submerged rectangular body with lateral vertical faces located at distances $x_{l}$ and
$x_{r}$ from the left side of the pool ($0<x_l<x_r<l$), where $x=0$ and $x=l$ are coordinates of the left and right side vertical walls of the basin. With the assumptions made, the flow region diagram looks like it is shown in Fig.~\ref{scheme_of_task}.

For the simplified case under consideration, we will use the same notation as in the general case \cite{Khakimzyanov2018a}. Thus, $\Omega(t)=\Omega_e(t)\cup\Omega_i$, ${\cal D}={\cal D}_e\cup{\cal D}_i$, where
\begin{equation*}
{\cal D}_e=[0, x_l]\cup [x_r, l],\quad {\cal D}_i=(x_l, x_r),
\end{equation*}
\begin{equation*}
{\Omega}_e(t)=\left\{(x,y)\in R^2\Big{|}\ x\in {\cal D}_e, \ -h_0\le y\le \eta(x,t)\right\},
\end{equation*}
\begin{equation*}
{\Omega}_i=\left\{(x,y)\in R^2\Big{|}\ x\in {\cal D}_i, \ -h_0\le y\le d_0\right\},
\end{equation*}
$y=\eta(x,t)$ ($x\in{\cal D}_e$) is the free surface equation.
Also, $\Gamma_0 =\left\{0, l\right\}$, $\Gamma=\left\{x_l, x_r\right\}$.

In that way, we solve the Euler equations, assuming the potentiality of the flow, in the two-dimensional domain $\Omega(t)$ and the shallow water equations in the one-dimensional domain ${\cal D}$. Below we present these equations using the notation introduced in \cite{Khakimzyanov2018a}.

\subsection{Potential flow model}
The formulation of the problem for the nonlinear model of potential (for the $\Po$ model) 2D flows differs from the one given in \cite{Khakimzyanov2018a}: instead of the three-dimensional Laplace operator, a two-dimensional one is used: $\Delta=\partial/\partial x^2+ \partial/\partial y^2$. Therefore, the equations can be written as:
\begin{equation}
\Phi_{xx}+\Phi_{yy}=0,
\label{2D_11_1.14}
\end{equation}
\begin{equation}
\left(\eta_t+U \eta_x-V\right)\big|_{y=\eta(x,t)} =0,\qquad {x}\in  {\cal D}_e,
\label{2D_11_1.5phi}
\end{equation}
\begin{equation}
\big(\Phi_t +\frac{U^2 + V^2}{2} +  g\eta\big)\big|_{y=\eta( x,t)}=0,\qquad {x}\in  {\cal D}_e,
\label{2D_11_1.16}
\end{equation}
where
\begin{equation}
U=\Phi_x, \quad V=\Phi_y.
\label{2D_11_1.1.71}
\end{equation}

The impermeability conditions are simplified to the following:
\begin{equation}
\Phi_y\big|_{y=-h_0}=0,\qquad {x}\in {\cal D},
\label{2D_11_1.15}
\end{equation}
\begin{equation}
\Phi_y\big|_{y=d_0} =0,\qquad {x}\in {\cal D}_i,
\label{2D_11_bodybottom_2}
\end{equation}
\begin{equation}
\Phi_x=0, \qquad  x\in \Gamma_0,\quad -h_0\le y\le \eta(x,t),
\label{2D_Gamma_0}
\end{equation}
\begin{equation}
\Phi_x=0, \qquad x\in \Gamma,\quad  d_0\le y\le \eta(x,t).
\label{2D_Gamma}
\end{equation}

\subsection{Fully nonlinear weakly dispersive shallow water equations}

In the one-dimensional case with a horizontal bottom, the $\SGN$ equations in the outer region ${\cal D}_e$ \cite{Khakimzyanov2018a} can be written as:
\begin{equation}
\H_t +  (\H u)_x=0,
\label{Pt_2_cont_eq1}
\end{equation}
\begin{equation}
(\H u)_t+(\H u^2)_x+\frac{p_x}{{\rho}}=0,
\label{Pt_2_mov_eq1}
\end{equation}
where $\H=h_0+\eta$, ${\rho}={\textrm {const}}$ is the fluid density,
\begin{equation}
p=\rho g\frac{H^2}{2}-\Pnh,
\label{1D_pDe}
\end{equation}
$\Pnh$ is the dispersion component of the column-integrated pressure $p$,
\begin{equation}
\Pnh=\rho\frac{\H^3}{3}R_1,  \quad R_1=u_{xt}+uu_{xx}-u_x^2.
\label{1D_p_disp}
\end{equation}
The numerical algorithm also uses the non-divergent form of the equation of motion
\begin{equation}
u_t+uu_x+\frac{1}{{\rho}\H}p_x=0,
\label{Pt_2_mov_eq2}
\end{equation}
and the equation for the dispersion component of the pressure \cite{Khakimzyanov2016}, which in the one-dimensional case with a horizontal bottom has a very simple form
\begin{equation}
\left( k{\Pnh}_x\right)_x- k_0 {\Pnh}= F,
\label{curve_phi_1}
\end{equation}
where $k=1/\H$, $k_0={3}/{\H^3}$, $F={\rho}g\eta_{xx}+2{\rho}u_x^2$.

The boundary conditions \cite{Khakimzyanov2018a} on the outer boundary $\Gamma_0$ can be simplified:
\begin{equation}
u=0, \quad  \eta_x=0, \quad \Pnh_x=0, \qquad x\in \Gamma_0.
\label{Pt_2_Gamma_0}
\end{equation}

In the inner region ${\cal D}_i$, which has a common boundary $\Gamma$ with the outer region ${\cal D}_e$, the system of $\SGN$ equations turns into a system of equations for the intrachannel flow of an ideal incompressible fluid \cite{Khakimzyanov2018a}. For the one-dimensional problem with horizontal and fixed bottom and bottom of the body, this system can be written as:
\begin{equation}
u_x=0, \quad u_t+uu_x+\frac{1}{{\rho}S_0}p_x=0, \qquad x\in {\cal D}_i,
\label{1D_NLD_to_NSWE_a}
\end{equation}
where $S_0=h_0+d_0$. The first equation means the independence of the velocity under the body from the coordinate $x$, which is natural for the flow of an incompressible fluid in a channel of constant cross-section. Thus, the velocity under the body depends only on the time $t$. Using the $Q$ designation introduced in \cite{Khakimzyanov2018a} for mass flow, one can rewrite system (\ref{1D_NLD_to_NSWE_a}):
\begin{equation*}
{\rho}S_0 u(t)=Q(t), \quad \dot{Q}+p_x=0, \qquad x\in {\cal D}_i.
\end{equation*}
Integrating the second equation over the domain ${\cal D}_i$ we obtain the ODE
\begin{equation}
\dot{Q}(t)=-\frac{1}{L}\left(p\big|_{x_{r}-0}-p\big|_{x_{l}+0}\right),
\label{1D_NLD_to_NSWE_b}
\end{equation}
where $L=x_r-x_l$ is the length of the body in the horizontal direction, $p|_{x_{l}+0}$ and $p|_{x_{r}-0}$ are the limits of the internal region ${\cal D}_i$ 
pressure $p$ at the points $\Gamma$. Equation (\ref{1D_NLD_to_NSWE_b}) indicates that the change in fluid flow under the body is due to the difference in pressure values at the boundary of the inner region ${\cal D}_i$. The pressure itself is a linear function of the variable $x$ at each moment of time $t$ in ${\cal D}_i$:
\begin{equation}
p(x,t)=\frac{x_r-x}{L}p(x_{l}+0,t)+\frac{x-x_l}{L}p(x_{r}-0,t), \qquad x\in {\cal D}_i, \quad \forall t\ge 0.
\label{1D_p_in_Di}
\end{equation}

At the $\Gamma$ boundary, the condition \cite{Khakimzyanov2018a} for the flow in the outer region is used:
\begin{equation}
\eta_x\big|_{x_{l}-0}=0, \quad \eta_x\big|_{x_{r}+0}=0
\label{1D_NLD_Deta_dx}
\end{equation}
and also the conditions for the coupling of fluid flows in the external and internal regions.
In the formula (\ref{1D_NLD_Deta_dx}) and further, the designations $\ \cdot\ |_{x_{l}-0}$, $\ \cdot\ |_{x_{r}+0}$ are used for the values of dependent variables and their derivatives at points $\Gamma$ that are limits from the side of the external region ${\cal D}_e$.

Two types of compatibility conditions are proposed in \cite{Khakimzyanov2018a}. Both types contain the same condition for the mass flow rate $Q$. In the model one-dimensional problem considered here, this condition is written as:
\begin{equation}
\rho \H u\big|_{x_{l}-0}=Q=\rho \H u\big|_{x_{r}+0}
\label{1D_mass_conj_1}
\end{equation}
and expresses in mathematical form the fact that the mass of the incompressible fluid flowing in from the left under the body (flowing out from under the body on the left) is equal to
the mass of the fluid flowing out from under the body on the right (flowing in from the right under the body) and both of these quantities are equal to the mass flow rate $Q(t)$ of the fluid under the body.

In the first type of compatibility conditions, in addition to (\ref{1D_mass_conj_1}) and (\ref{1D_NLD_to_NSWE_b}), pressure continuity conditions  \cite{Khakimzyanov2018a} are also used on the common boundary $\Gamma$, which can be written in the one-dimensional case as:
\begin{equation}
p\Big|_{x_{l}+0} = {\rho}gS_0\Big(h_0-\frac{S_0}{2}+\eta\Big|_{x_{l}-0}\Big) + S_0\frac{(S_0^2-3\H^2)\Pnh}{2\H^3}\Big|_{x_{l}-0},
\label{Pt_2_conj_p_left}
\end{equation}
\begin{equation}
p\Big|_{x_{r}-0} = {\rho}gS_0\Big(h_0-\frac{S_0}{2}+\eta\Big|_{x_{r}+0}\Big) + S_0\frac{(S_0^2-3\H^2)\Pnh}{2\H^3}\Big|_{x_{r}+0}.
\label{Pt_2_conj_p_right}
\end{equation}
For brevity, the set of compatibility conditions (\ref{1D_NLD_to_NSWE_b}),  (\ref{1D_mass_conj_1}), (\ref{Pt_2_conj_p_left}), (\ref{Pt_2_conj_p_right}) will be referred to as compatibility conditions (C1).

In the second type of compatibility conditions (in \cite{Khakimzyanov2018a} they are presented as alternative compatibility conditions), the values of total energy in the outer and inner regions are connected on the common boundary of $\Gamma$:
\begin{equation}
\Big({\cal E}+\frac{p}{S_0}\Big)\Big|_{x_{l}+0}=\Big({\cal E}+\frac{p}{\H}\Big)\Big|_{x_{l}-0}, \quad
\Big({\cal E}+\frac{p}{S_0}\Big)\Big|_{x_{r}-0}=\Big({\cal E}+\frac{p}{\H}\Big)\Big|_{x_{r}+0},
\label{Alter_NLD_conj_1}
\end{equation}
where
\begin{equation*}
{\cal E}\Big|_{x_{l}+0}={\rho}\Big(\frac{1}{2}u^2\Big|_{x_{l}+0}+g\frac{d-h_0}{2}\Big)={\rho}\Big(\frac{1}{2}u^2\Big|_{x_{r}-0}+g\frac{d-h_0}{2}\Big) = {\cal E}\Big|_{x_{r}-0}\equiv {\cal E}_i(t),
\end{equation*}
\begin{equation*}
{\cal E}\Big|_{x_{l}-0}={\rho}\Big(\frac{u^2}{2}+\frac{\H^2}{6}u_x^2+g\frac{\eta-h_0}{2}\Big)\Big|_{x_{l}-0},
\qquad {\cal E}\Big|_{x_{r}+0}={\rho}\Big(\frac{u^2}{2}+\frac{\H^2}{6}u_x^2+g\frac{\eta-h_0}{2}\Big)\Big|_{x_{r}+0}.
\end{equation*}
Here ${\cal E}_i(t)$ is the total energy of the fluid particles in the flow under the body. It was stated above that the velocity under the body is constant, it does not depend on the variable $x$ but changes with time. Therefore, the value of ${\cal E}_i$ depends only on $t$.

Using (\ref{Alter_NLD_conj_1}) and the expression (\ref{1D_pDe}) we obtain
\begin{equation}
p\Big|_{x_{l}+0}={\rho}S_0\Big(\frac{u^2}{2}+\frac{\H^2}{6}u_x^2+g\eta-\frac{\Pnh}{{\rho}\H}\Big)\Big|_{x_{l}-0} - S_0{\cal E}_i(t),
\label{Alter_NLD_conj_2}
\end{equation}
\begin{equation}
p\Big|_{x_{r}-0}={\rho}S_0\Big(\frac{u^2}{2}+\frac{\H^2}{6}u_x^2+g\eta-\frac{\Pnh}{{\rho}\H}\Big)\Big|_{x_{r}+0} - S_0{\cal E}_i(t).
\label{Alter_NLD_conj_3}
\end{equation}
Further, the set of compatibility conditions (\ref{1D_NLD_to_NSWE_b}), (\ref{1D_mass_conj_1}), (\ref{Alter_NLD_conj_2}), (\ref{Alter_NLD_conj_3}) will be succinctly denoted as (C2).

\subsection{Dispersionless shallow water equations}

For $\SW$, equations (\ref{Pt_2_cont_eq1}), (\ref{Pt_2_mov_eq1}), boundary conditions (\ref{Pt_2_Gamma_0}), (\ref{1D_NLD_Deta_dx}) and compatibility conditions (C1) and (C2) retain their form, while everywhere one should put $\Pnh\equiv 0$ and neglrct the dispersive terms in expressions (\ref{Alter_NLD_conj_2}), (\ref{Alter_NLD_conj_3}). Thus, conditions (C1) become
\begin{equation}
p\big|_{x_{l}+0} = {\rho}gS_0\Big(h_0-\frac{S_0}{2}+\eta\big|_{x_{l}-0}\Big),\quad
p\big|_{x_{r}-0} = {\rho}gS_0\Big(h_0-\frac{S_0}{2}+\eta\big|_{x_{r}+0}\Big).
\label{Pt_2_conj_p_SW}
\end{equation}
For the alternative approach, we can rewrite the relations (\ref{Alter_NLD_conj_2}), (\ref{Alter_NLD_conj_3}) in the following form:
\begin{equation}
p\Big|_{x_{l}+0}={\rho}S_0\Big(\frac{u^2}{2}+g\eta\Big)\Big|_{x_{l}-0} - S_0{\cal E}_i(t),
\quad p\Big|_{x_{r}-0}={\rho}S_0\Big(\frac{u^2}{2}+g\eta\Big)\Big|_{x_{r}+0} - S_0{\cal E}_i(t).
\label{Alter_SW_conj_5}
\end{equation}

\subsection{Consistent initial conditions for models of different spatial dimensions}

For the equations of $\Po$ and shallow water equations described above, it is necessary to set initial conditions.
To compare rationally the numerical results obtained within the framework of different hierarchical chain models, it is necessary to set the same initial conditions for them. However, the initial conditions at $t=0$ for the $\SW$ and $\SGN$ equations must be set for the velocity and shape (elevation) of the initial wave:
\begin{equation}
u(x,0)=u_0(x), \quad \eta(x, 0)=\eta_0(x),
\label{init_cond_1K}
\end{equation}
while for the two-dimensional $\Po$ model, the velocity vector field ${\vect U}_0=\left(U_0, V_0\right){}^{\top}$ and the initial wave elevation must be set:
\begin{equation}
U(x,y,0)=U_0(x,y), \quad V(x,y,0)=V_0(x,y), \quad \eta(x, 0)=\eta_0(x).
\label{initial_PFK}
\end{equation}
As can be seen from formulas~(\ref{init_cond_1K}), (\ref{initial_PFK}), the initial data differ (elevation and velocity in (\ref{init_cond_1K}), elevation and the two components of the velocity vector in (\ref{initial_PFK})), so we cannot talk about a complete coincidence of the initial data, we can only talk about the desirability of some agreement of the initial data for the considered one-dimensional and two-dimensional models.

Let us explain what we mean by consistency of initial data for models of different spatial dimensions and how these consistent initial data are constructed.
In this subsection, we will not consider the presence of a semi-submerged body, moving the discussion of this issue to the next subsection, where it will be shown how to adjust the initial data for the shallow water models ($\SGN$ and $\SW$) to take into account the presence of a body (so that the initial data satisfy the compatibility conditions). Moreover, we will consider the notion of initial data consistency for an infinite region, i.e., at $x\in (-\infty, \infty)$.

So, let the initial data (\ref{init_cond_1K}) for the one-dimensional model be given. We will say that the initial data (\ref{init_cond_1K}), (\ref{initial_PFK}) are consistent if:

1)~for the $\Po$ model, the function $\eta_0(x)$ in (\ref{initial_PFK}) is the same as in~(\ref{init_cond_1K});

2)~the velocity vector field ${\vect U}_0(x,y)$ in (\ref{initial_PFK}) is potential (vortex-free);

3)~after averaging, the horizontal component of velocity $U_0(x,y)$ coincides with $u_0(x)$, i.e.
\begin{equation}
\frac{1}{\H_0(x)}\int\limits_{-h_0}^{\eta_0(x)} {U_0}(x,y)\; dy=u_0(x),
\label{init_cond_5}
\end{equation}
where $\H_0(x)=h_0+\eta_0(x)$.

To construct consistent initial data, we will use formulas that allow us to reconstruct (restore) \cite{Khakimzyanov2018a,Khakimzyanov2016c,Khakimzyanov2016} components of the velocity vector ${\vect U}_0(x,y)$ of the two-dimensional problem from the initial data for the one-dimensional $\SGN$ model with a certain accuracy:
\begin{equation}
U_0(x,y)=u_0(x)+\Big(\frac{\H_0(x)^2}{6}-\frac{(y+h_0)^2}{2}\Big)u_0^{\prime\prime}(x), \quad  V_0(x,y)=-(y+h_0)u_0^{\prime}(x).
\label{init_cond_3}
\end{equation}
Obviously, with this reconstruction the requirement (\ref{init_cond_5}) is satisfied. In addition, the velocity vector field ${\vect U}_0(x,y)$ is potential:
\begin{equation*}
\frac{\partial V_0}{\partial x}(x,y)-\frac{\partial U_0}{\partial y}(x,y) \equiv 0.
\end{equation*}
Thus, the initial condition (\ref{initial_PFK}) with the components (\ref{init_cond_3}) of the velocity vector is consistent with (\ref{init_cond_1K}), i.e.
the initial data are consistent for models with different spatial dimensions.
Here are some simple examples of consistent initial data.
\\

{\bf Example 1.} Let the initial functions in the condition (\ref{init_cond_1K}) be given as:
\begin{eqnarray}
 &&\displaystyle
  \eta_0(x)=a_0{\textrm {sech}}^2(X), \label{Full_an_sol_eta} \\[2mm]
 &&\displaystyle
u_0(x)=c_0\frac{\eta_0(x)}{\H_0(x)}, \label{Full_an_sol_u}
\end{eqnarray}
where $X= k(x-x_0)$, $a_0$ is the initial wave amplitude,  $x=x_0$ is the position of its peak, $0<x_0<x_l$,
\begin{equation*}
c_0=\sqrt{g(a_0+h_0)}, \quad  k=\frac{1}{h_0}\sqrt{\frac{3a_0}{4(a_0+h_0)}}.
\end{equation*}
Then $\SGN$ equations (\ref{Pt_2_cont_eq1}), (\ref{Pt_2_mov_eq1}) have exact solution $\eta(x,t)=\eta_0(x-c_0t)$, $u(x,t)=u_0(x-c_0t)$ describing the solitary wave propagating with constant speed $c_0$ over the horizontal bottom. Since for the functions (\ref{Full_an_sol_eta}), (\ref{Full_an_sol_u}) the following formulas are true:
\begin{equation}
u_0^\prime(x)=\frac{h_0c_0}{\H_0^2(x)}\eta_0^\prime(x), \quad u_0^{\prime\prime}(x)=\frac{h_0c_0}{\H_0^3(x)}\left(\H_0(x)\eta_0^{\prime\prime}(x)-2\left(\eta_0^{\prime}(x)\right)^2\right),
\label{formulas_for_u0}
\end{equation}
\begin{equation}
\eta_0^\prime(x)=-2k\eta_0(x)\tanh (X), \quad \eta_0^{\prime\prime}(x)=2k^2\eta_0(x)\left(2-\frac{3\eta_0(x)}{a_0}\right),
\label{formulas_for_eta0}
\end{equation}
then, according to (\ref{init_cond_3}), consistent initial data are obtained when we set the velocity components in (\ref{initial_PFK}) as:
\begin{equation}
\begin{array}{c}
\displaystyle
U_{0}(x,y)  =  u_0(x) \times \\[3mm]
\displaystyle
\times  \left[1+\Big(\frac{1}{4}-\frac{3}{4}\frac{(y+h_0)^2}{\H_0^2(x)}\Big)
\frac{\H_0(x)\big(2a_0-3\eta_0(x)\big)+ 4\big(\eta_0(x)-a_0\big)\eta_0(x)}{h_0(a_0+h_0)}\right],\\[3mm]
\displaystyle
V_{0}(x,y)  =  \sqrt{3a_0g}\;\frac{\eta_0(x)}{\H_0^2(x)}\;(y+h_0) \tanh(X)\;.
\end{array}
\label{init_cond_6_new}
\end{equation}

Note that after elementary transformations, formulas (\ref{init_cond_6_new}) for calculating the initial velocity components in the $\Po$ model coincide with those used earlier in~the work~\cite{Khakimzyanov2018}. As shown by numerical calculations within the $\Po$ model, initial data (\ref{Full_an_sol_eta}), (\ref{init_cond_6_new}) give at $t>0$ the solitary wave moving at a constant speed, with the shape of the moving wave being slightly different from initial shape (\ref{Full_an_sol_eta}).
\\

{\bf Example 2.} Let the initial data (\ref{init_cond_1K}) have a finite support and describe the single wave of length $\lambda>0$ with elevation
\begin{equation}
\eta_0(x)=\left \{\begin{array}{cc}
\displaystyle
\frac{a_0}{2} \Big( 1+\cos (X) \Big),  & \displaystyle \left|x-x_0\right|\le {\lambda}/{2},\\
0, & \displaystyle \left|x-x_0\right|>{\lambda}/{2}
\end{array} \right.
\label{test_eta_0}
\end{equation}
and velocity (\ref{Full_an_sol_u}).
Here we use the same notations as in formulas (\ref{Full_an_sol_eta}), (\ref{Full_an_sol_u}), except for one: $k=2\pi/\lambda$. At $\left|x-x_0\right|\le {\lambda}/{2}$  formulas (\ref{formulas_for_u0}) are valid, and instead of (\ref{formulas_for_eta0}) we should use expressions
\begin{equation*}
\eta_0^\prime(x)=-\frac{k}{2}a_0\sin (X), \quad \eta_0^{\prime\prime}(x)=\frac{k^2}{2}\big(a_0-2\eta_0(x)\big).
\end{equation*}
Thus, the consistency of conditions (\ref{init_cond_1K}) with (\ref{initial_PFK}) will take place if the components of the initial velocity in the $\Po$ model are calculated by the formulas
\begin{equation}
\begin{array}{c}
\displaystyle
U_{0}(x,y)  =  u_0(x) +  \\[3mm]
\displaystyle
+c_0\frac{h_0}{\H_0(x)}\Big(\frac{1}{3}-\frac{(y+h_0)^2}{\H_0^2(x)}\Big)
\frac{\H_0(x)\big(a_0-2\eta_0(x)\big)+ 4\big(\eta_0(x)-a_0\big)\eta_0(x)}{\left(\lambda/\pi\right)^2},\\[4mm]
\displaystyle
V_{0}(x,y)=c_0\frac{\sqrt{\big(a_0-\eta_0(x)\big)\eta_0(x)}}{\H_0^2(x)}\;\frac{2\pi h_0}{\lambda}\;(y+h_0)\; {\textrm {sgn}}(X).
\end{array}
\label{init_cond_single_wave}
\end{equation}

{\bf Example 3.} In this example, the initial data for the shallow water models, also as in Example 2, are set on the final support. The elevation of the free boundary at $t=0$ is still given as ``raised cosine'' (\ref{test_eta_0}), and the initial velocity is calculated by another formula:
\begin{equation}
u_0(x)=2\sqrt{g\H_0(x)}-2c_0,
\label{test_u_0}
\end{equation}
where $c_0=\sqrt{gh_0}$.
The advantage of the initial data in form (\ref{test_eta_0}), (\ref{test_u_0}) is that now the $\SW$ equations have the exact solution \cite{Pelinovsky2008,Khakimzyanov2019c}:
\begin{equation}
\eta(x,t)=\frac{1}{9g}\big({2c_0+\xi(x,t)}\big)^2-h_0, \quad u(x,t)=2\sqrt{g\H(x,t)}-2c_0,
\label{test_eta_u}
\end{equation}
until the gradient catastrophe comes. Here $\H(x,t)=\eta(x,t)+h_0$, $\xi(x,t)$ is the root $\xi$ of the nonlinear equation
\begin{equation}
\xi=3\sqrt{g\Big[h_0+\eta_0(x-{\xi}t)\Big]}-2c_0.
\label{l30_p_exact}
\end{equation}
Solution (\ref{test_eta_u}) of the $\SW$ equations describes the wave moving to the right with the constant speed $c_0$, its amplitude and length remain constant and equal to the corresponding values of the initial wave (\ref{test_eta_0}). The profile of the moving wave deforms over the time so that its leading edge steepens while its trailing one flattens. Thus, the exact solution will have the rarefaction wave and the compression wave ahead of it, leading to the gradient catastrophe.

Initial data (\ref{test_eta_0}), (\ref{test_u_0}) for shallow water models lead, according to (\ref{init_cond_3}), to the following consistent initial data for the $\Po$ model:
\begin{equation}
\begin{array}{c}
\displaystyle
U_{0}(x,y)  =  u_0(x) +  \\[3mm]
\displaystyle
+\sqrt{g\H_0(x)}\Big(\frac{1}{3}-\frac{(y+h_0)^2}{\H_0^2(x)}\Big)
\frac{\H_0(x)\big(a_0-2\eta_0(x)\big)+ \big(\eta_0(x)-a_0\big)\eta_0(x)}{\left(\lambda/\pi\right)^2},\\[4mm]
\displaystyle
V_{0}(x,y)=c_0\sqrt{\frac{\big(a_0-\eta_0(x)\big)\eta_0(x)}{h_0\H_0(x)}}\;\frac{2\pi}{\lambda}\;(y+h_0)\; {\textrm {sgn}}(X).
\end{array}
\label{init_cond_singleSV_wave}
\end{equation}

{\bf Remark}. We can go the other way and set the initial data for the $\Po$ model and obtain on their basis the consistent initial data for the $\SGN$ model.
When using the $\Po$ model in problems with a soliton wave propagating over a horizontal bottom, it is desirable to set the initial data~(\ref{initial_PFK}) so that at $t>0$ the wave moves as a soliton: with constant speed, without changing its shape, without a ``dispersion tail''. However, exact solutions for the soliton wave in the form of finite formulas containing only elementary functions are not known for the $\Po$ model, so the soliton wave is defined approximately with some error \cite{Laitone_1960,Tanaka1986}. If you want to set the initial solitary wave for the $\Po$ model with the highest accuracy, you can use the results of studies \cite{Clamond2012b,Dutykh2013b}.

\subsection{Initial conditions in the presence of a semi-submerged body}

In the presence of the semi-submerged body, the initial conditions at $t=0$ for the $\SGN$ and $\SW$ shallow water equations are set for the velocity in the entire flow domain $D$ and for the elevation of the initial wave in the subdomain $D_e$ outside the body:
\begin{equation}
\begin{array}{cl}
\displaystyle
u(x,0)=u_0(x),  & \  x \in {\cal D}; \\[2mm]
\displaystyle
\eta(x, 0)=\eta_0(x), & \ x \in {\cal D}_e.
\end{array}
\label{init_cond_1}
\end{equation}
For the two-dimensional $\Po$ model, the velocity vector field and the free surface are set at the initial moment of time:
\begin{equation}
\begin{array}{cl}
\displaystyle
U(x,y,0)=U_0(x,y), \quad V(x,y,0)=V_0(x,y), & \ (x,y) \in \Omega(0); \\[2mm]
\displaystyle
\eta(x, 0)=\eta_0(x), & \ x \in {\cal D}_e.
\end{array}
\label{initial_PF}
\end{equation}
If the initial velocity vector field ${\vect U}_0=(U_0, V_0){}^{\top}$ is potential, then we can uniquely determine the initial values for the potential $\Phi(x,y,0)$ from it \cite{Khakimzyanov2018}.

In the presence of a semi-submerged body, it is necessary to adjust the consistent initial data for the shallow water equations obtained in the previous section so that the fluid velocity under the body is constant over $x$ (see first equation (\ref{1D_NLD_to_NSWE_a})).  In this section, we denote these adjusted initial functions by ${\tilde{\eta}}_0(x)$ and ${\tilde{u}}_0(x)$. In addition, for the shallow water equations it is necessary to set at $t=0$ the flow under the body $Q(0)$ and the rate of change of flow $\dot{Q}(0)$, and for the $\SGN$ equations to set the dispersion component of the pressure $\Pnh(x,0)=\Pnh_0(x)$, $x\in {\cal D}$.

If the initial data are given on a finite support such as in Examples 2 and 3, where it is assumed that $\lambda<x_l$, $\ \lambda/{2}<x_0<x_l-\lambda/{2}$, i.e. the initial data are concentrated on the final support $(x_0-{\lambda}/{2}, \; x_0+{\lambda}/{2})$ contained in the interval $(0, x_l)$, the initial data need not be corrected, since under the body the speed automatically is constant, namely zero. Accordingly, both $Q(0)=0$ and the rate of change of flow $\dot{Q}(0)=0$. The initial values of $\Pnh_0(x)$ are determined numerically by solving at $x\in {\cal D}_e$ the equation (\ref{curve_phi_1}).

If the initial data support is infinite, we can propose two approaches that take into account the presence of the semi-submerged body. Let us explain their essence for the initial data considered in Example 1. One can set the elevation and velocity of the fluid on the left side of the body by formulas (\ref{Full_an_sol_eta}), (\ref{Full_an_sol_u}), i.e.
\begin{equation}
{\tilde{\eta}}_0(x)=\eta_0(x), \quad {\tilde{u}}_0(x)=u_0(x), \qquad 0\le x\le x_l
\label{init_cond_2}
\end{equation}
and require the compatibility condition (\ref{1D_mass_conj_1}) to be satisfied. Then the velocity under the body will be constant, and
\begin{equation*}
{\tilde{u}}_0(x)=\frac{1}{S_0}\Big(h_0+{\tilde{\eta}}_0(x_{l}-0)\Big){\tilde{u}}_0(x_{l}-0)={\textrm {const}}, \quad x_l<x<x_r.
\end{equation*}
Again based on (\ref{1D_mass_conj_1}) for the initial data on the right side of the body, we can set
\begin{equation*}
{\tilde{\eta}}_0(x)=\eta_0(x-L), \quad  {\tilde{u}}_0(x)=u_0(x-L), \qquad x_r\le x\le l.
\end{equation*}
Then for the corrected functions ${\tilde{\eta}}_0(x)$ and ${\tilde{u}}_0(x)$ the compatibility condition (\ref{1D_mass_conj_1}) will be satisfied and the fluid velocity under the body will be constant.

In the second approach, which we will use in numerical calculations within the $\SGN$ and $\SW$ models, we do not require the compatibility condition (\ref{1D_mass_conj_1}) for the initial data, setting the initial functions to the left of the body by formulas (\ref{init_cond_2}), and setting the rest of the domain with a rest state
\begin{equation*}
{\tilde{\eta}}_0(x)\equiv 0, \quad x_r\le x\le l; \qquad {\tilde{u}}_0(x) \equiv 0, \quad x_l< x\le l.
\end{equation*}
Of course, when using the second approach, small perturbations of the solution may arise at the first moments of time, caused by the aspiration of the solution to adjust to the given compatibility conditions. Note here that minor perturbations of the flow may arise at the very first moments of $t>0$ for other reasons as well, for example, because the initial functions (\ref{Full_an_sol_eta}), (\ref{Full_an_sol_u}) of the boundary conditions (\ref{Pt_2_Gamma_0}) are not exactly satisfied.

\section{Numerical algorithms}\setcounter{equation}{0}\label{sec3}

To investigate numerically the problem of interaction of surface waves with a semi-immersed body of rectangular cross section (Fig.~\ref{scheme_of_task}), we will use the algorithms described in \cite{Khakimzyanov2018}, \cite{Khakimzyanov2016} and \cite{Khakimzyanov2016} for the $\Po$, $\SGN$  and $\SW$ models correspondingly. These papers present numerical algorithms for calculating surface waves in basins with moving or stationary walls and its bottom fragments (see also \cite{Khakimzyanov2015b,Khakimzyanov2019c}), as well as for calculating wave runup on the shore using the new algorithm  for calculation of the motion of the shoreline point \cite{Khakimzyanov2016d}. At the same time, no obstacles crossing (piercing) the free boundary were contained within the basin. The presence of a semi-immersed body requires some modification of these algorithms, and in this section we focus on these modifications only.

\subsection{Some features of the numerical algorithm for the Pot model}

\begin{figure}[h!]
\centering
\includegraphics[width=0.4\textwidth]{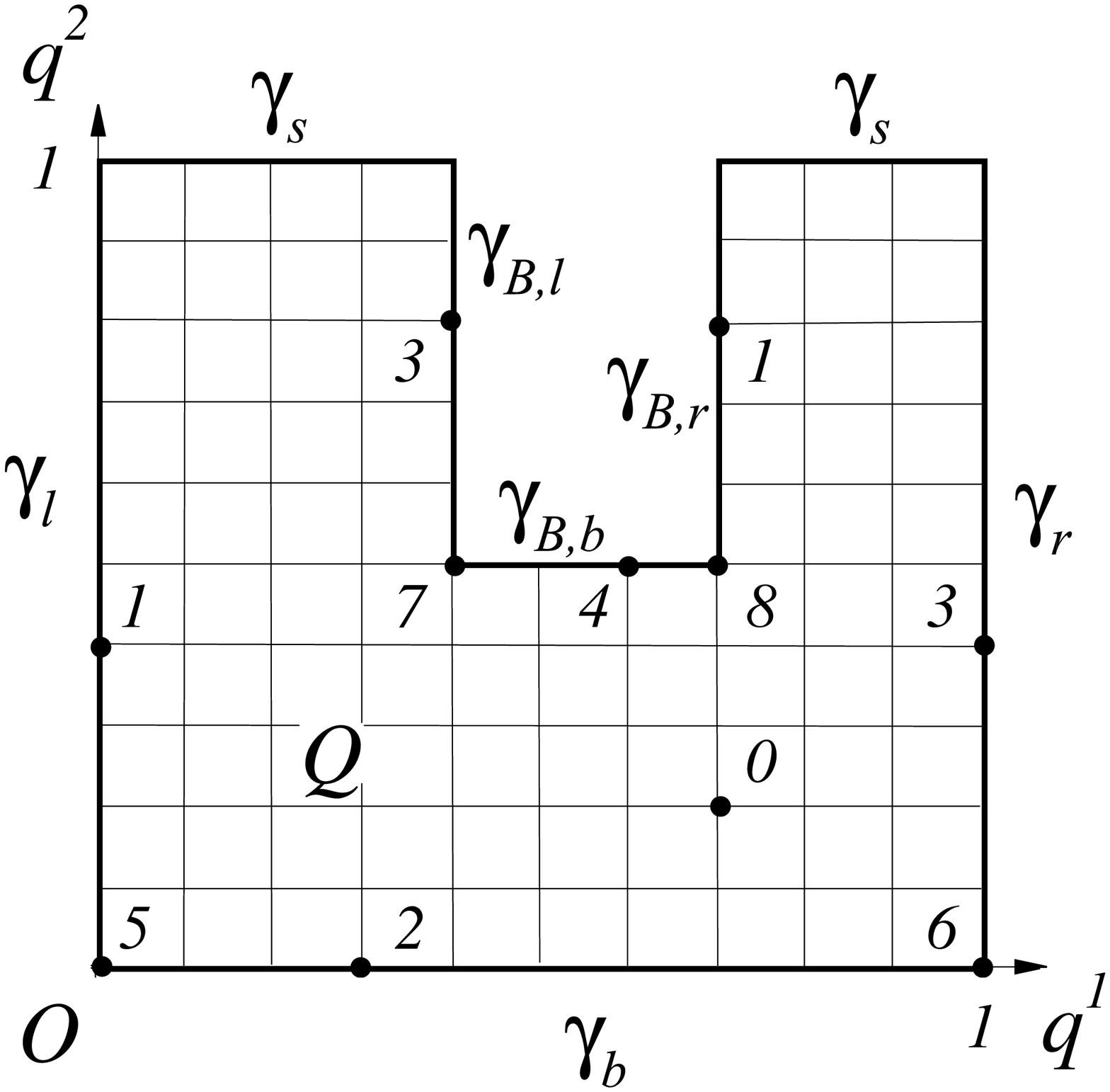}
\hspace*{-5mm}\includegraphics[width=0.32\textwidth]{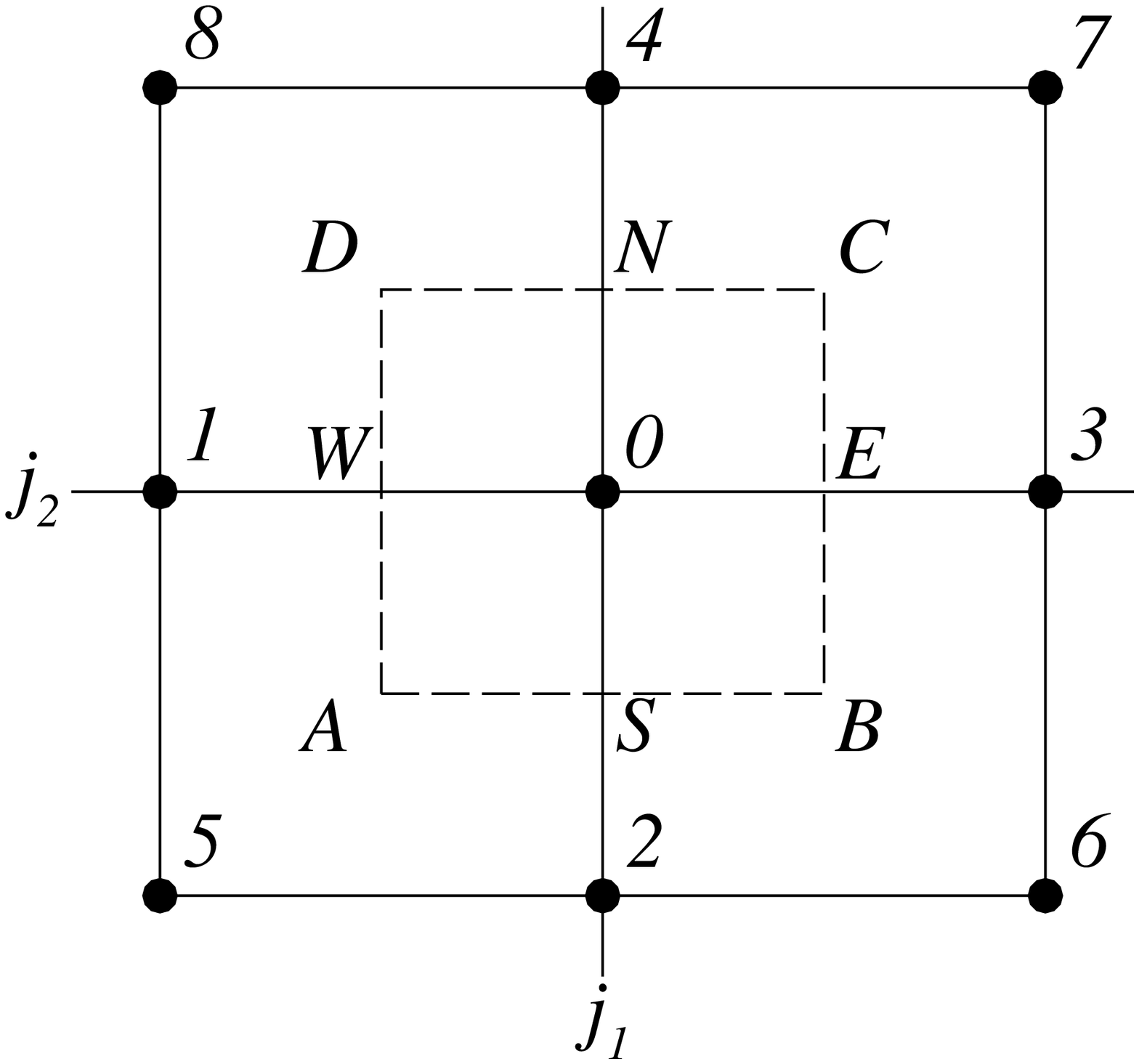}
\hspace*{-5mm}\includegraphics[width=0.32\textwidth]{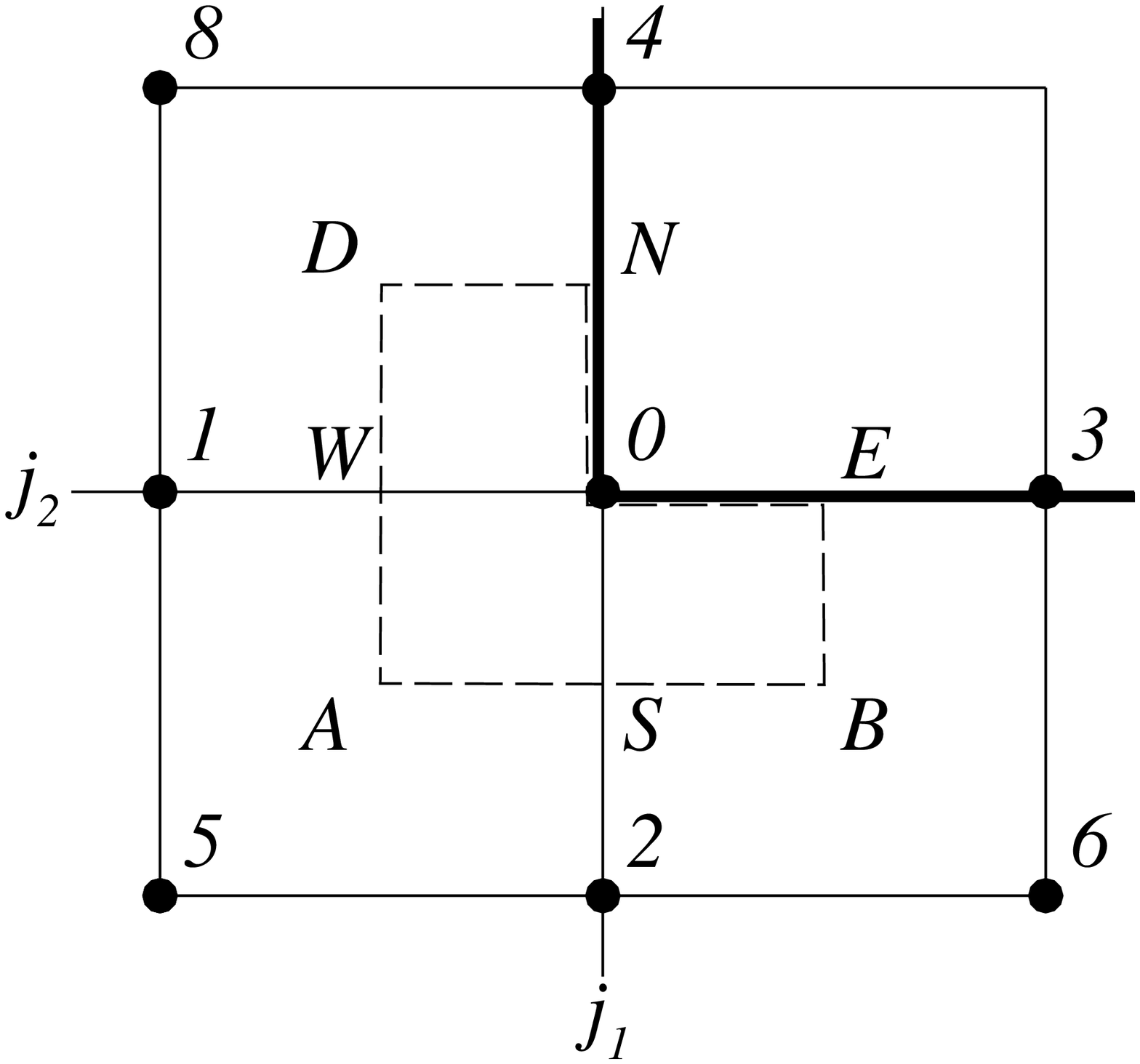}\\
\parbox[t]{0.4\textwidth}{\centering ({\it a})}
\hspace*{-5mm}\parbox[t]{0.32\textwidth}{\centering ({\it b})}\hfill
\hspace*{-5mm}\parbox[t]{0.32\textwidth}{\centering ({\it c})}\\
\vspace*{-1mm}
{\caption{The grid in the computational domain $Q$ ({\it a}); the pattern and the integration contour in the node of type~{\it 0}~({\it b}) and in the boundary node of type~{\it 7}~({\it c})} 
\label{Calcul_domain_Q}}
\end{figure}

The flow domain $\Omega(t)$ transforms with time, so the
moving meshes are used for the calculations. In order to construct the finite-difference scheme on the movable curvilinear grid, we first make the transition to the new formulation of the problem in the movable curvilinear coordinate system, in which all parts of the boundary of ${\Omega}(t)$ lie on the coordinate lines of the first or second family. Let the coordinate transformation
\begin{equation}
x=x(q^1, q^2, t), \quad z=z(q^1, q^2, t),
\label{2.1}
\end{equation}
establishes the one-to-one continuously differentiable correspondence at each moment of time $t$  between the initial (physical) domain ${\Omega}(t)$ and~the stationary computational domain $Q$ of simple form in the space of variables $q^1$, $q^2$. In contrast to \cite{Khakimzyanov2018}, the present study will use a unit square with a rectangle cut out from above (see Fig.~\ref{Calcul_domain_Q}({\it a})).

We assume that the sides $\gamma_l$ and $\gamma_r$ of the computational domain $Q$ are mapped in the transformation (\ref{2.1}) to the vertical side walls of the pool shown in Fig.~\ref{scheme_of_task}, the lower side $\gamma_b$~--- to the horizontal bottom of the pool, the sides $\gamma_{B,l}$, $\gamma_{B,r}$, $\gamma_{B,b}$ of the cutout rectangle~--- to the vertical walls of the body and its bottom, respectively. At that, $\gamma_{B,l}=\left\{ {\vect q}\big|\; q^1=q^1_l,\, q^2_b\le q^2\le 1\right\}$, $\ \gamma_{B,r}=\left\{{\vect q}\big|\; q^1=q^1_r, \, q^2_b\le q^2\le 1\right\}$,  $\gamma_{B,b}=\left\{{\vect q}\big|\; q^1_l\le q^1\le q^1_r,\, q^2=q^2_b\right\}$, where ${\vect q}=(q^{1}, q^{2})$, $0<q^1_l<q^1_r<1$, $0<q^2_b<1$.
Note that in the new coordinates the free surface is stationary and represents $\gamma_s$ which is the combination of two segments lying on the upper side $q^2=1$ of the unit square. In addition, the assumption $q^2_b<1$ means that when constructing the numerical algorithm, it is assumed that the bottom of the semi-immersed body is always under water, or, in other words, the bottom is never partially or completely drained and the points of intersection of the free surface with the body faces always lie above the object bottom:
\begin{equation}
\eta(x_l,t)>d_0, \quad \eta(x_r,t)>d_0, \quad \forall t\ge 0.
\label{wet_body_bottom}
\end{equation}

Laplace equation (\ref{2D_11_1.14}), kinematic (\ref{2D_11_1.5phi}) and dynamic (\ref{2D_11_1.16}) conditions are written in the new coordinate system and solved numerically on the rectangular grid $\overline{Q}_h={Q}_h\cup\gamma_h$, covering $Q$. This grid have steps $h_{1}=1/N_1$ and $h_{2}=1/N_2$ and number of nodes $N_{1}$, $N_{2}$ in the direction of axes $Oq^{1}$ and $Oq^{2}$, respectively, and consists of internal nodes ${\vect q}_{\vect j}\in {Q}_h\subset Q$ and boundary ${\vect q}_{\vect j}\in \gamma_h\subset\gamma=\partial Q$, where ${\vect j}=(j_1, j_2)$.
It is assumed that the boundaries  $\gamma_{B,l}$, $\gamma_{B,r}$, $\gamma_{B,b}$ lie on the grid coordinate lines, i.~e. $q^1_l=j_lh_1$, $q^1_r=j_rh_1$, $q^2_b=j_bh_2$, where $0<j_l<j_r<N_1$, $0<j_b<N_2$.

The numerical algorithm for investigating surface waves in a~basin with a fixed bottom and with moving or fixed sidewalls is presented in sufficient detail in \cite{Khakimzyanov2018}. Therefore, here we focus only on some differences from the algorithm \cite{Khakimzyanov2018} related to the presence of the semi-immersed body.

Let the curvilinear grid ${\vect x}^n$ is constructed at $t=t^n$ and the values of the grid functions $\eta^n$ and $\Phi^n$ are calculated there. The computation of the solution $\eta^{n+1}$, $\Phi^{n+1}$ on the $(n+1)$ layer by time, i.e., at the~time moment $t^{n+1}=t^{n}+\tau_n$, consists of several steps. First, the potential values $\Phi_{\vect j}^{n+1}$ at the grid nodes ${\vect q}_{\vect j}\in \gamma_{\textrm {s,h}}$ are computed. This uses a finite-difference approximation of dynamic condition (\ref{2D_11_1.16}) rewritten in new coordinates. The only difference between this step and the one presented in \cite{Khakimzyanov2018} is that the prototype $\gamma_s$ of the free boundary here consists of two segments, so the values $\Phi_{\vect j}^{n+1}$ are defined in the nodes of the $\gamma_{s,h}$ having numbers $j_2=N_2$, $j_1=0,\ldots, j_l$, $j_1=j_r,\ldots,N_1$.

After calculating the potential in the nodes of the $\gamma_{s,h}$, the new values of the potential $\Phi^{n+1}_{\vect j}$ in all other nodes of the grid ${\vect{q}}_{\vect j} \in \overline{Q}_h\setminus \gamma_{s,h}$ are computed. For this purpose, we use the finite-difference analog of Laplace equation (\ref{2D_11_1.14}) in curvilinear coordinates:
\begin{equation}
\frac{\partial}{\partial q^{1}}\left ( k_{11} \frac{\partial \Phi}{\partial q^{1}}+ k_{12} \frac{\partial \Phi}{\partial q^{2}} \right )+
\frac{\partial}{\partial q^{2}}\left ( k_{21} \frac{\partial \Phi}{\partial q^{1}}+ k_{22} \frac{\partial \Phi}{\partial q^{2}} \right ) = 0,
\quad {\vect q}\in Q,
\label{2.34}
\end{equation}
where
\begin{equation}
k_{11}=\frac{g_{22}}{J}, \quad k_{12}=k_{21}=-\frac{g_{12}}{J}, \quad k_{22}=\frac{g_{11}}{J},
\label{2.35}
\end{equation}
\begin{equation}
g_{11}= x^2_{q^1}+z^2_{q^1}, \quad g_{12}=g_{21}=x_{q^1}x_{q^2}+z_{q^1}z_{q^2}, \quad g_{22}= x^2_{q^2}+z^2_{q^2},
\label{2.25}
\end{equation}
$J=x_{q^1}z_{q^2}-x_{q^2}z_{q^1}$ is the Jacobian of transformation (\ref{2.1}),  $J>0$.
In these coordinates, boundary conditions (\ref{2D_11_1.15})---(\ref{2D_Gamma}) are used on the boundary of the computational domain:
\begin{equation}
k_{21}\frac{\partial\Phi}{\partial q^1}+k_{22}\frac{\partial\Phi}{\partial q^2} \biggm|_{{\vect q}\in \gamma_b\cup \gamma_{B,b}}=0,
\label{Pal_contr_5}
\end{equation}
\begin{equation}
k_{11}\frac{\partial\Phi}{\partial q^1}+k_{12}\frac{\partial\Phi}{\partial q^2} \biggm|_{\gamma_l\cup \gamma_r\cup \gamma_{B,l}\cup \gamma_{B,r}}=0.
\label{Pal_contr_6}
\end{equation}

The finite-difference equations for the velocity potential are obtained by the integro-interpolation method \cite{Khakimzyanov2018}, in which the differential equation (\ref{2.34}) is rewritten in the integral form
\begin{equation}
\oint\limits_{{\cal{C}}}\left(k_{11}\frac{\partial\Phi}{\partial
q^1}+ k_{12}\frac{\partial\Phi}{\partial q^2}\right)dq^2-
\left(k_{21}\frac{\partial\Phi}{\partial q^1}+
k_{22}\frac{\partial\Phi}{\partial q^2}\right)dq^1=0
\label{3.1}
\end{equation}
and some quadrature formula is used to approximate the integral. Depending on the choice of this formula, one or another finite-difference scheme for $\Phi$ will be obtained. The finite-difference analogues of the integral relations (\ref{3.1}) are written out for the computational nodes ${\vect{q}}_{\vect j} \in \overline{Q}_h\setminus \gamma_{s,h}$. These nodes are divided into non-intersecting classes, each of which is assigned a unique number (type) depending on whether the nodes in this class are internal or belong to certain parts of the boundary. 
Internal nodes are assigned type $0$, boundary nodes ${\vect q}_{\vect j}\in \gamma_h\setminus \gamma_{s,h}$ may have type $1$ to $8$ depending on which part of the boundary they belong to (see Table ~\ref{tabular:Node_types} and Figure ~\ref{Calcul_domain_Q}({\it a})).
The type of the node determines the integration contour $\cal{C}$, which is the boundary of an elementary internal or boundary cell, including the pattern of the finite-difference equation in that node. For the internal nodes (type 0) the template of the finite-difference equation for the potential is 9-point. For boundary nodes (type 1-8) the template include from 4 to 8 mesh nodes.
\begin{table}[h!]
\caption{The node types ${\vect{q}}_{\vect j} \in \overline{Q}_h\setminus \gamma_{s,h}$}
\label{tabular:Node_types}
\begin{center}
{\tabcolsep=3.2mm
\begin{tabular}{c|c|ll} \hline
 Type of the node ${\vect{q}}_{\vect j}$  & Node displacement    & \multicolumn{2}{|c}{Indexes $(j_1, j_2)$ for the node ${\vect{q}}_{\vect j}$}\\ \hline
   &                             & $0< j_1<j_l$,         & $0< j_2< N_2$;\\
 0 & ${\vect{q}}_{\vect j}\in Q_h$ & $j_r< j_1<N_1$,       & $0< j_2< N_2$;\\
   &                             &  $j_l\le  j_1\le j_r$,& $0< j_2< j_b$\\ \hline
 1 & ${\vect{q}}_{\vect j}\in \gamma_l\cup \gamma_{B,r}$ & $j_1=0$,   & $0< j_2< N_2$;\\
   &                             &  $j_1=j_r$,           & $j_b< j_2< N_2$\\ \hline
 2 & ${\vect{q}}_{\vect j}\in \gamma_b$ &  $0< j_1<N_1$,   & $j_2=0$ \\  \hline
 3 & ${\vect{q}}_{\vect j}\in \gamma_r\cup \gamma_{B,l}$ & $j_1=N_1$, & $0< j_2< N_2$;\\
   &                             &  $j_1=j_l$,           & $j_b< j_2< N_2$\\ \hline
 4 & ${\vect{q}}_{\vect j}\in \gamma_{B,b}$ &  $j_l< j_1<j_r$,   & $j_2=j_b$ \\  \hline
 5 & ${\vect{q}}_{\vect j}=\gamma_l\cap  \gamma_b$ &  $j_1=0$,   & $j_2=0$ \\  \hline
 6 & ${\vect{q}}_{\vect j}=\gamma_r\cap  \gamma_b$ &  $j_1=N_1$,   & $j_2=0$ \\  \hline
 7 & ${\vect{q}}_{\vect j}=\gamma_{B,l}\cap  \gamma_{B,b}$ &  $j_1=j_l$,   & $j_2=j_b$ \\  \hline
 8 & ${\vect{q}}_{\vect j}=\gamma_{B,r}\cap  \gamma_{B,b}$ &  $j_1=j_r$,   & $j_2=j_b$ \\
\end{tabular}
}
\end{center}
\end{table}

Figure~\ref{Calcul_domain_Q}({\it b}) shows the integration contour (dashed line $ABCD$) in the case when the finite-difference equation is written in the inner nodes of the grid ${\vect{q}}_{\vect j} \in {Q}_h$. In this case, the contour is the rectangle whose sides are parallel to the coordinate axes and divide in half the distances to the nodes adjacent to ${\vect{q}}_{\vect j}$. The integration contour is the boundary of the unit cell associated with the inner node ${\vect{q}}_{\vect j} \in {Q}_h$. Applying the quadrature formula of rectangles $ABCD$ for the integrals over the sides of the rectangle, we obtain the  finite-difference equation \cite{Khakimzyanov2018}
\begin{equation}
\Big( \sum_{k=0}^8\alpha_k\Phi_k^{n+1}\Big)_{\vect j} = 0
\label{3.4.eq2}
\end{equation}
on a nine-point pattern consisting of the nodes with local numbers $k=0, \ldots , 8$. Here $\Phi_k^{n+1}$ is the value of the grid function $\Phi$ in the node having local number~$k$. The local numbering of the pattern nodes corresponding to ${\vect q}_{\vect j}$ is introduced here to shorten the record. Thus, according to Table~\ref{tabular:kinds_of_nodes} the local number $k=0$ is used instead of the global number $(j_1,j_2)$, $k=1$ instead of $(j_1-1,j_2)$, $k=2$ instead of $(j_1,j_2-1)$, etc. The coefficients $\alpha_k$ ($k=1, \ldots , 8$) of equations (\ref{3.4.eq2}) for the internal nodes
are given in the first row of Table~\ref{Koef_dif_eq_p65_eq_sum1}. The following notations are used in this table:
\begin{equation*}
\xi_1=\frac{h_2}{h_1}k_{11}(W), \quad  \xi_2=\frac{h_1}{h_2} k_{22}(S), \quad
\xi_3=\frac{h_2}{h_1} k_{11}(E), \quad \xi_4=\frac{h_1}{h_2}k_{22}(N),
\end{equation*}
\begin{equation*}
\xi_5=\frac{k_{12}(1)+k_{12}(2)}{4}, \quad \xi_6=-\frac{k_{12}(2)+k_{12}(3)}{4},
\end{equation*}
\begin{equation*}
\xi_7=\frac{k_{12}(3)+k_{12}(4)}{4}, \quad \xi_8=-\frac{k_{12}(4)+k_{12}(1)}{4},
\end{equation*}
\begin{equation*}
\zeta_m=\frac{k_{12}(0)-k_{12}(m)}{4}, \quad m=1,2,3,4.
\end{equation*}
The coefficient $\alpha_0$ is defined as
\begin{equation*}
\alpha_0=-\sum_{k=1}^8 \alpha_k.
\end{equation*}
\begin{table}[h!]
\caption{Correspondence between the global numbers ${\vect j}$ of the grid nodes and the local numbers $k$ of the pattern nodes}
\label{tabular:kinds_of_nodes}
\begin{center}
{\tabcolsep=5mm
\begin{tabular}{c|c||c|c||c|c}
\hline
${\vect j}$     &  $k$   &  ${\vect j}$       &  $k$  & ${\vect j}$         & $k$  \\ \hline
$(j_1,j_2)$    &  0     &  $(j_1+1,j_2)$    &  3    &  $(j_1+1,j_2-1)$   & 6\\ \hline
$(j_1-1,j_2)$  &  1     &  $(j_1,j_2+1)$    &  4    &  $(j_1+1,j_2+1)$   & 7\\ \hline
$(j_1,j_2-1)$  &  2     &  $(j_1-1,j_2-1)$  &  5    &  $(j_1-1,j_2+1)$   & 8\\
\end{tabular}
}
\end{center}
\end{table}
\begin{table}[h!]
\caption{Coefficients  $\alpha_k$ of equation (\ref{3.4.eq2}) for the internal and boundary nodes}
\label{Koef_dif_eq_p65_eq_sum1}
\begin{center}
{\tabcolsep=3mm
\begin{tabular}{c|c|c|c|c|c|c|c|c} \hline
 Node type & $\alpha_1$     & $\alpha_2$            & $\alpha_3$            & $\alpha_4$            &$\alpha_5$&$\alpha_6$&$\alpha_7$&$\alpha_8$\\ \hline
 0 & $\xi_1$               & $\xi_2$               & $\xi_3$               & $\xi_4$               & $\xi_5$ & $\xi_6$ & $\xi_7$ & $\xi_8$\\ \hline
 1 & 0                     & ${\xi_2}/{2}-\zeta_2$ & $\xi_3$               & ${\xi_4}/{2}+\zeta_4$ & 0       & $\xi_6$ & $\xi_7$ & 0      \\ \hline
 2 & ${\xi_1}/{2}-\zeta_1$ & 0                     & ${\xi_3}/{2}+\zeta_3$ & $\xi_4$               & 0       & 0       & $\xi_7$ & $\xi_8$\\ \hline
 3 & $\xi_1$               & ${\xi_2}/{2}+\zeta_2$ & 0                     & ${\xi_4}/{2}-\zeta_4$ & $\xi_5$ & 0       & 0       & $\xi_8$\\ \hline
 4 & ${\xi_1}/{2}+\zeta_1$ & $\xi_2$               & ${\xi_3}/{2}-\zeta_3$ & 0                     & $\xi_5$ & $\xi_6$ & 0       & 0      \\ \hline
 5 & 0                     & 0                     & ${\xi_3}/{2}+\zeta_3$ & ${\xi_4}/{2}+\zeta_4$ & 0       & 0       & $\xi_7$ & 0      \\ \hline
 6 & ${\xi_1}/{2}-\zeta_1$ & 0                     & 0                     & ${\xi_4}/{2}-\zeta_4$ & 0       & 0       & 0       & $\xi_8$\\ \hline
 7 & $\xi_1$               & $\xi_2$               & ${\xi_3}/{2}-\zeta_3$ & ${\xi_4}/{2}-\zeta_4$ & $\xi_5$ & $\xi_6$ & 0       & $\xi_8$\\ \hline
 8 & ${\xi_1}/{2}+\zeta_1$ & $\xi_2$               & $\xi_3$               & ${\xi_4}/{2}+\zeta_4$ & $\xi_5$ & $\xi_6$ & $\xi_7$ & 0      \\
\end{tabular}
}
\end{center}
\end{table}

As elementary cells for the boundary nodes ${\vect q}_{j}\in \gamma_h\setminus \gamma_{s,h}$ we take that part of the rectangle $ABCD$ which is contained in $Q$. For example, for the node ${\vect{q}}_{\vect j}=\gamma_{B,l}\cap \gamma_{B,b}$ of type 7, which coincides with the prototype of the intersection point between the face of the body and its bottom, the unit cell is the figure with boundary $ABE0ND$ (see Fig.~\ref{Calcul_domain_Q}({\it c})), with fragments $0E$ and $0N$ of this boundary lying on the prototype boundary of the semi-immersed body. With (\ref{3.1}), the $ABE0ND$ integrals on the sides of $0E$ and $0N$ is zero due to boundary conditions (\ref{Pal_contr_5}), (\ref{Pal_contr_6}). Therefore, integral relation (\ref{3.1}) becomes
\begin{equation*}
\int\limits_{(BE)}\left(k_{11}\frac{\partial\Phi}{\partial
q^1}+ k_{12}\frac{\partial\Phi}{\partial q^2}\right)dq^2-
\int\limits_{(AD)}\left(k_{11}\frac{\partial\Phi}{\partial
q^1}+ k_{12}\frac{\partial\Phi}{\partial q^2}\right)dq^2+
\end{equation*}
\begin{equation*}
+\int\limits_{(DN)}\left(k_{12}\frac{\partial\Phi}{\partial
q^1}+ k_{22}\frac{\partial\Phi}{\partial q^2}\right)dq^1-
\int\limits_{(AB)}\left(k_{12}\frac{\partial\Phi}{\partial
q^1}+ k_{22}\frac{\partial\Phi}{\partial q^2}\right)dq^1=0,
\end{equation*}
and its finite-difference analogue can be written as
\begin{equation*}
\left[k_{11}(E)\frac{\Phi_3-\Phi_0}{h_1}+\frac{1}{2}\left(k_{12}(0)\frac{\Phi_0-\Phi_2}{h_2}+k_{12}(3)\frac{\Phi_3-\Phi_6}{h_2}\right)\right]\frac{h_2}{2}-
\end{equation*}
\begin{equation*}
-\left[k_{11}(W)\frac{\Phi_0-\Phi_1}{h_1}+\frac{1}{2}\left(k_{12}(0)\frac{\Phi_4-\Phi_2}{2h_2}+k_{12}(1)\frac{\Phi_8-\Phi_5}{2h_2}\right)\right]h_2+
\end{equation*}
\begin{equation*}
+\left[k_{22}(N)\frac{\Phi_4-\Phi_0}{h_2}+\frac{1}{2}\left(k_{12}(0)\frac{\Phi_0-\Phi_1}{h_1}+k_{12}(4)\frac{\Phi_4-\Phi_8}{h_1}\right)\right]\frac{h_1}{2}-
\end{equation*}
\begin{equation*}
-\left[k_{22}(S)\frac{\Phi_0-\Phi_2}{h_2}+\frac{1}{2}\left(k_{12}(0)\frac{\Phi_3-\Phi_1}{2h_1}+k_{12}(2)\frac{\Phi_6-\Phi_5}{2h_1}\right)\right]h_1=0.
\end{equation*}
So we obtain the finite-difference equation on the 8-point pattern shown in Fig.~\ref{Calcul_domain_Q}({\it c}). A similar 8-point equation is obtained for a node of type 8.

In the boundary nodes of types 1, 2, 3, and 4, the pattern is six-point, and in the corner nodes (types 5 and 6) it is four-point. In the boundary nodes, the finite-difference equations can be written formally as nine-point equations (\ref{3.4.eq2}) by zeroing the coefficients $\alpha_k$ for those nodes of the nine-point template that are not part of the boundary node templates. Expressions for the $\alpha_k$ coefficients depending on the type 1---8 boundary node ${\vect q}_{\vect j}\in \gamma_h\setminus \gamma_{s,h}$ are given in Table~\ref{Koef_dif_eq_p65_eq_sum1}. The system of finite-difference equations (\ref{3.4.eq2}) is solved by the iterative method of successive over-relaxation as in \cite{Khakimzyanov2018}.

The next step of the computational algorithm determines the new position of the free boundary $\eta_{j_1}^{n+1}$ (${\vect q}_{\vect j}\in \gamma_{s,h}$) by approximating kinematic condition (\ref{2D_11_1.5phi}) written in the coordinates $q^1$, $q^2$, $t$ \cite{Khakimzyanov2018}.

All the above calculations are performed on the grid ${\vect x}_{\vect j}^n$ corresponding to the $n$-layer by time $t=t^n$. Therefore, the new grid ${\vect x}_{\vect j}^{n+1}=\left(x_{\vect j}^{n+1}, y_{\vect j}^{n+1}\right)$ is to be constructed next. Compared to \cite{Khakimzyanov2018}, this study uses a simpler computational grid, namely the grid with the fixed vertical coordinate lines of the second family, i.e., with time-invariant node abscissa, so ${\vect x}_{\vect j}^{n+1}=\left(x_{\vect j}, y_{\vect j}^{n+1}\right)$.  Moreover, at $y\le d_0$ the grid is rectangular, uniform along the axis $Oy$ and does not change when going from one layer in time to another. In the subdomain $\Omega_i$ under the body (see Fig.~\ref{scheme_of_task}), the grid is uniform in both the horizontal and vertical directions with steps $\Delta x=L/(j_r-j_l)$ and $\Delta y=|d_0|/j_b$. The grid is movable only in the outer subdomain $\Omega_e$ at $y> d_0$, and the nodes of the grid move only in the vertical direction and $y_{\vect j}^{n+1}=-h_0$ at ${\vect q}_{\vect j}\in \gamma_{b,h}$, $y_{\vect j}^{n+1}=\eta_{j_1}^{n+1}$ at ${\vect q}_{\vect j}\in \gamma_{s,h}$. In $\Omega_e$, the grid is non-uniform in the horizontal direction: the grid steps $\Delta x_{j_1+1/2}\equiv x_{j_1+1,j_2}-x_{j_1,j_2}=x_{j_1+1,0}-x_{j_1,0}$ increase monotonically  with the distance from the body by the law of geometric progression. Thus,
\begin{equation}
\Delta x_{j_1+1/2}=(\Delta x)\cdot z_l^{j_l-1-j_1}, \quad j_1=0,  \ldots , j_l-1,
\label{grid_Omega_e_left}
\end{equation}
\begin{equation}
\Delta x_{j_1+1/2}=(\Delta x)\cdot z_r^{j_1-j_r}, \quad j_1=j_r,  \ldots , N_1-1.
\label{grid_Omega_e_right}
\end{equation}
The denominators $z_l$ and $z_r$ of these progressions are the roots, respectively, of the following nonlinear equations:
\begin{equation}
x_l=\Delta x\frac{1-z_l^{j_l}}{1-z_l}, \qquad l-x_r=\Delta x\frac{1-z_r^{N_1-j_r}}{1-z_r}.
\label{grid_3}
\end{equation}
If the the condition 
\begin{equation}
\frac{L}{j_r-j_l}< \min\left\{\frac{x_l}{j_l};\ \frac{l-x_r}{N_1-j_r}\right\},
\label{3.51}
\end{equation}
is satisfied, each of equations (\ref{grid_3}) will have a single solution, with $z_l>1$ and $z_r>1$.
Using formulas (\ref{grid_Omega_e_left}), (\ref{grid_Omega_e_right}), a smooth coupling of the meshes outside and under the body is achieved and the steps $\Delta x_{j_1+1/2}$ decrease smoothly when approaching the body.
The use of the finer mesh in the vicinity of the obstacle makes it possible to increase the accuracy of calculation of the wave-body interaction.

After constructing the new grid, it is necessary to repeat the calculations in the previous steps in order to match the values $\varphi^{n+1}$ and $\eta^{n+1}$ with grid ${\vect x}_{\vect j}^{n+1}$. Some details of the recalculation step were described in \cite{Khakimzyanov2018}.

\subsubsection{Some results of calculations with the developed algorithm for the Pot model}

\begin{figure}[h!]
\centering
\includegraphics[width=0.9\textwidth]{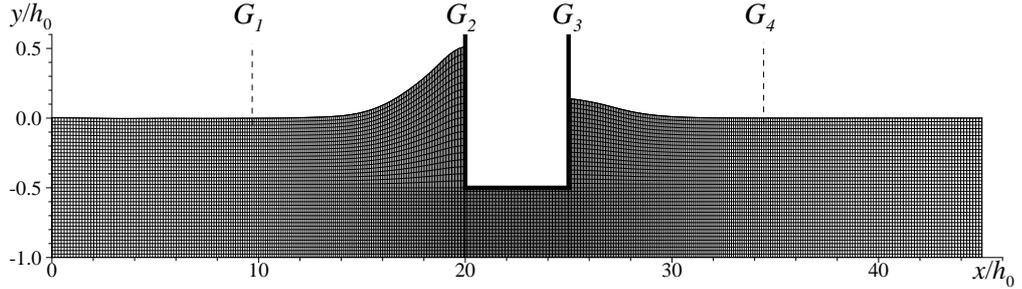}

{\caption{$\Po$ model. Computation grid at $t\sqrt{g/h_0}=10$. $G_i$ ($i=1, 2, 3, 4$) are waves gauges.  $a_0/h_0=0.4$, $L/h_0=5$, $d_0/h_0=-0.5$ }
\label{Grid_Pot}}
\end{figure}

Figure~\ref{Grid_Pot} shows an example of a typical grid used in the $\Po$ model calculations of the interaction of solitary wave (\ref{Full_an_sol_eta}), (\ref{init_cond_6_new}) with the semi-immersed stationary body having the length $L$ in the $Ox$ direction. In this example, the following input values are taken:
\begin{equation}
\frac{a_0}{h_0}=0.4, \quad  \frac{L}{h_0}=5,\quad \frac{d_0}{h_0}=-0.5, \quad \frac{x_l}{h_0}=20, \quad x_0=\frac{x_l}{2},\quad x_r=x_l+L,\quad l=x_r+x_l,
\label{input_Pot}
\end{equation}
\begin{equation*}
N_1=400, \quad N_2=40, \quad j_l=160, \quad j_r=240,\quad j_b=20.
\end{equation*}
Obviously, condition (\ref{3.51}) is satisfied for these data, so the mesh is thickened in the vicinity of the body.

\begin{figure}[h!]
\centering
\includegraphics[width=0.49\textwidth]{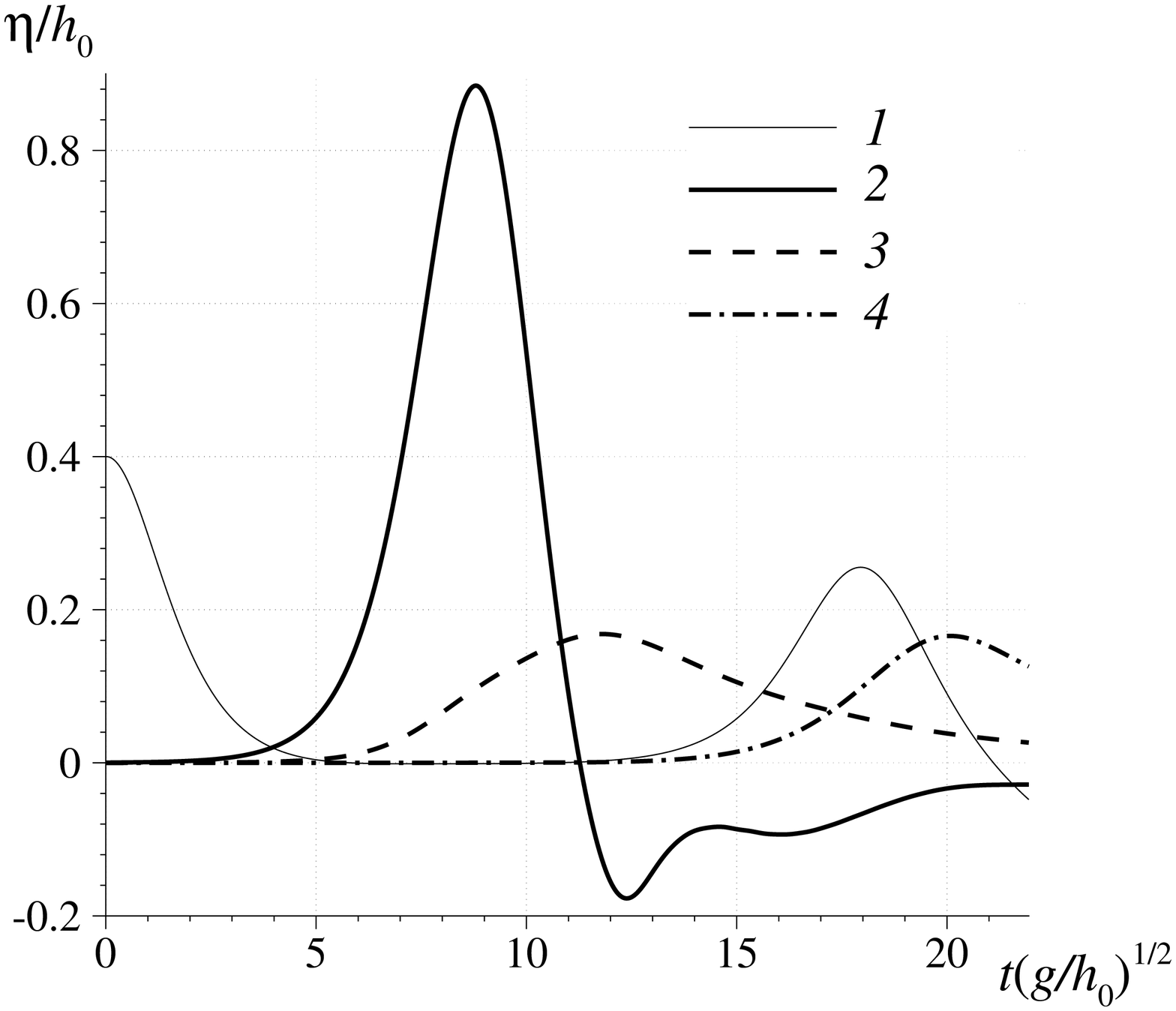}
\hfill
\includegraphics[width=0.49\textwidth]{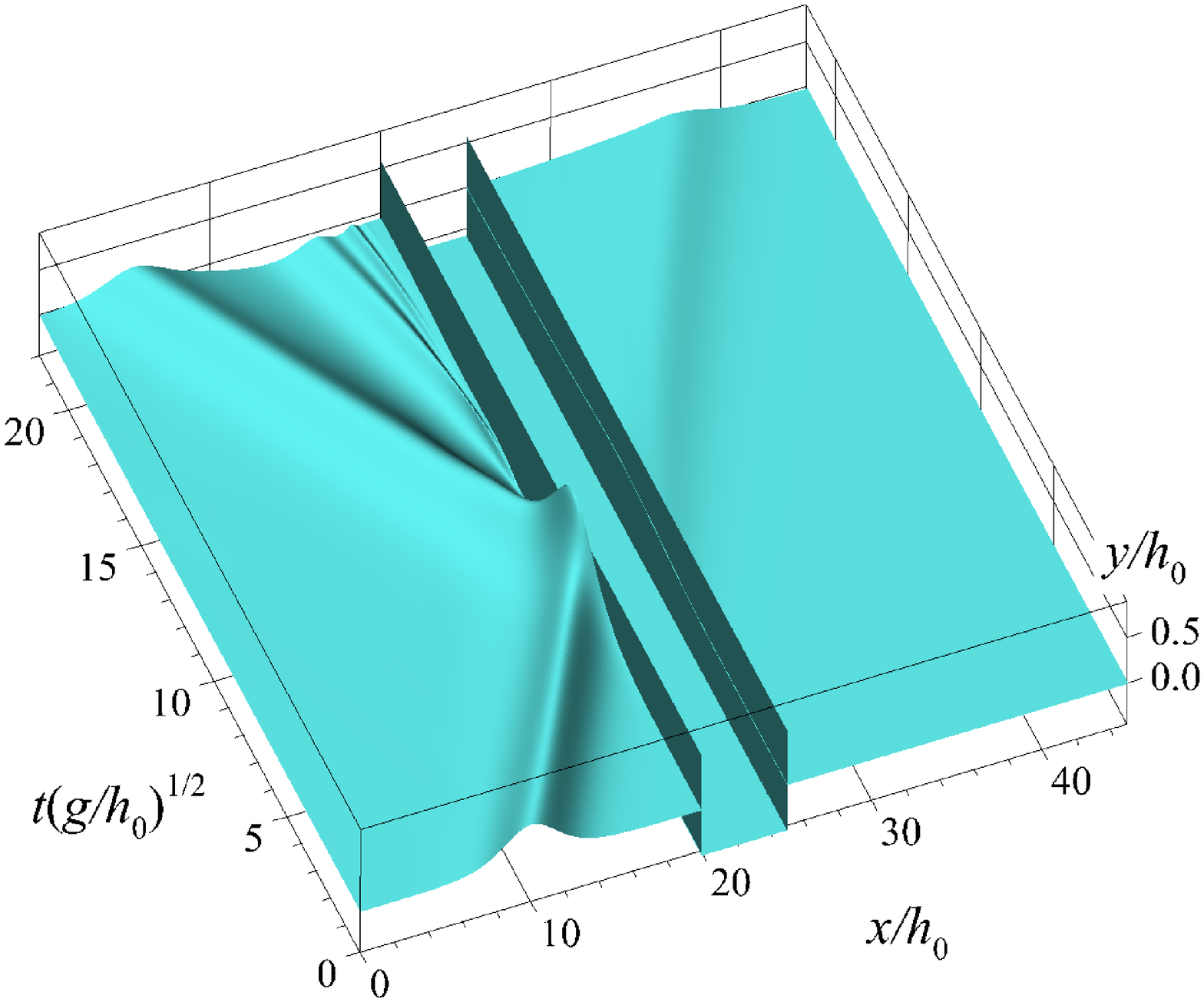}\\
\parbox[t]{0.49\textwidth}{\centering ({\it a})}
\hfill
\parbox[t]{0.49\textwidth}{\centering ({\it b})}\\
\vspace*{-1mm}
{\caption{$\Po$ model: computed time histories of the free surface at the gauges  $G_i$ ($i=1, 2, 3, 4$) ({\it a});  space–time plots of the free surface evolution ({\it b}).  $a_0/h_0=0.4$, $L/h_0=5$, $d_0/h_0=-0.5$
\label{Mar+Surf_Pot}}}
\end{figure}

Virtual wave gauges $G_i$ ($i=1, \ldots , 4$) are installed (see Fig.~\ref{Grid_Pot}) to measure the amplitude of the wave reflected from the body, the runup on the left and right sides of the body (the front and back faces of the body), and the amplitude of the wave that passed behind the body. These wave gauges record the level of the free surface at the points with the following abscissa values:
\begin{equation}
x_{G_1}=\frac{x_l}{2}, \quad x_{G_2}=x_l, \quad x_{G_3}=x_r, \quad x_{G_4}=\frac{x_r+l}{2}\;.
\label{x_mareogrs}
\end{equation}
Figure~\ref{Mar+Surf_Pot}({\it a}) shows the chronograms measured by these wave gauges. It can be seen that after the interaction of the incoming wave with the body, a reflected wave is formed (line {\sl 1} in Fig. ~\ref{Mar+Surf_Pot}({\it a})). The amplitude of this reflected wave is less than $a_0$ and it has a profile that differs from that of the soliton wave: the rising wave is followed by a falling wave (see also Fig.~\ref{Mar+Surf_Pot}({\it b})). The line {\sl 2} depicts the chronogram of the elevation of the free surface on the face of the body. This chronogram differs significantly from the chronogram of the soliton wave runup on the vertical wall, in particular, by the greater amplitude of the negative polarity wave, which occurs after the runup and is caused by the overflow of water under the body from the left side of the domain to the right side. The remaining two chronograms (lines {\sl 3, 4}) depict the elevation of the free boundary on the right side of the body and the elevation of the wave that passed behind the body. The latter has the smaller amplitude than the incoming wave.

\begin{figure}[h!]
\centering
\includegraphics[width=0.49\textwidth]{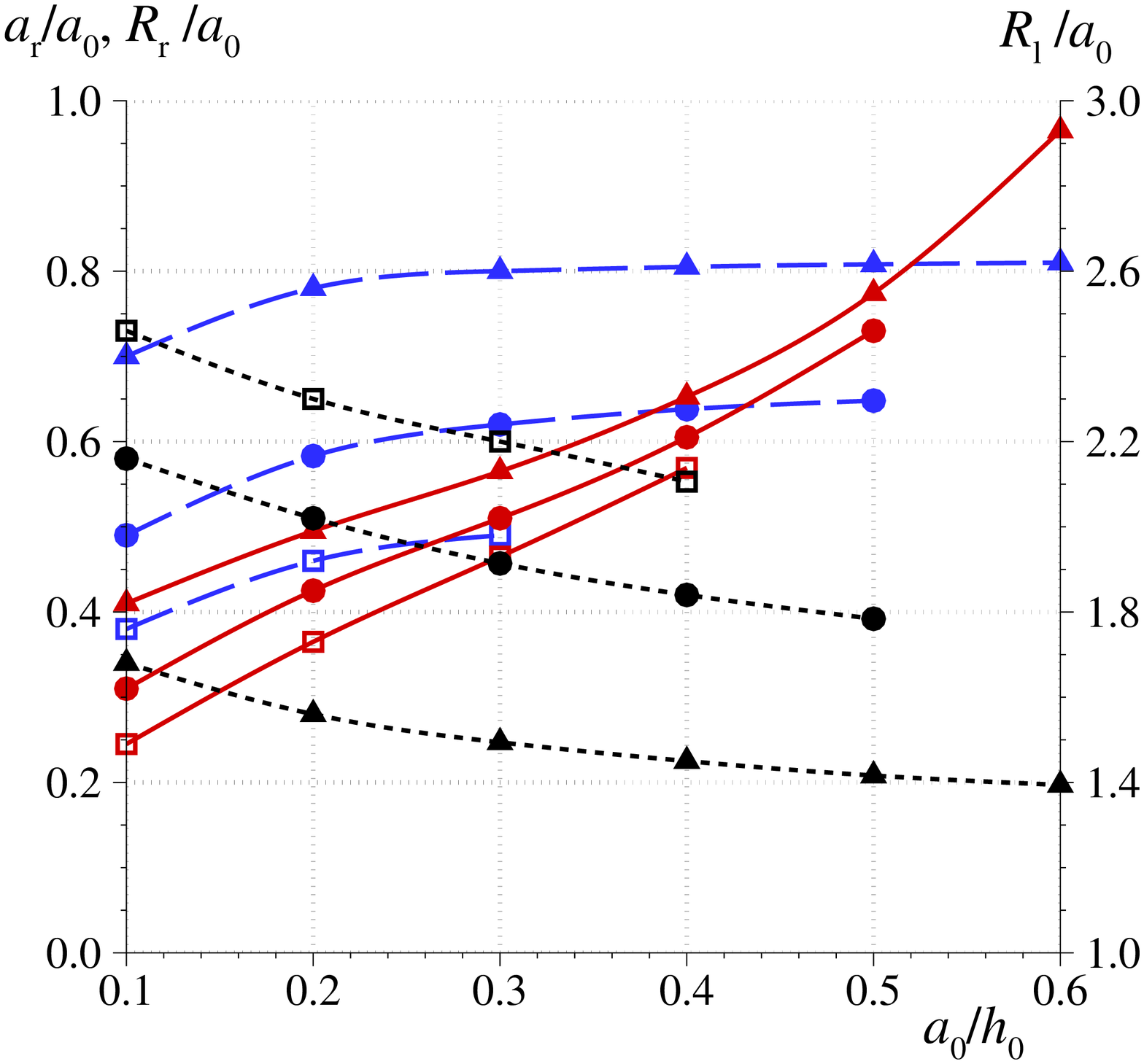}
\hfill
\includegraphics[width=0.49\textwidth]{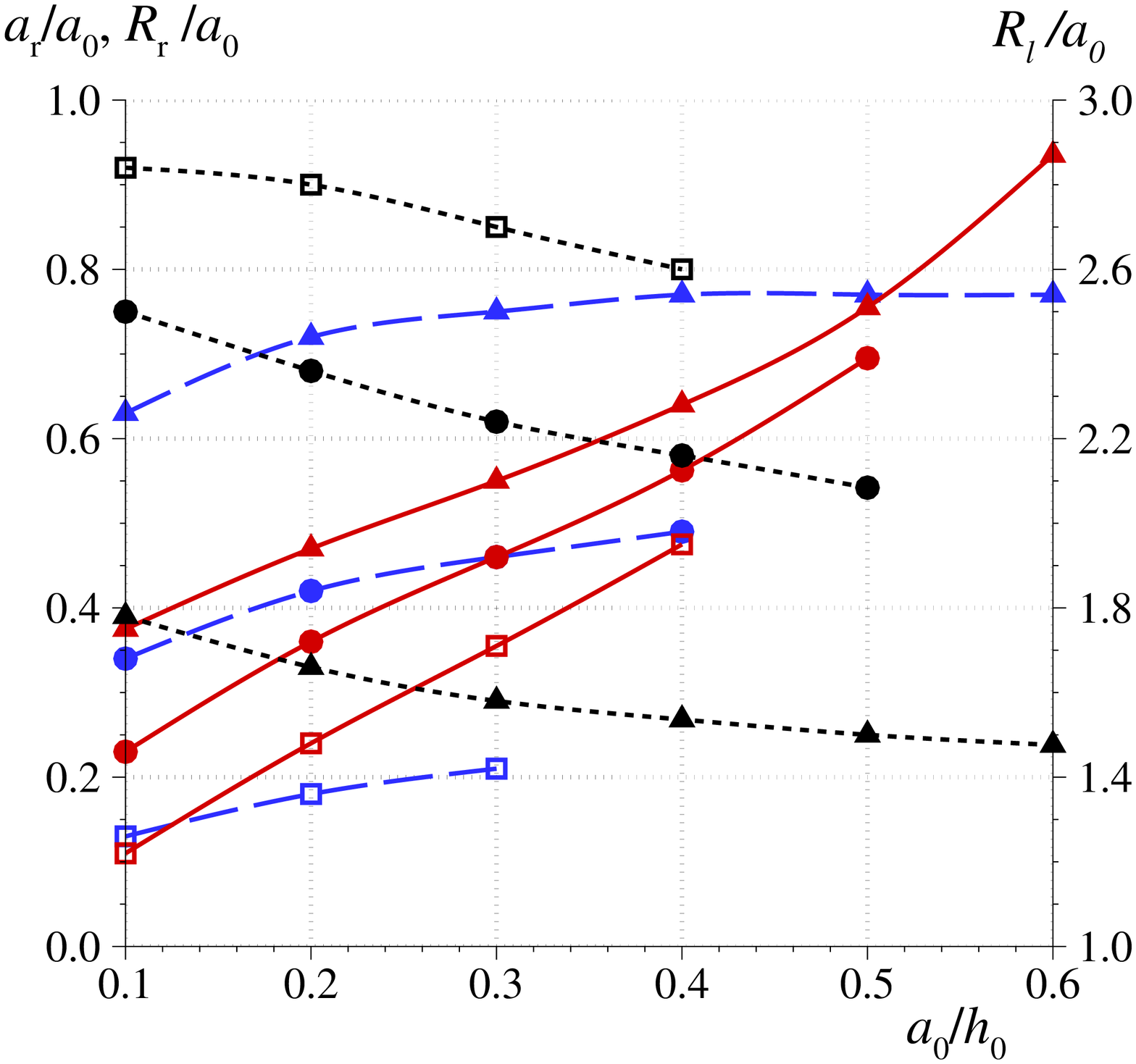}\\
\parbox[t]{0.49\textwidth}{\centering ({\it a})}
\hfill
\parbox[t]{0.49\textwidth}{\centering ({\it b})}\\
\vspace*{-1mm}
{\caption{$\Po$ model. Dependence of the amplitude of the reflected wave $a_r/a_0$ ({\bf ---\;---}), the maximum vertical runup on the front $R_l/a_0$ ({\bf ---}) 
and back $R_r/a_0$ ({\bf -\hspace*{0.3mm}-\hspace*{0.3mm}-}) faces of the body from the amplitude $a_0/h_0$ of the incoming wave at 
({\it a}): fixed body length $L/h_0=5$ and different values of its depth $d_0/h_0=-0.2$~($\Box$), $-0.5$~($\bullet$), $-0.8$~($\blacktriangle$); 
({\it b}): fixed body depth $d_0/h_0=-0.5$ and different values of its length $L/h_0=0.625$~($\Box$), $2.5$~($\bullet$), $10$~($\blacktriangle$)
\label{G123_vs_a_var_dL}}}
\end{figure}

\begin{figure}[h!]
\centering
\includegraphics[width=0.49\textwidth]{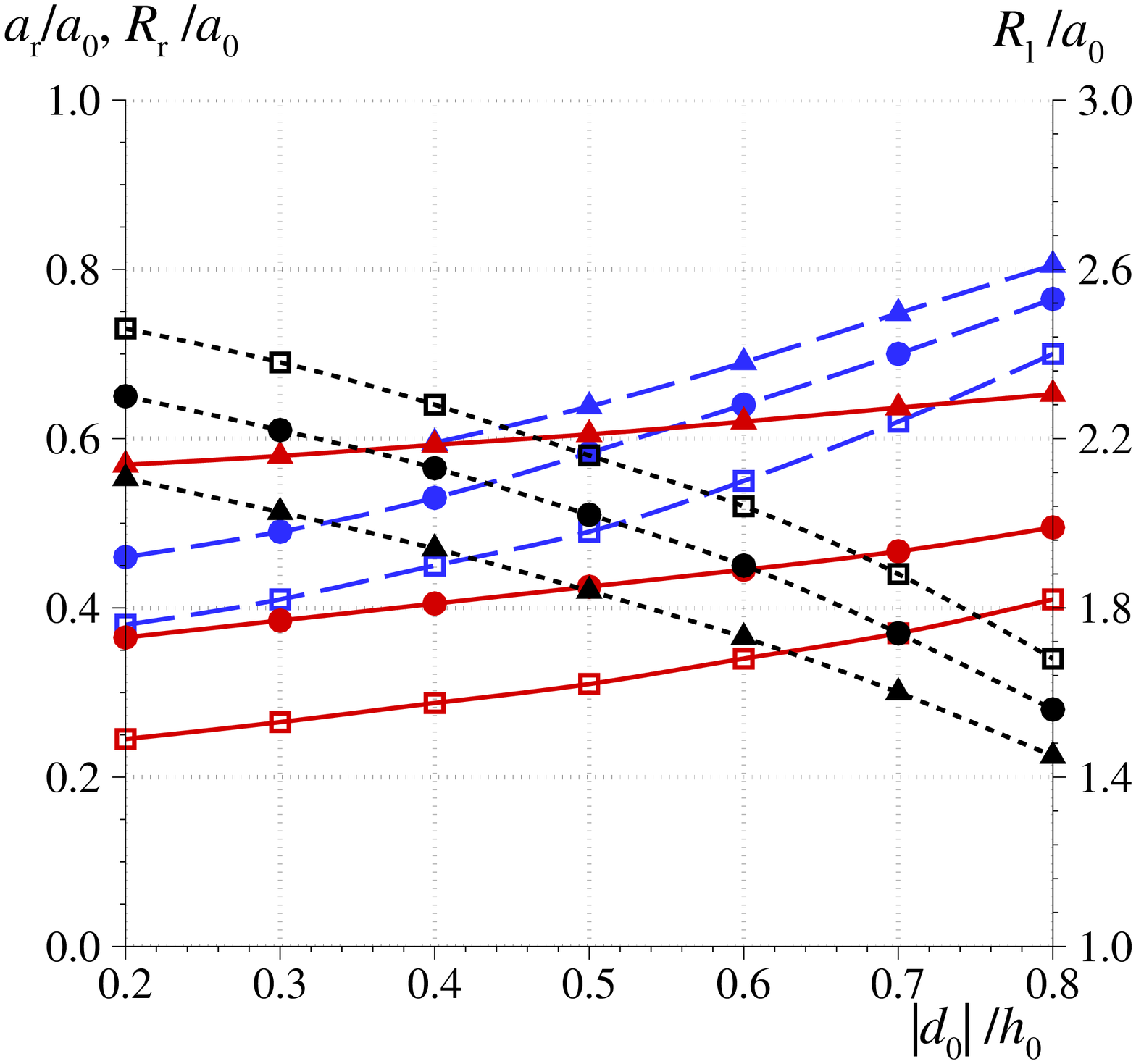}
\hfill
\includegraphics[width=0.49\textwidth]{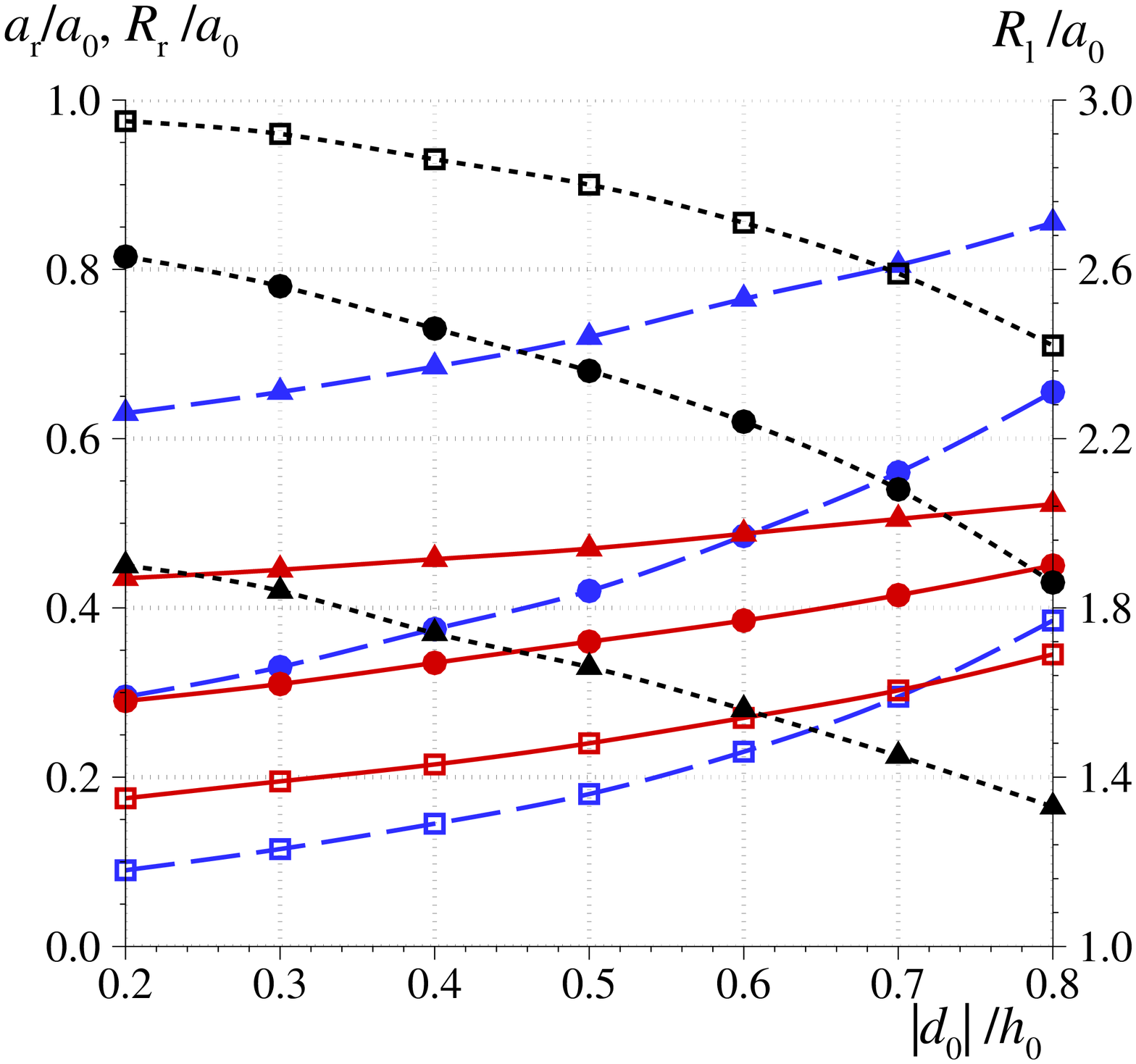}\\
\parbox[t]{0.49\textwidth}{\centering ({\it a})}
\hfill
\parbox[t]{0.49\textwidth}{\centering ({\it b})}\\
\vspace*{-1mm}
{\caption{$\Po$ model. Dependence of the amplitude of the reflected wave $a_r/a_0$ ({\bf ---\;---}), the maximum vertical runup on the front $R_l/a_0$ ({\bf ---}) 
and back $R_r/a_0$ ({\bf -\hspace*{0.3mm}-{\hspace*{0.3mm}}-}) faces of the body from the body submergence $|d_0|/h_0$ at 
({\it a}): fixed body length $L/h_0=5$ and different values of the amplitude of the incoming wave $a_0/h_0=0.1$~($\Box$), $0.2$~($\bullet$), $0.4$~($\blacktriangle$); 
({\it b}): fixed amplitude of the incoming wave $a_0/h_0=0.2$ and different values of the body length $L/h_0=0.625$~($\Box$), $2.5$~($\bullet$), $10$~($\blacktriangle$)}
\label{G123_vs_d_var_aL}}
\end{figure}

\begin{figure}[h!]
\centering
\includegraphics[width=0.49\textwidth]{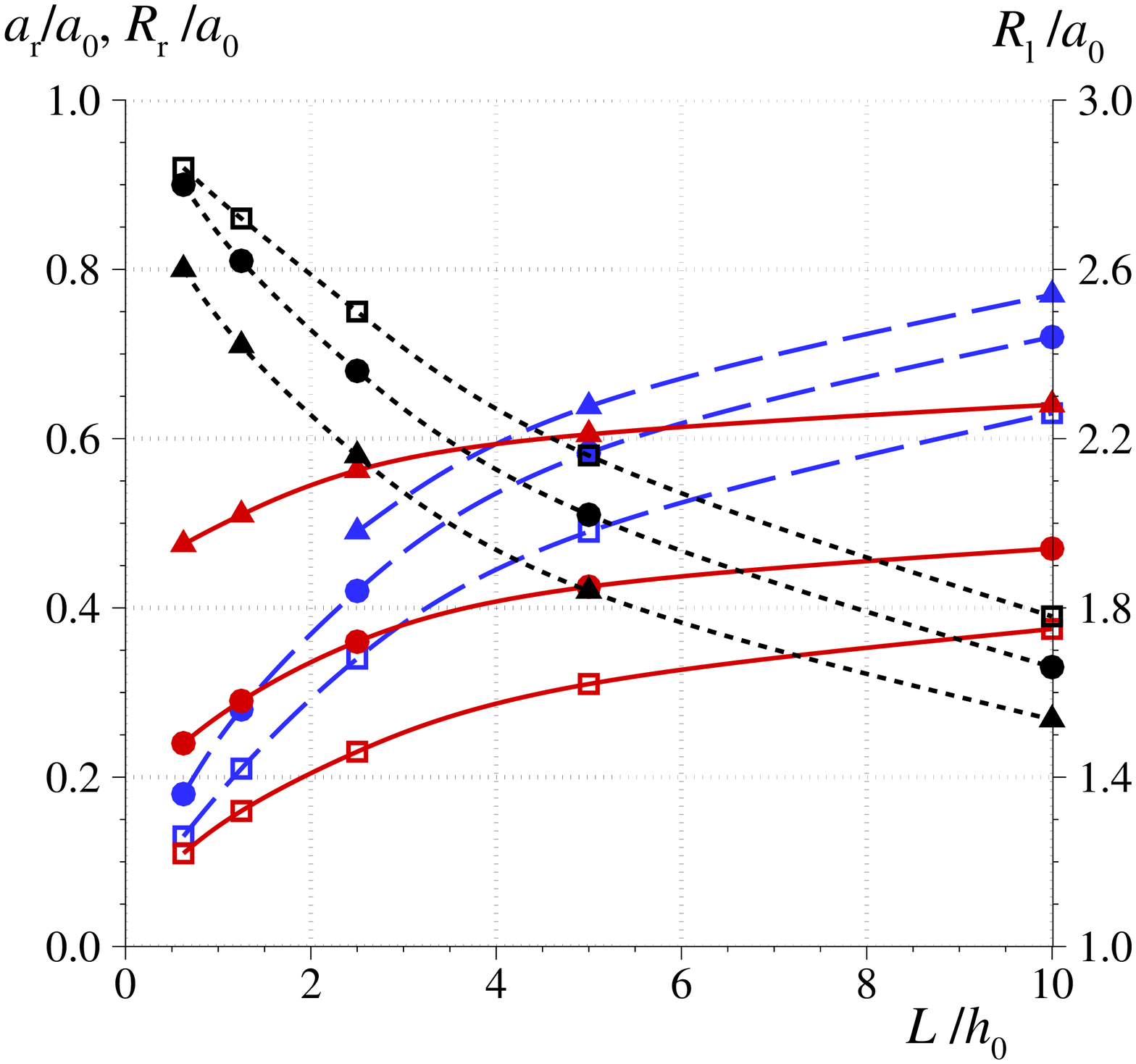}
\hfill
\includegraphics[width=0.49\textwidth]{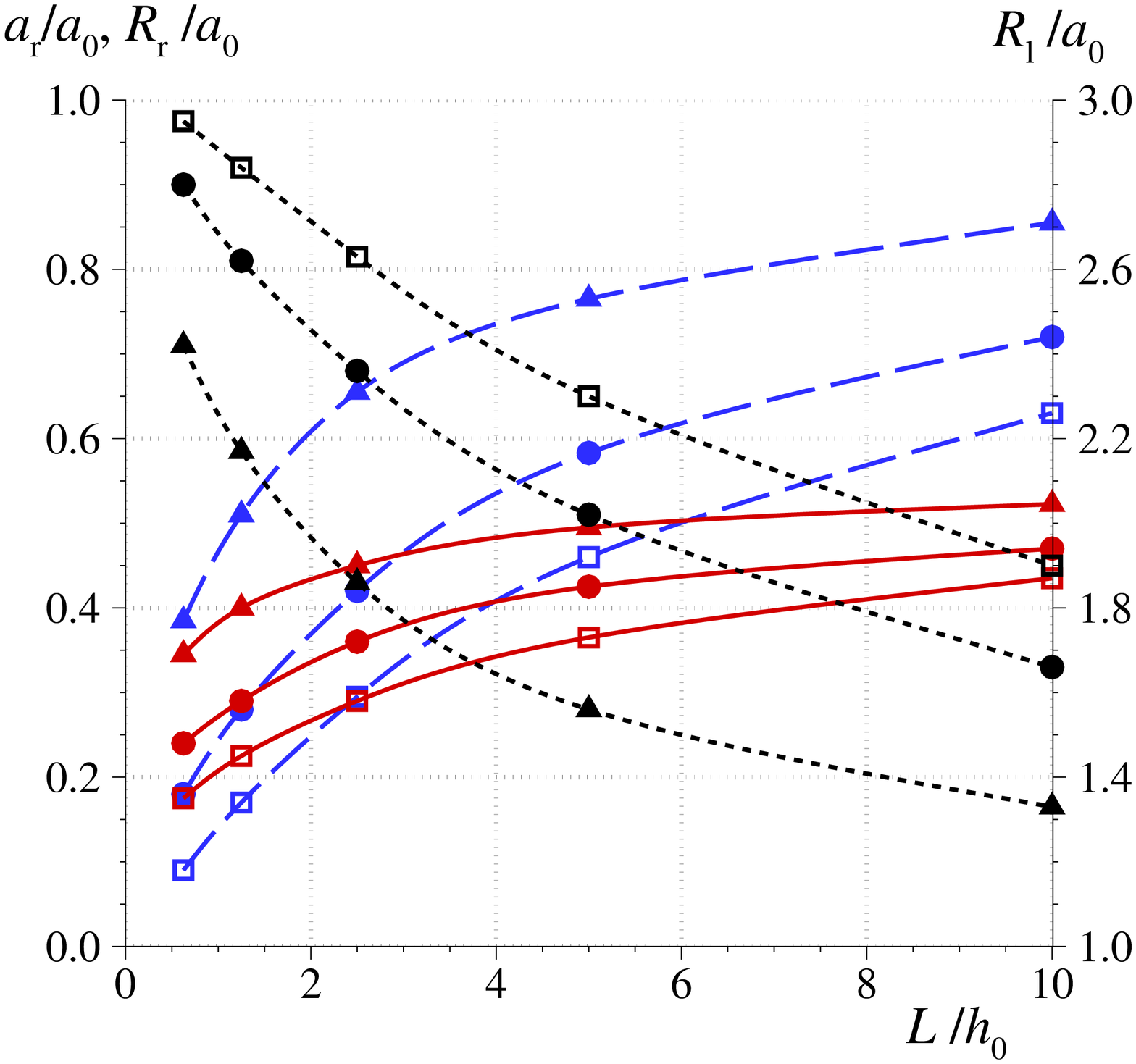}\\
\parbox[t]{0.49\textwidth}{\centering ({\it a})}
\hfill
\parbox[t]{0.49\textwidth}{\centering ({\it b})}\\
\vspace*{-1mm}
{\caption{$\Po$ model. Dependence of the amplitude of the reflected wave $a_r/a_0$ ({\bf ---\;---}), the maximum vertical runup on the front $R_l/a_0$ ({\bf ---}) 
and back $R_r/a_0$ ({\bf -\hspace*{0.3mm}-{\hspace*{0.3mm}}-}) faces of the body from the body length $L/h_0$ at 
({\it a}): fixed body depth $d_0/h_0=-0.5$ and different amplitudes of the incoming wave $a_0/h_0=0.1$~($\Box$), $0.2$~($\bullet$), $0.4$~($\blacktriangle$); 
({\it b}): fixed amplitude of the incoming wave $a_0/h_0=0.2$ and different body submergence $d_0/h_0=-0.2$~($\Box$), $-0.5$~($\bullet$), $-0.8$~($\blacktriangle$)
}
\label{G123_vs_L_var_ad}}
\end{figure}

Using the developed algorithm, large series of computational experiments were performed to study the wave pattern and the characteristics of the emerging waves when the initial data in (\ref{input_Pot}) changed, namely, the amplitude of the incoming wave $a_0$, the depth of the body $d_0$ and its length $L$.  Some results of these experiments are shown in Figures~\ref{G123_vs_a_var_dL}---\ref{G123_vs_L_var_ad} as plots for the reflection coefficient $a_r/a_0$ equal to the ratio of the maximum amplitude $a_r$ of the wave reflected from the body to the amplitude $a_0$ of the incoming wave, and the coefficients $R_l/a_0$ and $R_r/a_0$ of the relative maximum runups on the front  and back  faces of the body ($R_l$ and $R_r$, respectively).

With increasing $a_0$ and fixing other parameters, the reflection coefficient increases (less than the runup coefficient on the front face of the body), while the runup coefficient on the back face decreases (Fig.~\ref{G123_vs_a_var_dL}). Note that at large values of the relative amplitude of the incoming wave and small submergence or small body length, condition (\ref{wet_body_bottom}) is not hold, i.e., the body bottom is partially dried. In these cases, the presented algorithm fails, so some graphs in Fig.~\ref{G123_vs_a_var_dL} are not shown.
When the absolute value of the submergence $|d_0|$ and the length $L$ of the body in the horizontal direction increase, the behavior of the coefficients is the same as when the amplitude of the incoming wave increases (see Figures~\ref{G123_vs_d_var_aL} and~\ref{G123_vs_L_var_ad}). It is interesting that the amplitude of the wave passing behind the body is close to the value of the maximum vertical runup on the backside of the body.

The increase in the reflection and runup coefficients on the front face of the body and the decrease in the runup on the back face with increasing submergence $|d_0|$ and extent $L$ of the body can be explained by the fact that the body begins to act more and more as a vertical impermeable wall, i.e. the nature of the wave-body interaction is more and more like the wave-wall interaction.

\subsection{Some features of the numerical algorithm for solving nonlinear dispersive shallow water equations}

In contrast to the dispersionless $\SW$ equations, the equations of motion of the $\SGN$ model include mixed derivatives on time and space from the velocity vector components, which complicates the construction of the numerical algorithm. The original numerical algorithm for solving the $\SGN$ equations is described in detail in \cite{Khakimzyanov2016}. Therefore, here we briefly consider only those features of the algorithm that arise due to the presence of a semi-immersed body and the associated need to take into account the boundary condition (\ref{1D_NLD_Deta_dx}) and the compatibility conditions (C1) or (C2) on the common boundary $\Gamma$ of the outer ${\cal D}_e$ and inner ${\cal D}_i$ subdomains. If there is an obstacle in the form of a semi-immersed body, the $\SGN$ and $\SW$ equations should be solved separately at each time step for subdomains under and outside the body, coupling the obtained solutions by the compatibility conditions.

In the study \cite{Khakimzyanov2016} adaptive meshes were used to solve the $\SGN$ equations. However, we will use uniform meshes in the the algorithms for shallow water models, since the problem is one-dimensional and the desired accuracy can be achieved simply by increasing the number of nodes of the uniform grid. So, let us cover the region ${\cal D}$ with the uniform fixed grid $x_j$ ($j=0,\ldots,N$) with step $\Delta x=l/N$. We assume that $x_0=0$, $x_N=l$ and the boundaries of the body coincide with the grid nodes having numbers $j_l$ and $j_r$, i.~e.
\begin{equation*}
x_{l}=x_{j_l}, \quad x_{r}=x_{j_r}.
\end{equation*}
The nodes $x_{j_l}$ and $x_{j_r}$ are common for the sets ${\cal D}_e$ and ${\cal D}_i$. Note that in the one-dimensional case with the horizontal bottom, in contrast to the two-dimensional case, the calculation of values in the grid nodes under the body is not performed, because one ordinary differential equation (\ref{1D_NLD_to_NSWE_b}) is solved using compatibility conditions (C1) or (C2) instead of partial differential equations in the area under the body. Nevertheless, we will consider the grid also under the body, i.~e. use nodes $x_j$ ($j=j_l+1,\ldots, j_r-1$) that are not required for calculations in the problem considered here. This will help us compare the calculation results obtained within the $\SGN$ model and the $\Po$ model, as well as generalize the algorithm to the case of a non-horizontal moving bottom and a non-horizontal moving body bottom in the future.

Let us assume that on the time layer with number $n$ all the values are calculated. Thus, the free boundary $\eta^n_j$, velocity $u^n_j$ and dispersion component of pressure $\Pnh^n_j$ ($j=0,\ldots , j_l$, $j=j_r,\ldots , N$) are known outside the body and at the common boundary of the regions. The flow rate $Q^n$ is known under the body.

The predictor step first calculates the total depth $\H^*_{j+1/2}$ and velocity $u^*_{j+1/2}$ ($j=0,\ldots , j_l-1$, $j=j_r,\ldots , N-1$) in the centers of the grid cells covering the outer region ${\cal D}_e$. The description of the algorithm is available in \cite{Khakimzyanov2016}. Then the values of the dispersion component of the pressure $\Pnh^*_{j+1/2}$ are calculated. They are computed at the centers of the cells $x_{j+1/2}=x_j+\Delta x/2$ from the system of finite-difference equations approximating differential equation (\ref{curve_phi_1}). The integral form of equation (\ref{curve_phi_1}) and the finite-difference form of the compatibility conditions are used to obtain the finite-difference equations for $\Pnh^*_{j+1/2}$. The derivation of these finite-difference equations is given in Appendix.
The predictor step is completed by calculating the the rate of change of fluid flow under the body $\dot{Q}^*$.

In the corrector step, the total depth $\H^{n+1}_{j}$, velocity $u^{n+1}_{j}$ and dispersion component of the pressure $\Pnh^{n+1}_{j}$ are calculated. These grid functions are defined at the integer nodes $x_{j}$ of the grid covering the outer region ${\cal D}_e$ ($j=0,\ldots , j_l$, $j=j_r,\ldots , N$). The values of $\H^{n+1}_{j}$, $u^{n+1}_{j}$ in the inner nodes of this grid ($j=1,\ldots , j_l-1$, $j=j_r+1,\ldots , N_1$) are determined using the algorithm described in \cite{Khakimzyanov2016,Khakimzyanov2019c}.  The finite-difference approximation of the condition (\ref{Pt_2_Gamma_0}) is used at the outer boundary of $\Gamma_0$.
Condition (\ref{1D_NLD_Deta_dx}) is used at the common boundary of the $\Gamma$ subdomains ${\cal D}_e$ and ${\cal D}_i$. Next, the flow rate at the $(n+1)$ time step is determined using the predictor values of the rate of flow change $\dot{Q}^*$ under the body.
As in the predictor step, the finite-difference equations for $\Pnh^{n+1}_{j}$ are derived based on the integral form of equation (\ref{curve_phi_1}), but using different integration cells and other approximation formulas for the compatibility conditions. The detailed description of these formulas is given in Appendix.

Note that the algorithm described here has the property of rest state conservation as in the case without body \cite{Khakimzyanov2016}.

\section{Calculation results}\setcounter{equation}{0}\label{sec4}

Here we present the results of calculations obtained within the different mathematical models, their comparison between each other, and also with the experimental data. Finally, we investigate an interesting fact about the increase of runup on the vertical wall, ``protected'' by the semi-immersed body.

\subsection{Investigation of the wave-body interaction within the framework of the hierarchy of mathematical models}\label{num_results_1}

\begin{figure}[h!]
\centering
\includegraphics[width=0.45\textwidth]{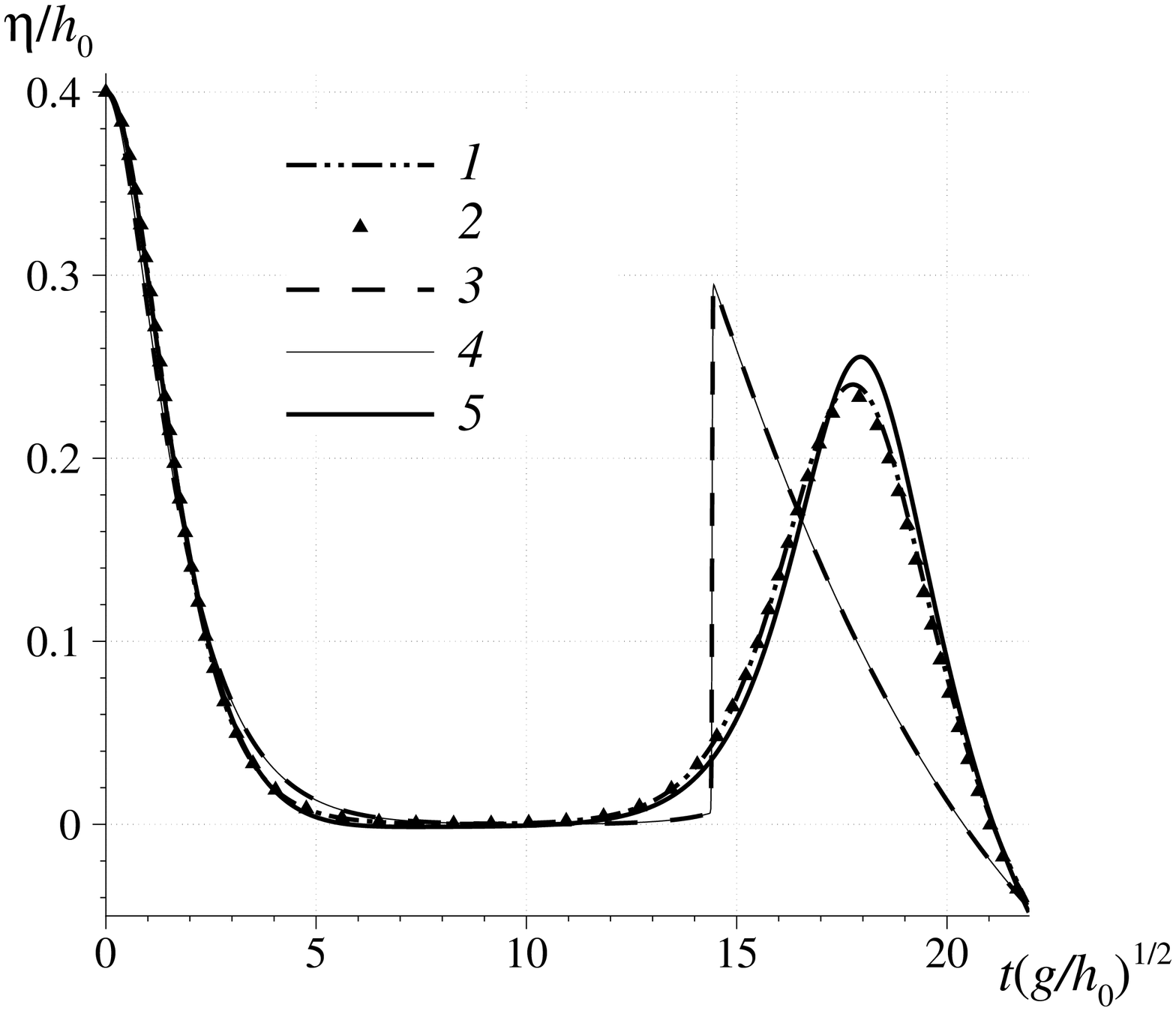}
\hfill
\includegraphics[width=0.45\textwidth]{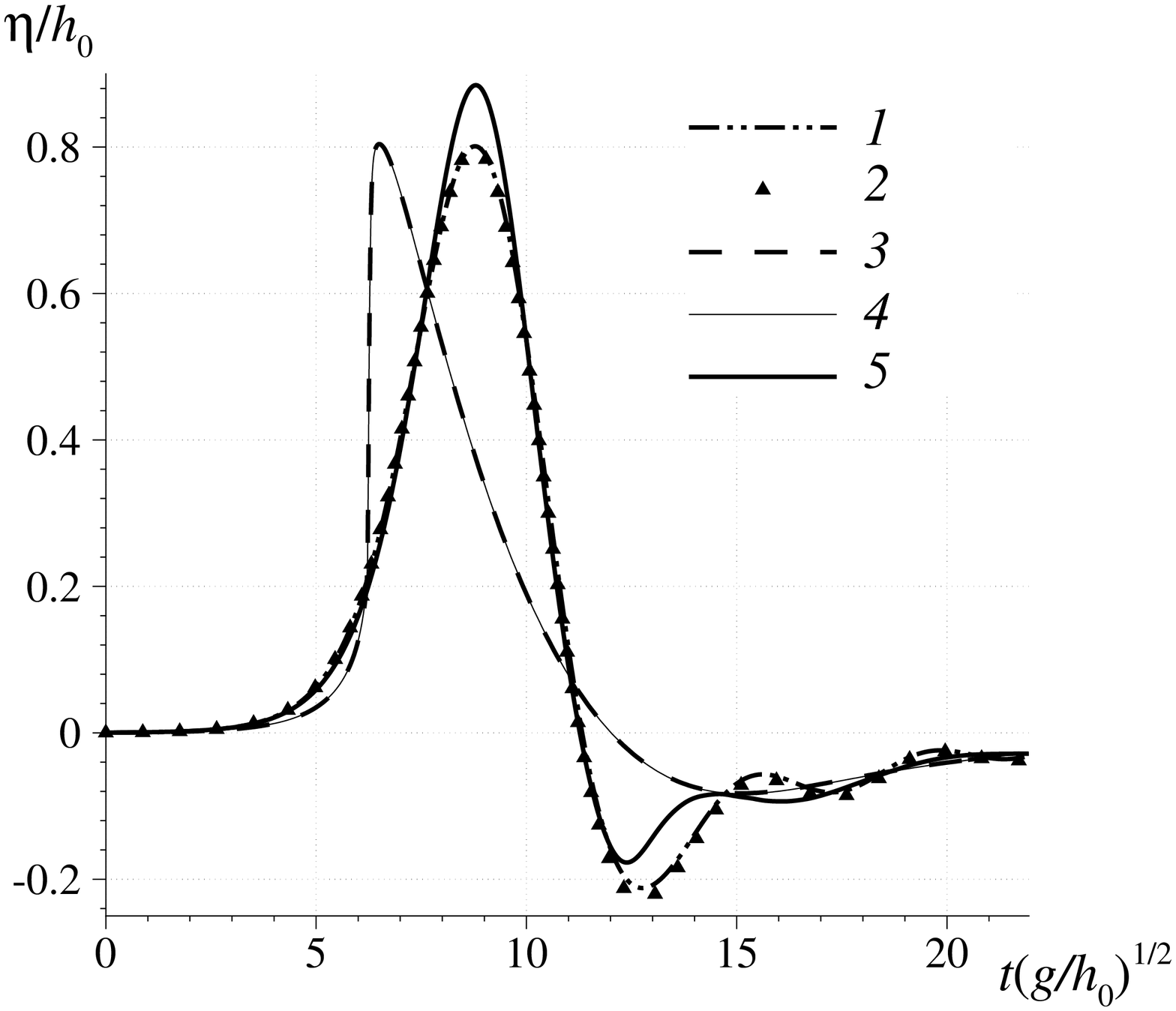}\\
\parbox[t]{0.45\textwidth}{\centering ({\it a})}
\hfill
\parbox[t]{0.45\textwidth}{\centering ({\it b})}\\
\includegraphics[width=0.45\textwidth]{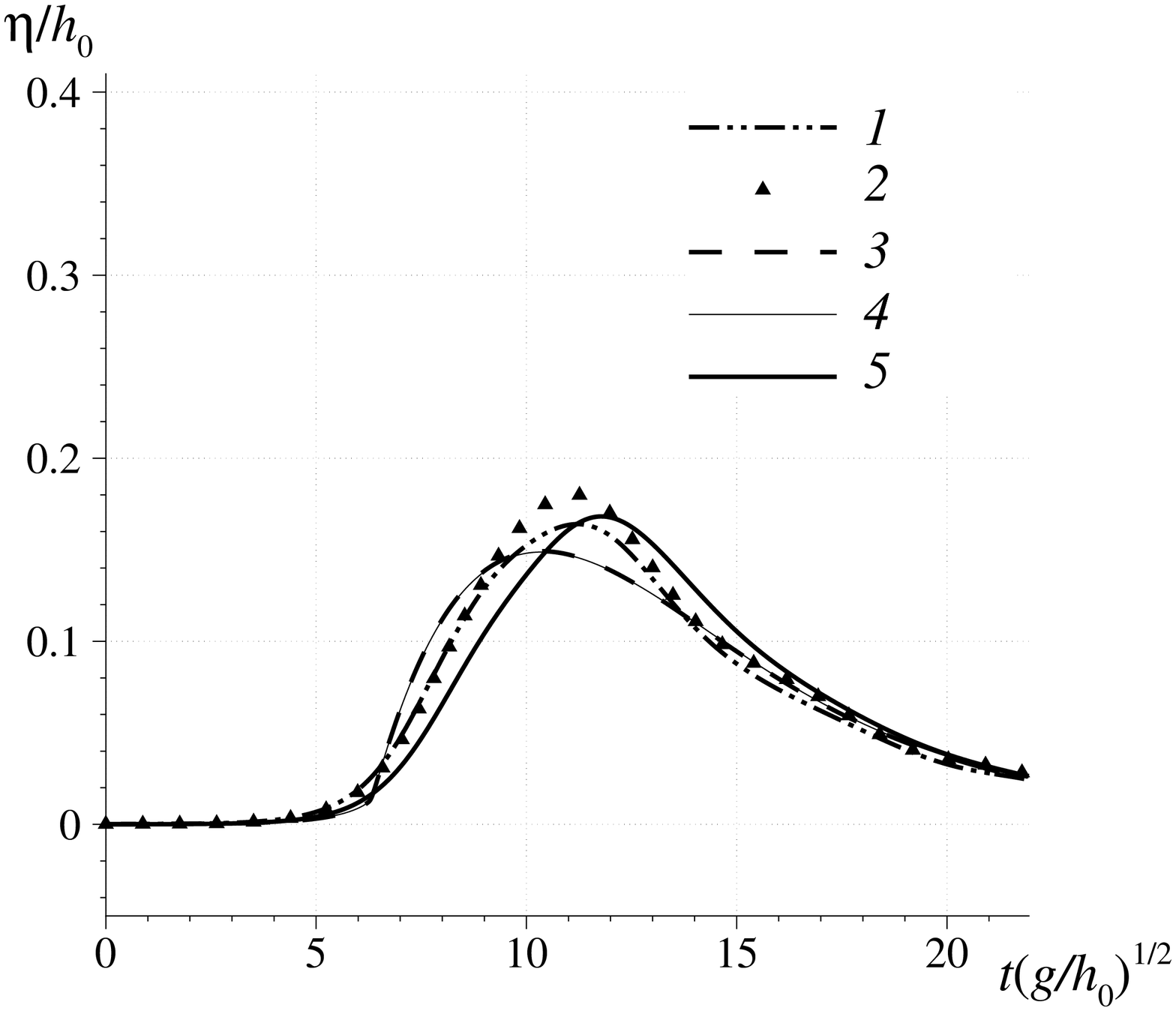}
\hfill
\includegraphics[width=0.45\textwidth]{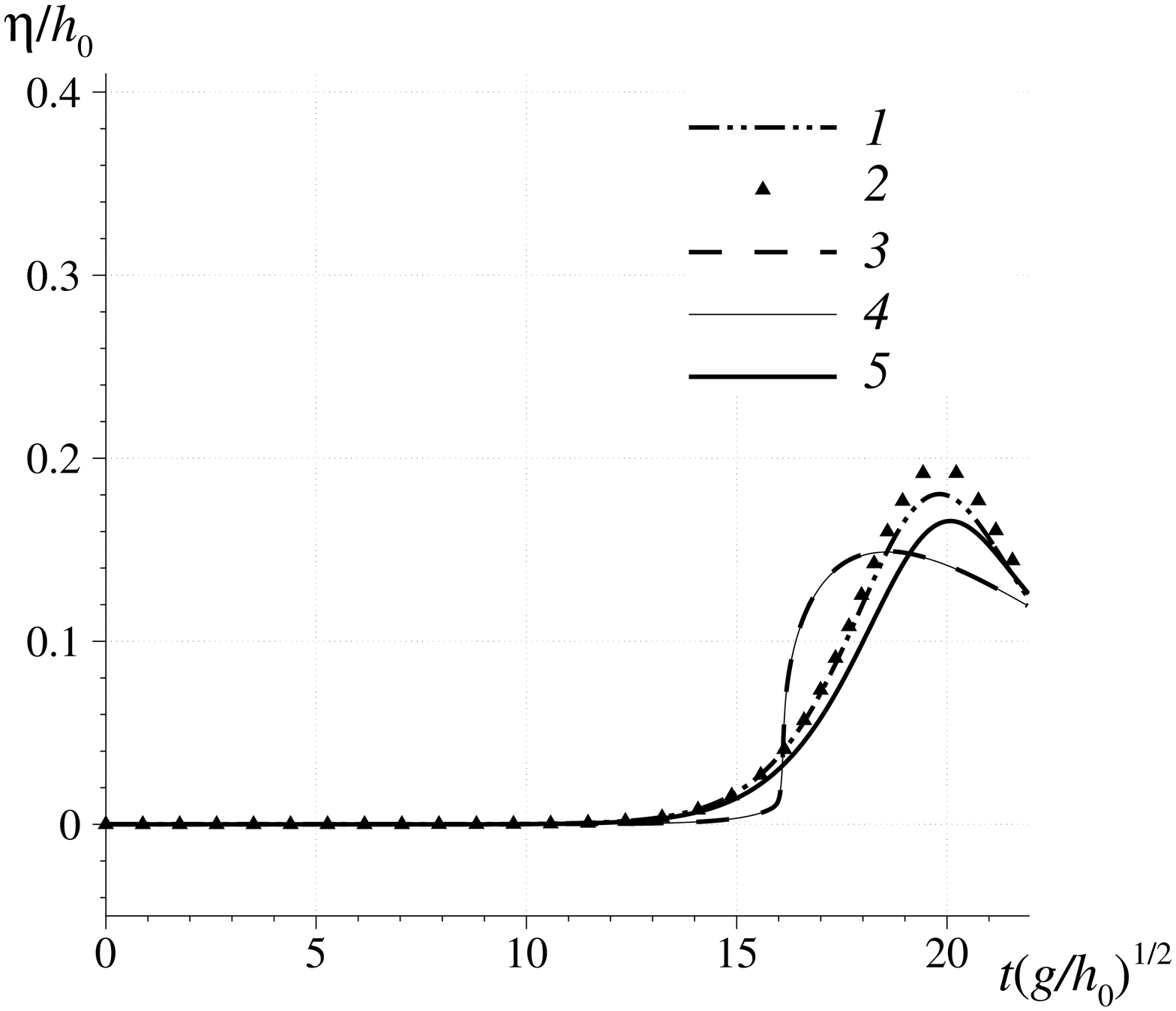}\\
\parbox[t]{0.45\textwidth}{\centering ({\it c})}
\hfill
\parbox[t]{0.45\textwidth}{\centering ({\it d})}\\
\vspace*{-1mm}

{\caption{Time histories of free surface at gauges $G_1$ ({\it a}), $G_2$ ({\it b}), $G_3$ ({\it c}), $G_4$ ({\it d}), calculated within: the $\SGN$ model with compatibility conditions (C1) ({\sl 1}), (C2) ({\sl 2}); $\SW$ model with compatibility conditions (C1) ({\sl 3}), (C2) ({\sl 4}); the $\Po$ model ({\sl 5}).  $a_0/h_0=0.4$, $L/h_0=5$, $d_0/h_0=-0.5$
\label{Mareogs_a04_Pot+NLD+NLSW}}}
\end{figure}

Figures \ref{Mareogs_a04_Pot+NLD+NLSW} show the free surface chronograms measured with virtual gauges (\ref{x_mareogrs}) during the $\Po$, $\SGN$ and $\SW$ model calculations for input data (\ref{input_Pot}). Two kinds of the compatibility conditions are used for the shallow water models: (C1) and (C2). For the dispersionless shallow water equations, both the conditions give identical results, indistinguishable in the plots. For the $\SGN$ model, there is a difference in the results when the different compatibility conditions are applied. Comparing with the results obtained in the $\Po$ model, we chose condition (C1) for further calculations.

\begin{figure}[h!]
\centering
\includegraphics[width=0.49\textwidth]{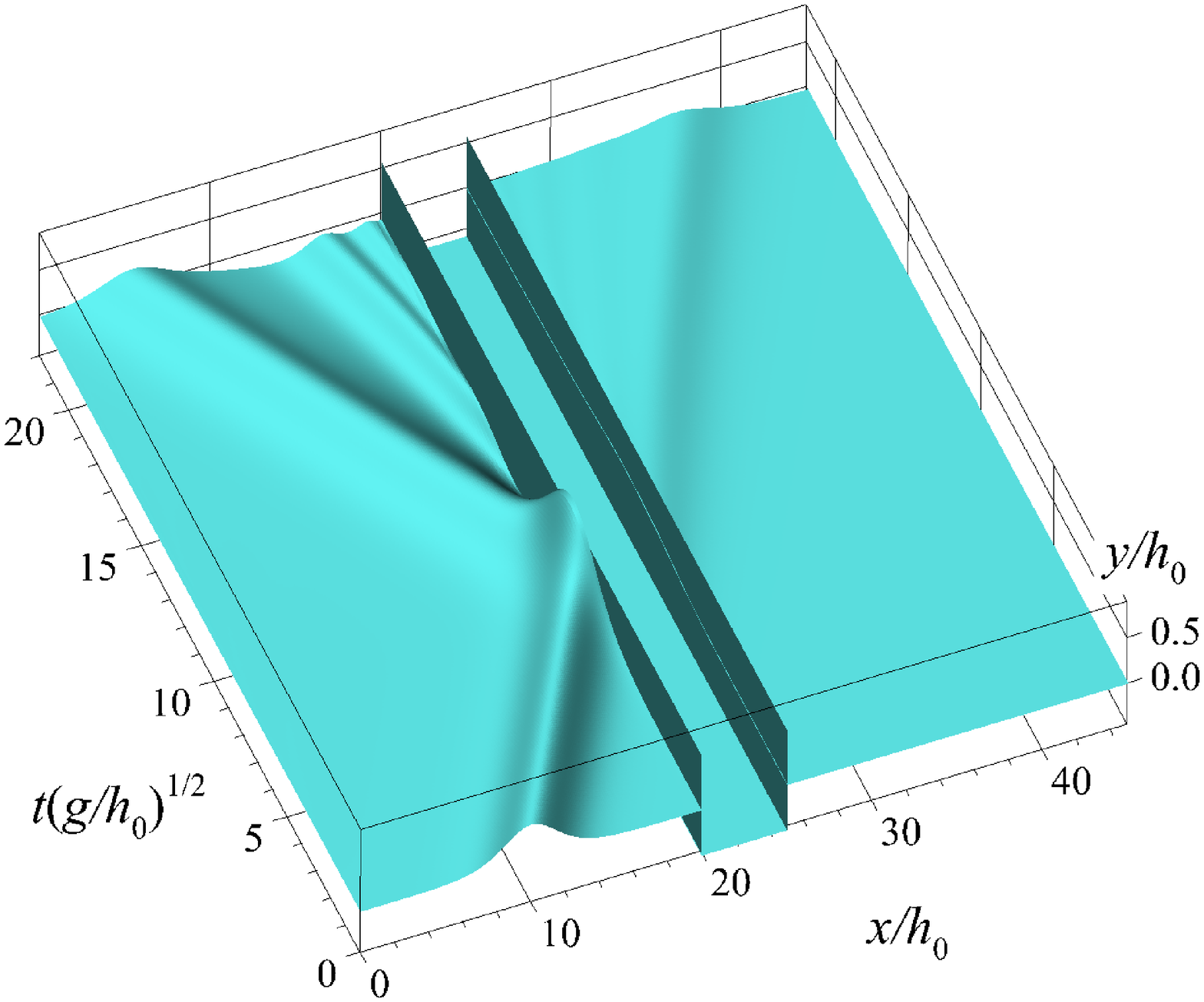}
\hfill
\includegraphics[width=0.49\textwidth]{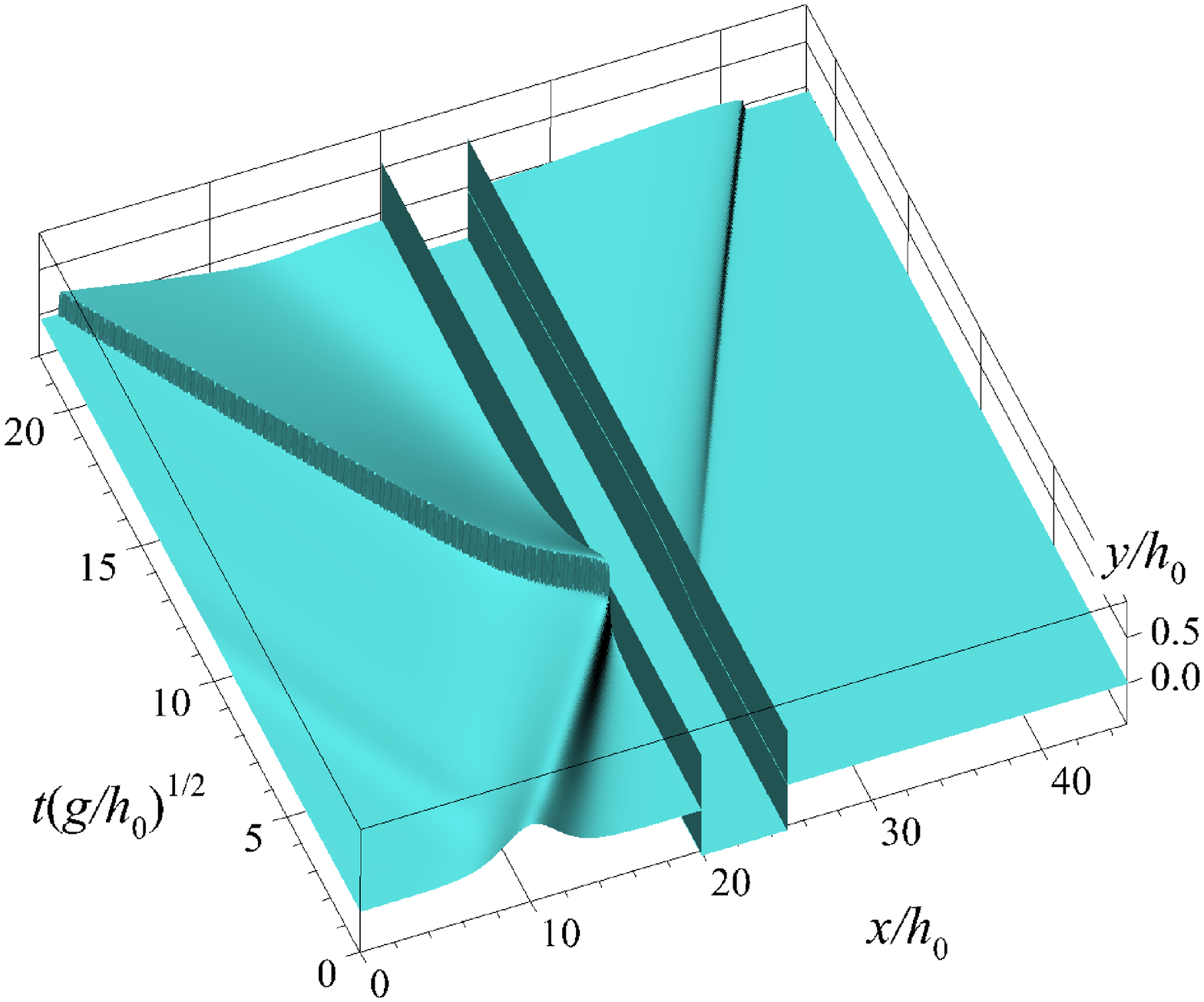}\\
\parbox[t]{0.49\textwidth}{\centering ({\it a})}
\hfill
\parbox[t]{0.49\textwidth}{\centering ({\it b})}\\
\vspace*{-1mm}
{\caption{Space–time plots of the free surface evolution, obtained within the $\SGN$ model ({\it a}) and the $\SW$ model ({\it b}).  $a_0/h_0=0.4$, $L/h_0=5$, $d_0/h_0=-0.5$
\label{Surf_NLD+NLSW}}}
\end{figure}

Comparing the results of calculations obtained with the $\SGN$ model (using compatibility condition (C1))  and the $\Po$ model, the largest differences are observed for the gauge $G_2$. It reaches the value of 11~\%. Note that for the smaller amplitude $a_0/h_0=0.2$ the differences do not exceed 5~\%. As for the $\SW$ model, it produces the larger amplitude of the reflected wave than the $\SGN$ model, the smaller amplitude of the transmitted wave, and the steeper leading fronts of the waves.
The differences in the interaction patterns are particularly well seen in Fig.~\ref{Surf_NLD+NLSW} depicting the dynamics of the free surface. Both the reflected and the passed waves in the $\SW$ model turn into bores. However, the interaction patterns computed within the $\SGN$ model (Fig.~\ref{Surf_NLD+NLSW}({\it a})) and the $\Po$ model (Fig.~\ref{Mar+Surf_Pot}({\it b}) ) are very similar: only with careful consideration one can notice the difference in reproducing water level fluctuations near the body face.

\begin{figure}[h!]
\centering
\includegraphics[width=0.49\textwidth]{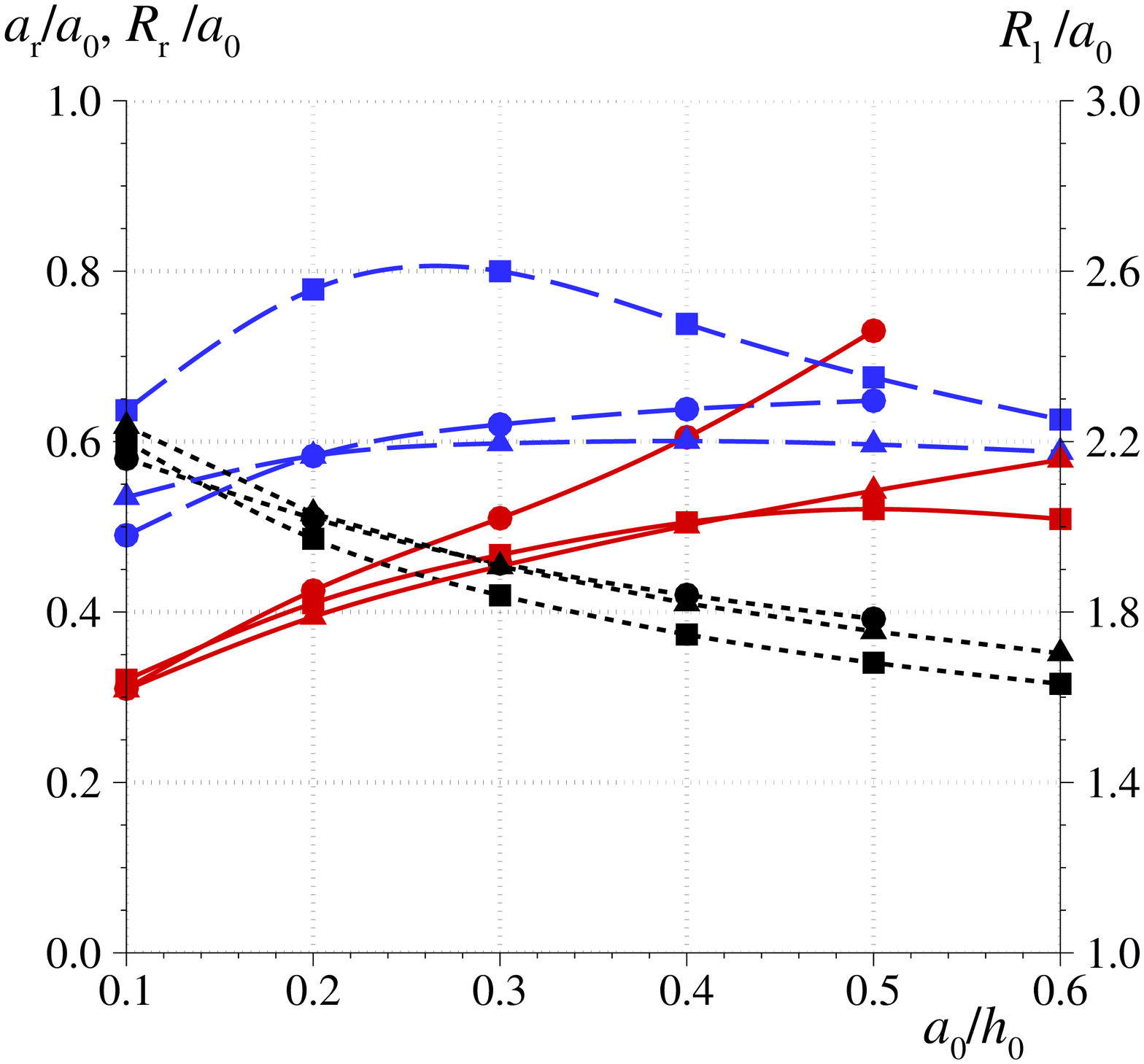}
\hfill
\includegraphics[width=0.49\textwidth]{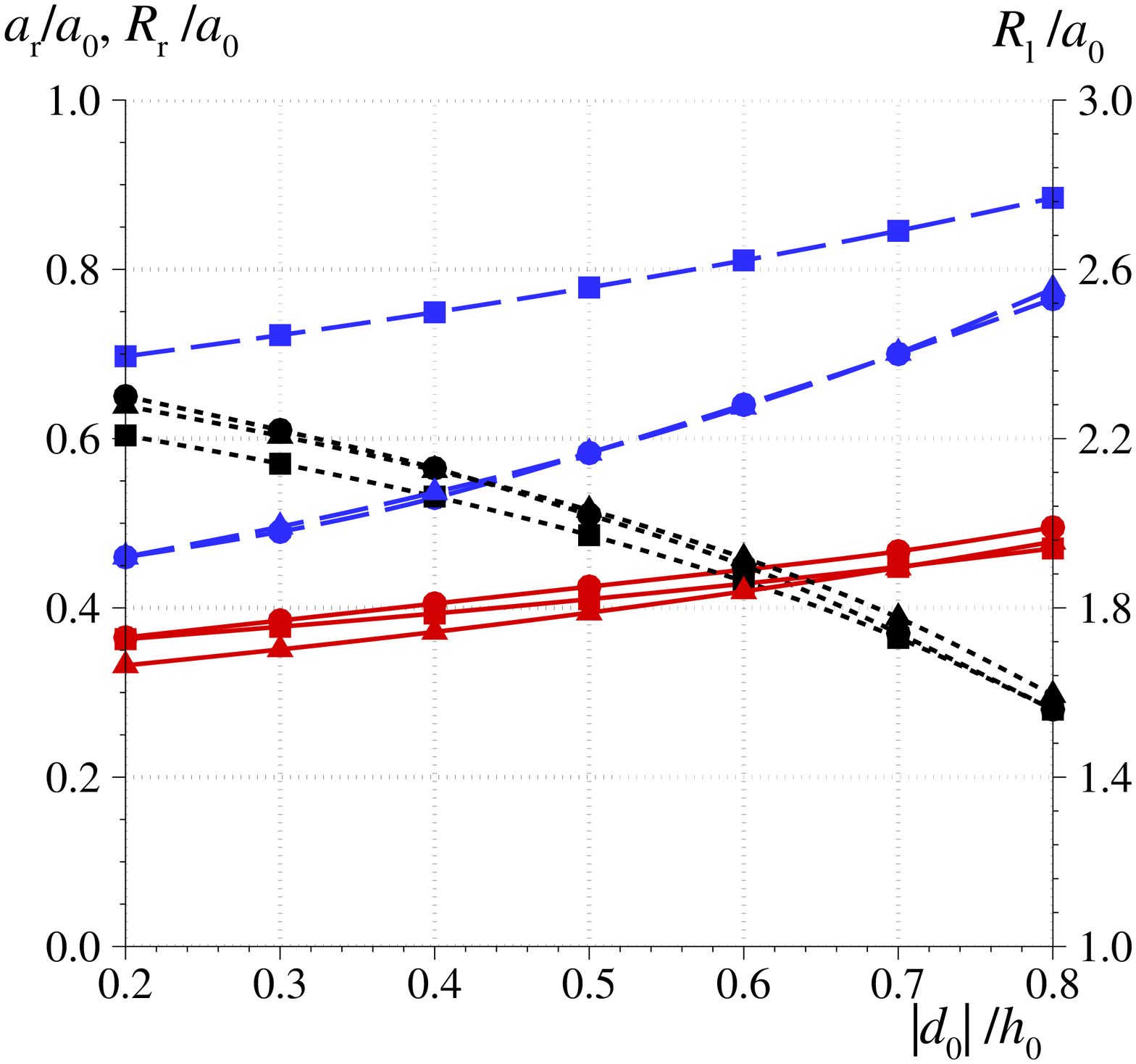}\\
\parbox[t]{0.49\textwidth}{\centering ({\it a})} \hfill \parbox[t]{0.49\textwidth}{\centering ({\it b})}\\
\includegraphics[width=0.49\textwidth]{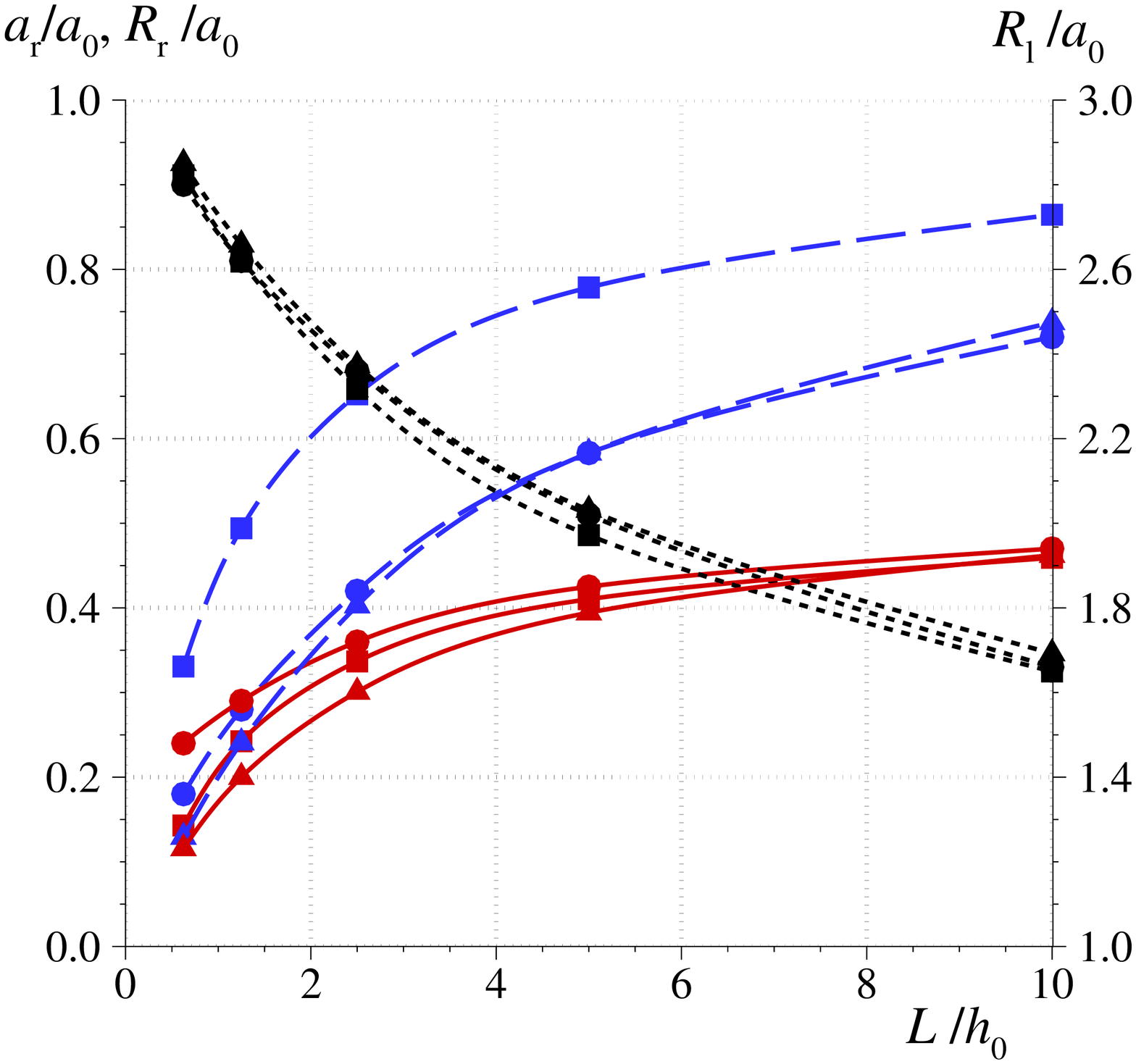}\\
\parbox[t]{0.49\textwidth}{\centering ({\it c})}
{\caption{Dependence of the amplitude of the reflected wave $a_r/a_0$ ({\bf ---\;---}), the maximum vertical runup on the front $R_l/a_0$ ({\bf ---}) and back $R_r/a_0$ ({\bf -\hspace*{0.3mm}-\hspace*{0.3mm}-}) faces of the body from:
the amplitude $a_0/h_0$ of the incoming wave with the fixed body length $L/h_0=5$ and submergence $d_0/h_0=-0.5$ ({\it a});
the body submergence $|d_0|/h_0$ with the fixed wave amplitude $a_0/h_0=0.2$ and body length $L/h_0=5$;
body length $L/h_0$ with the fixed wave amplitude $a_0/h_0=0.2$ and body submergence $d_0/h_0=-0.5$ ({\it c}). The calculations were made using the $\Po$ model~($\bullet$), $\SGN$ model~($\blacktriangle$) and $\SW$ model ($\blacksquare$)}
\label{G123_vs_adL_PT_SW_NLD}}
\end{figure}

Fig.~\ref{G123_vs_adL_PT_SW_NLD} shows plots of dependence of runup on the front and back faces of the body, as well as the reflected wave amplitude on the amplitude of the incoming solitary wave, body submergence and its length. These results were obtained numerically within the considered hierarchy of mathematical models and also confirm good agreement between the Pot and $\SGN$ models.
\begin{figure}[b!]
\centering
\includegraphics[width=0.49\textwidth]{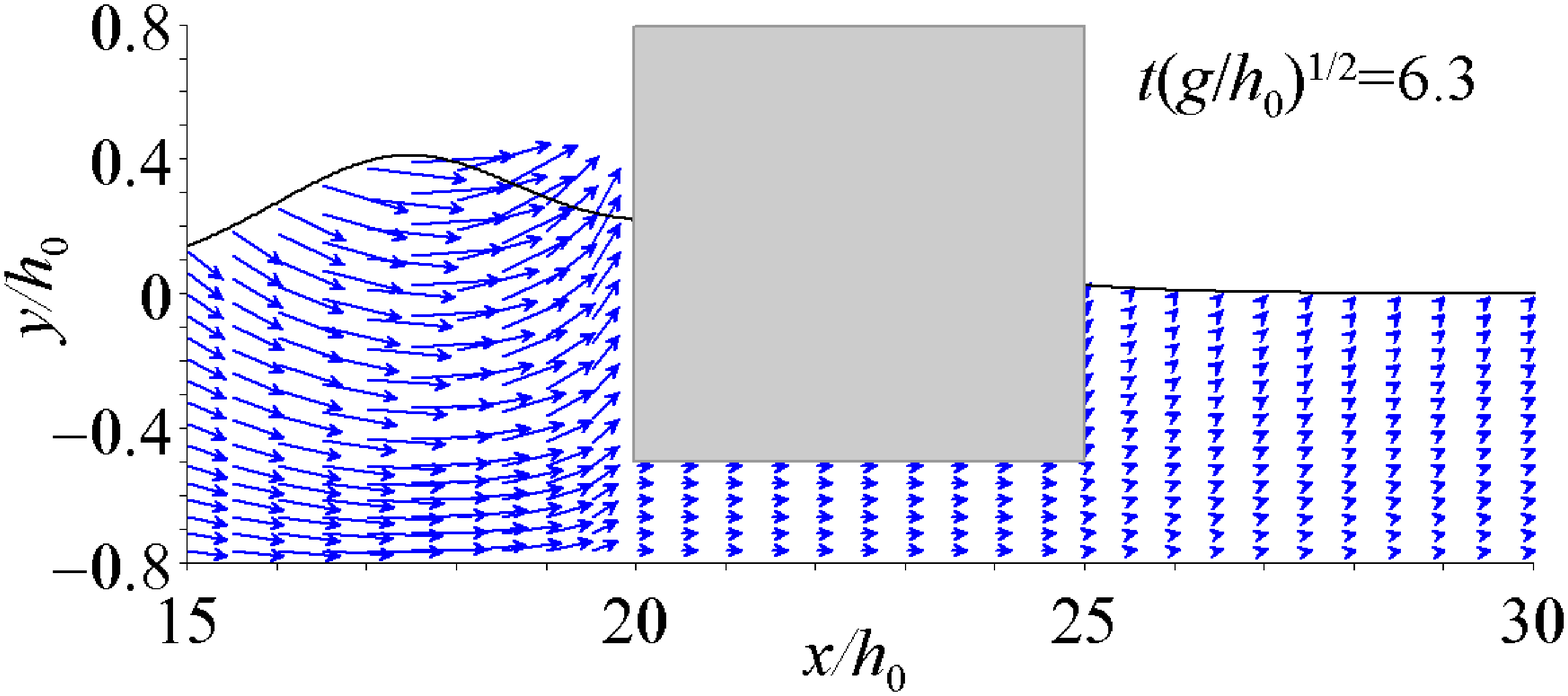}\hfill\includegraphics[width=0.49\textwidth]{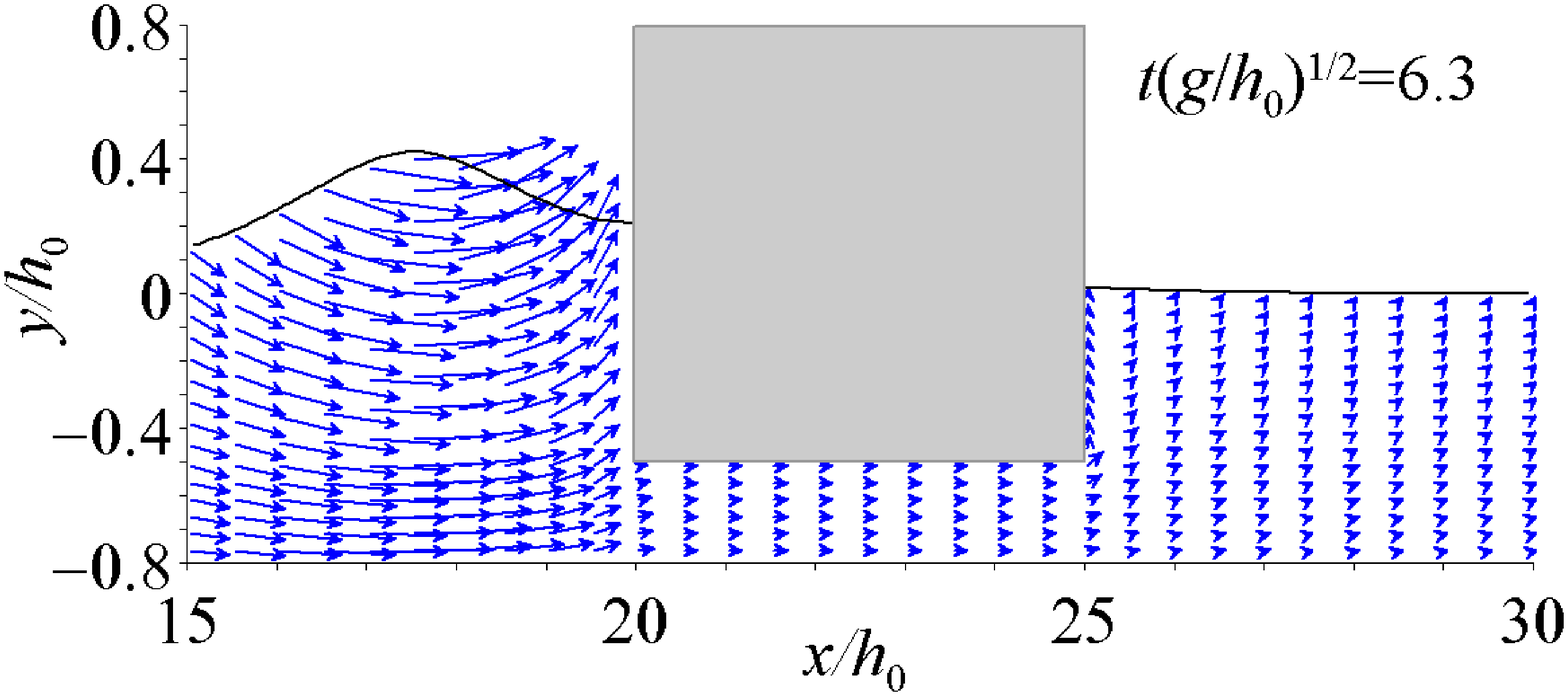}\\
\includegraphics[width=0.49\textwidth]{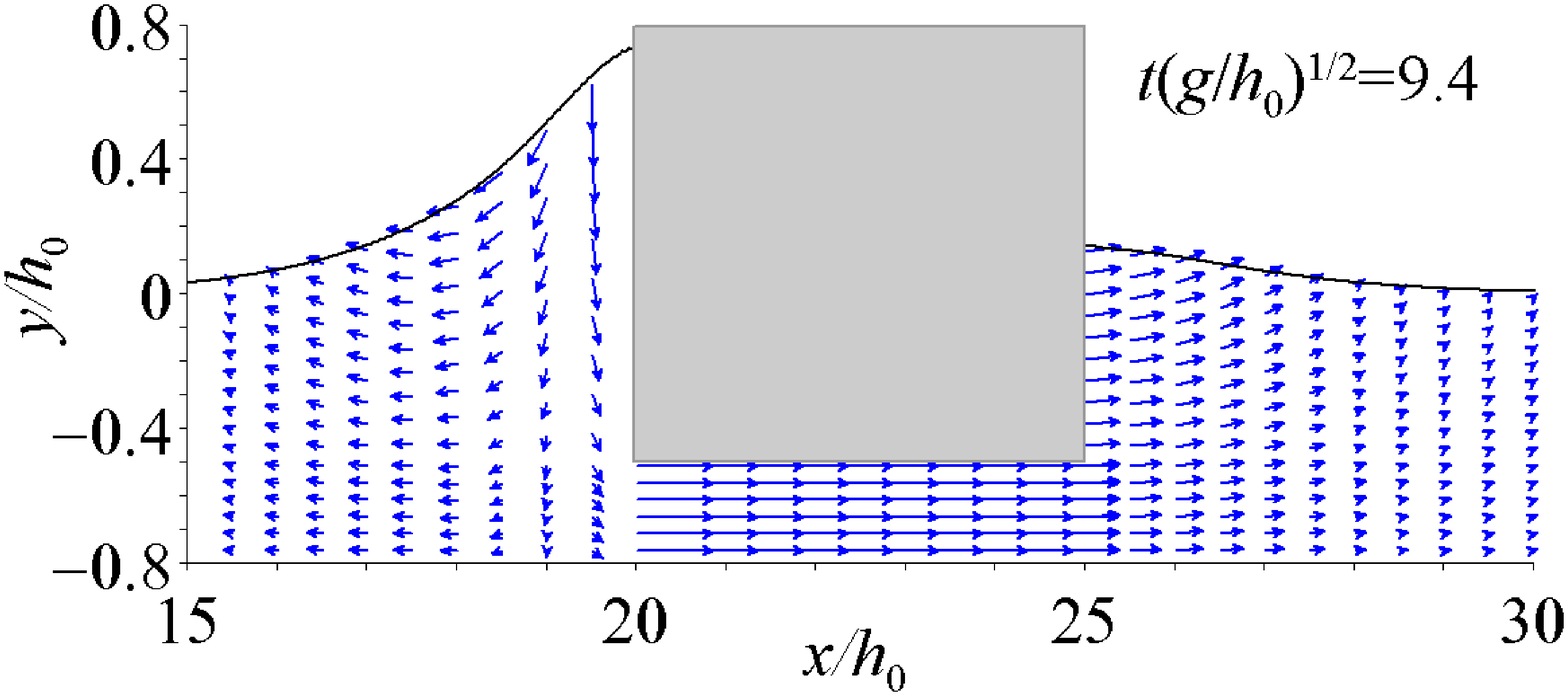}\hfill\includegraphics[width=0.49\textwidth]{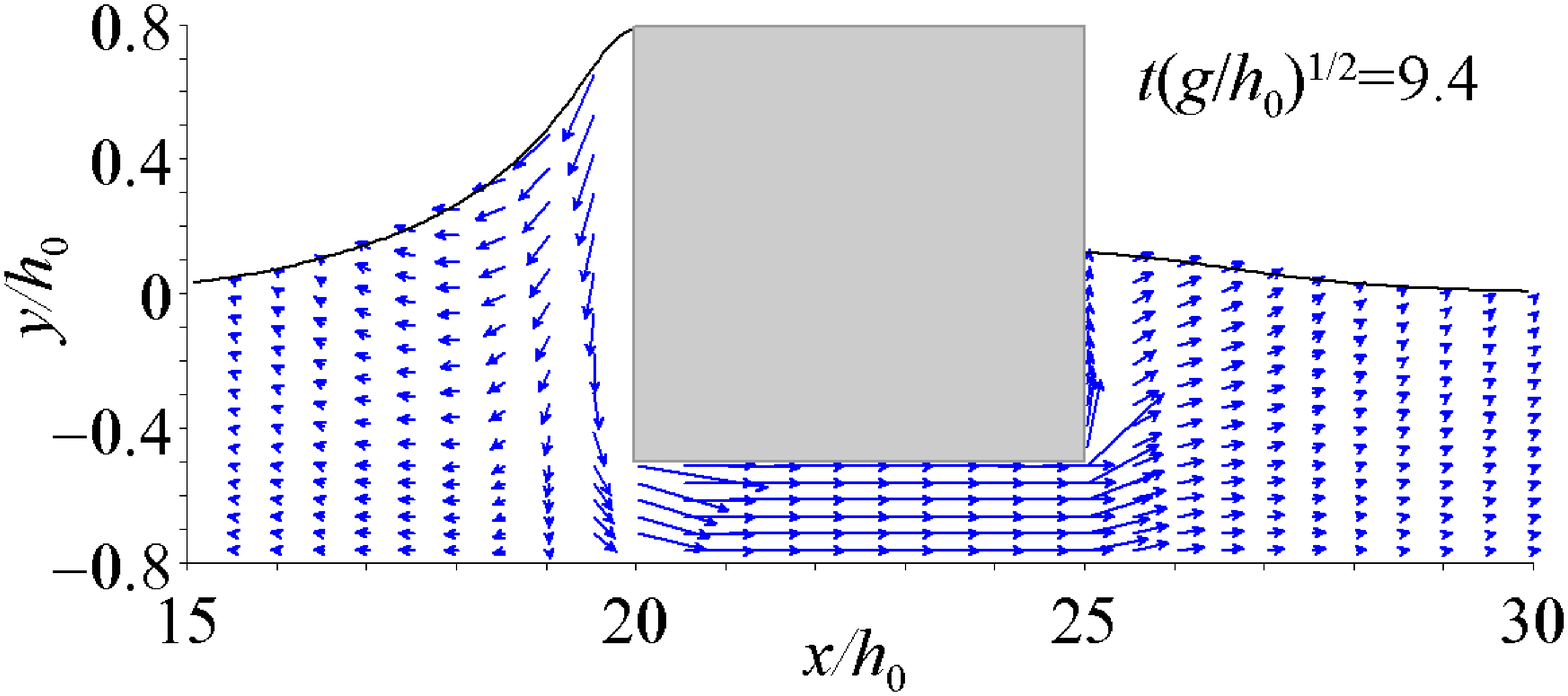}\\
\includegraphics[width=0.49\textwidth]{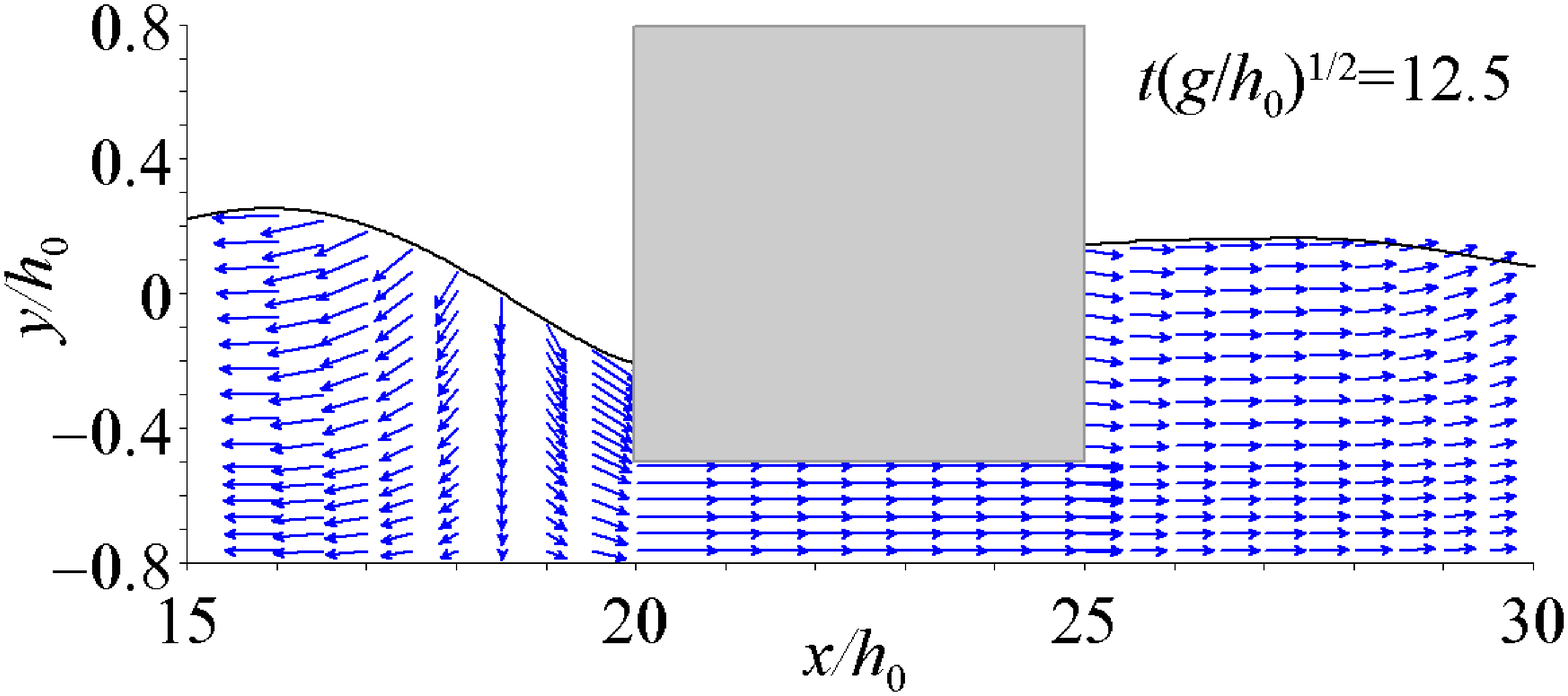}\hfill\includegraphics[width=0.49\textwidth]{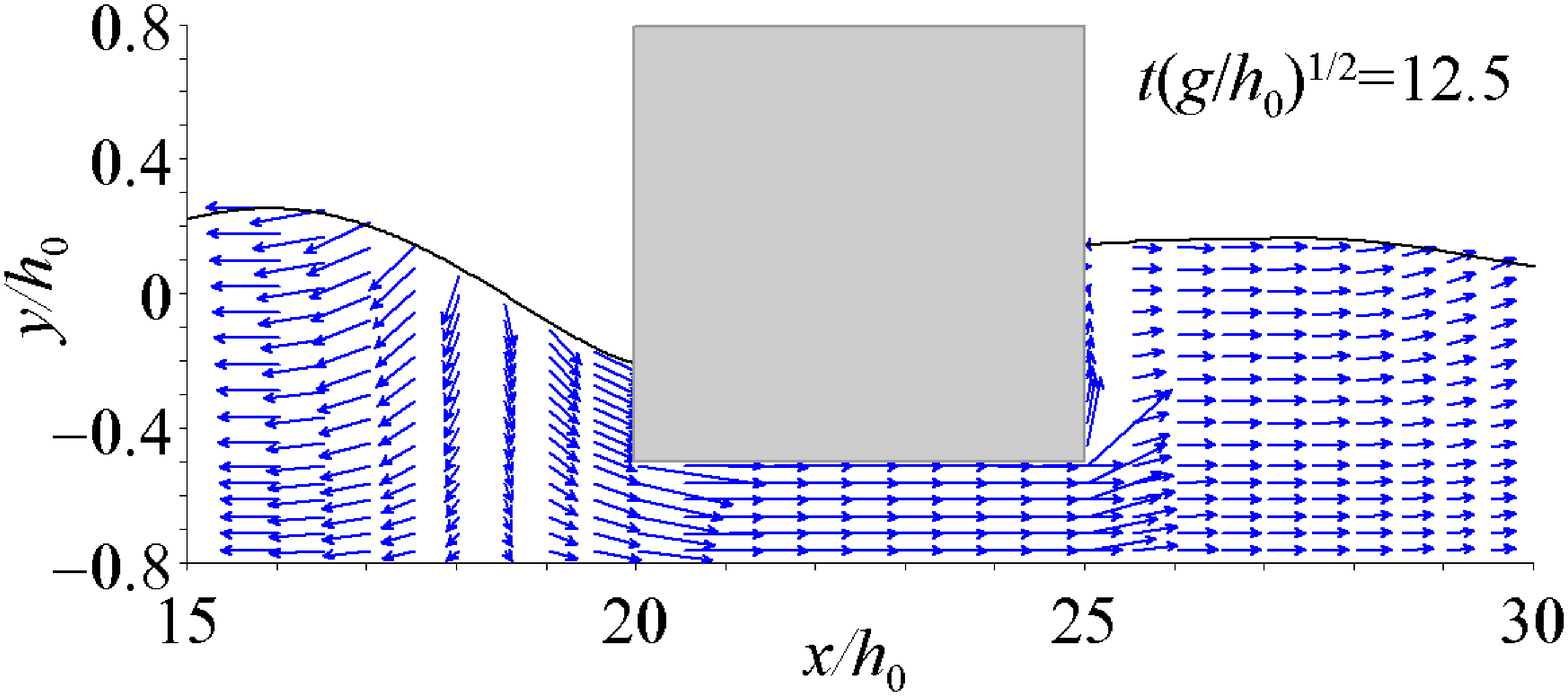}\\
\includegraphics[width=0.49\textwidth]{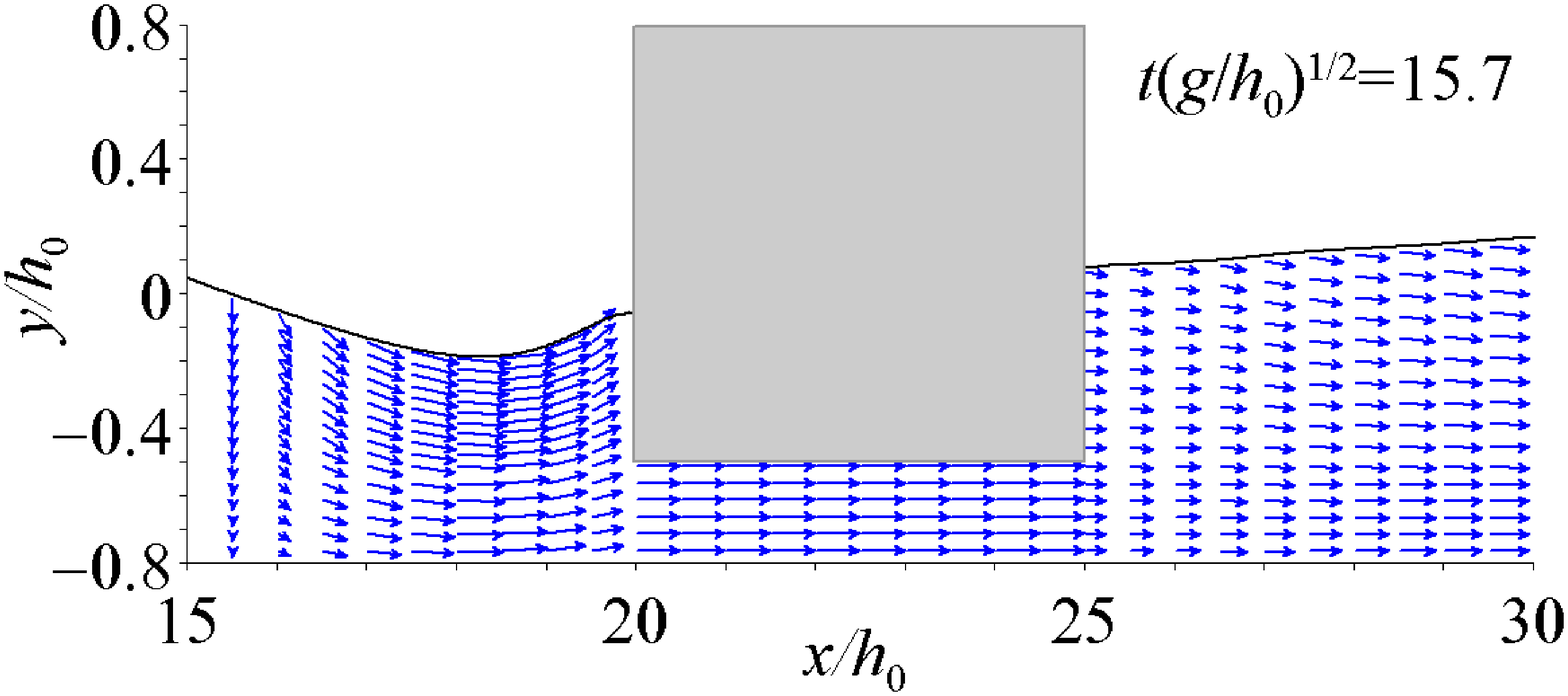}\hfill\includegraphics[width=0.49\textwidth]{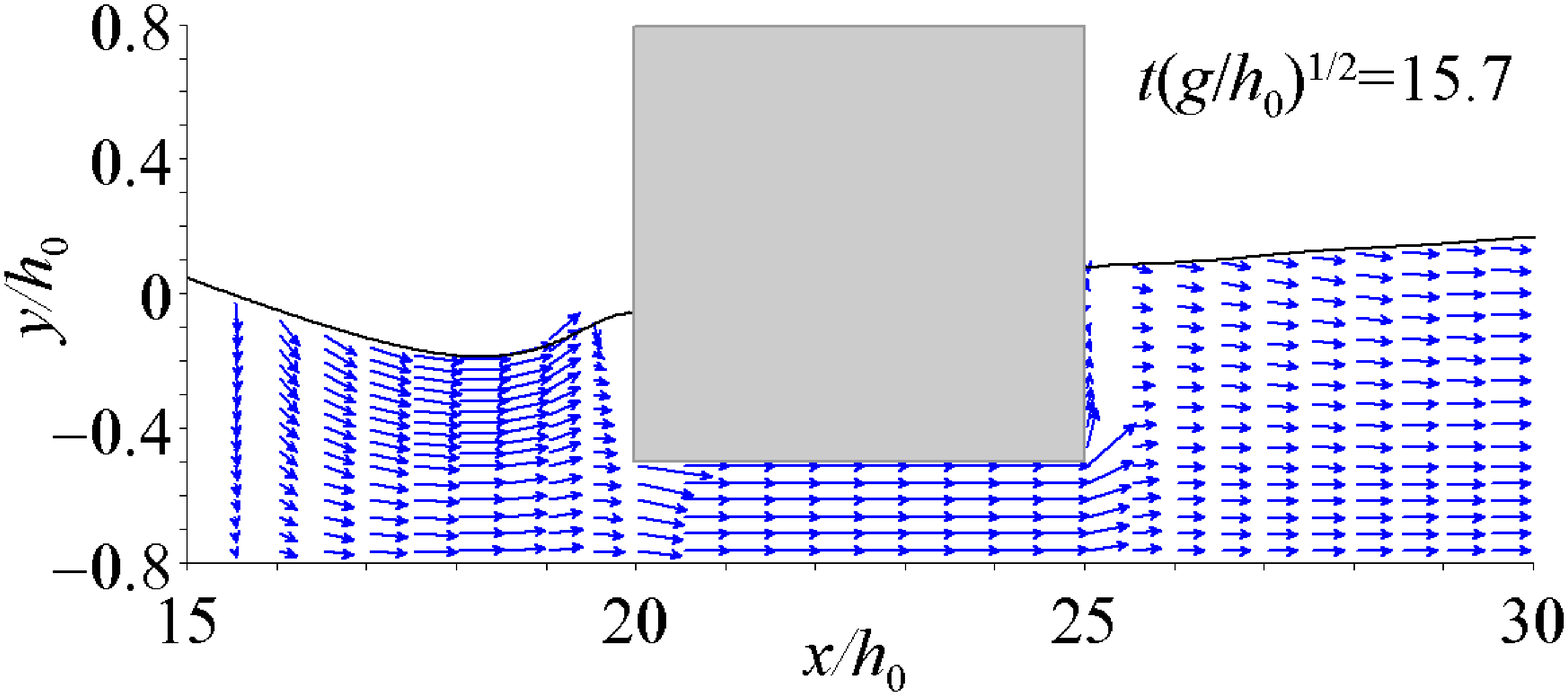}\\
\includegraphics[width=0.49\textwidth]{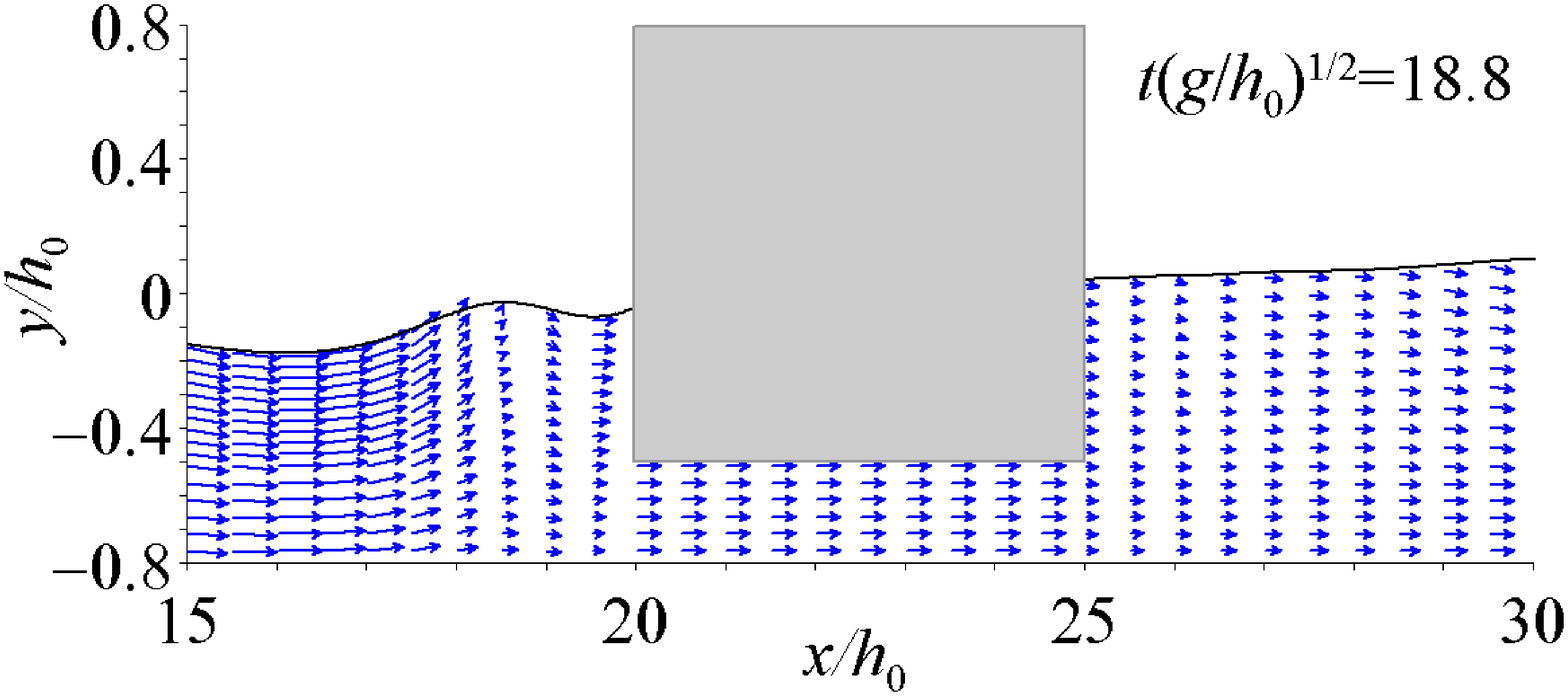}\hfill\includegraphics[width=0.49\textwidth]{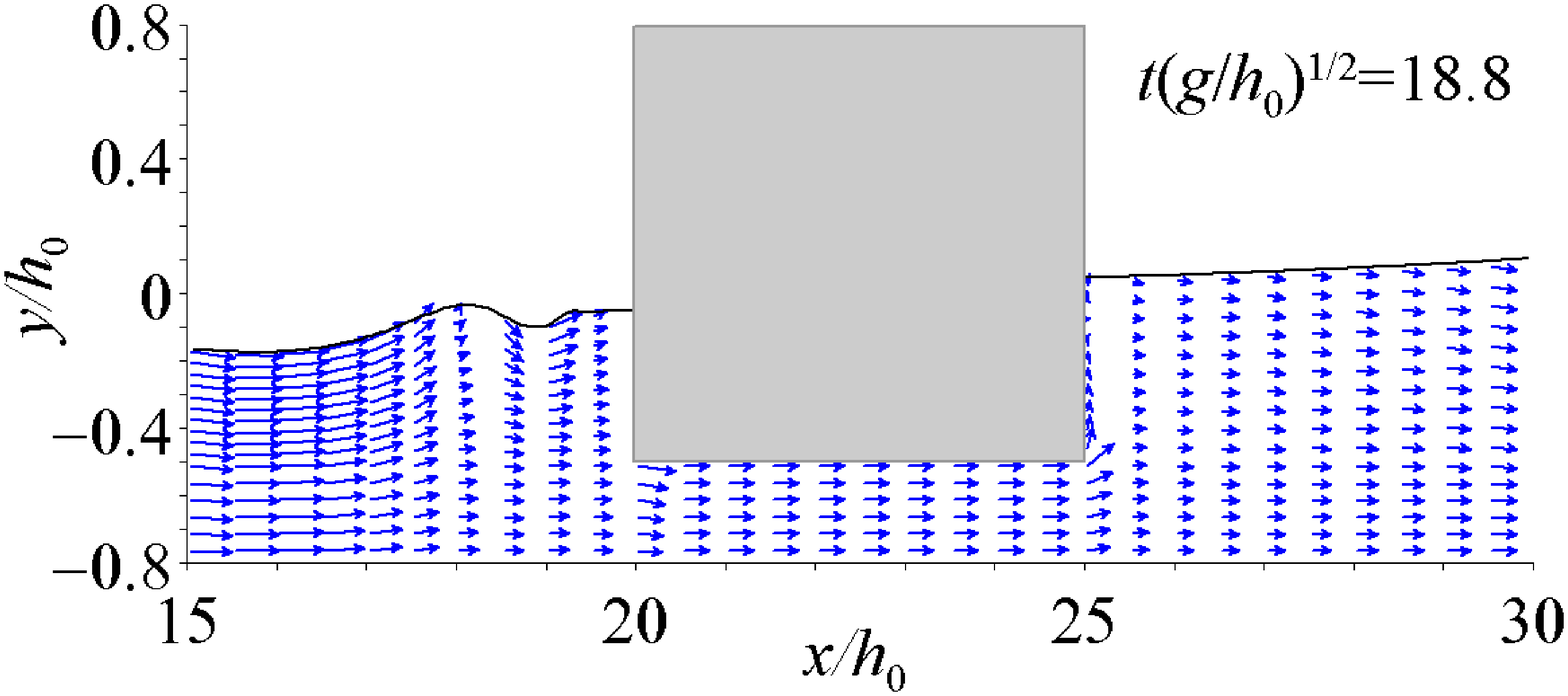}\\
{\caption{Runup of the solitary wave on the semi-immersed extended body. Velocity vector fields in the vicinity of the body at different moments of time calculated by the $\SGN$ model (left) and the $\Po$ model (right). $a_0/h_0=0.4$, $L/h_0=5$, $d_0/h_0=-0.5$
\label{Vel_Fields_a04_NLD+Pot}}}
\end{figure}
\begin{figure}[h!]
\centering 
\includegraphics[width=0.49\textwidth]{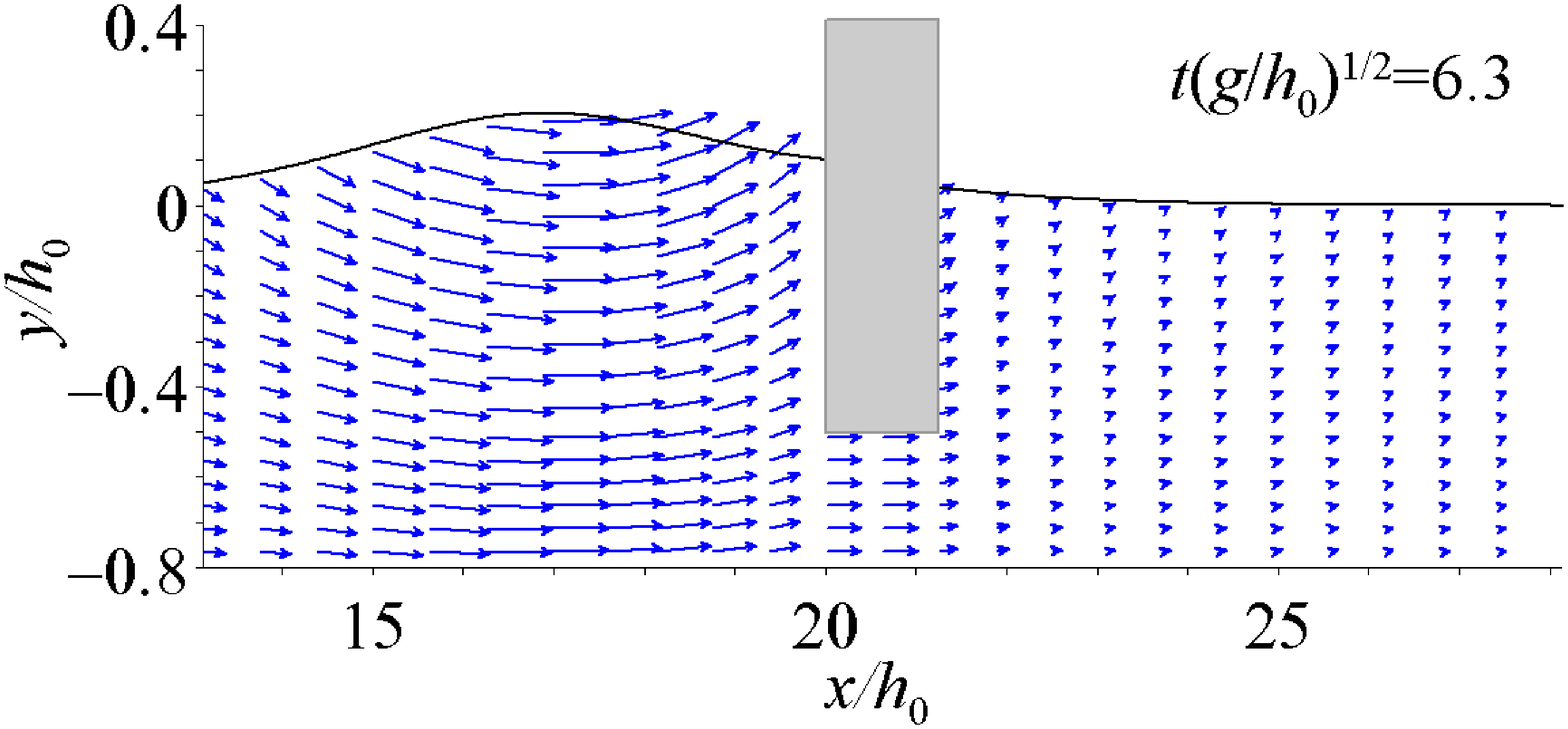}\hfill\includegraphics[width=0.49\textwidth]{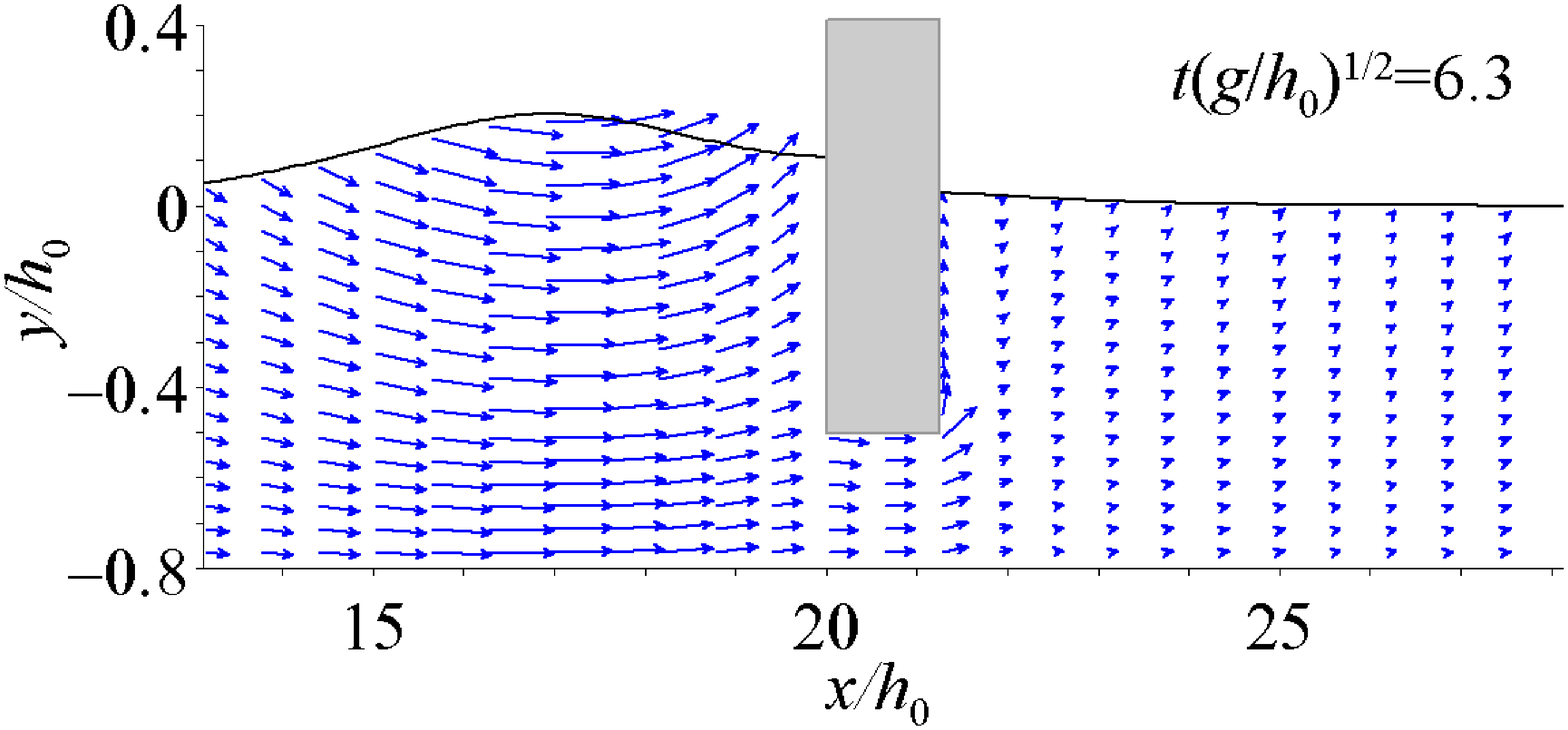}\\
\includegraphics[width=0.49\textwidth]{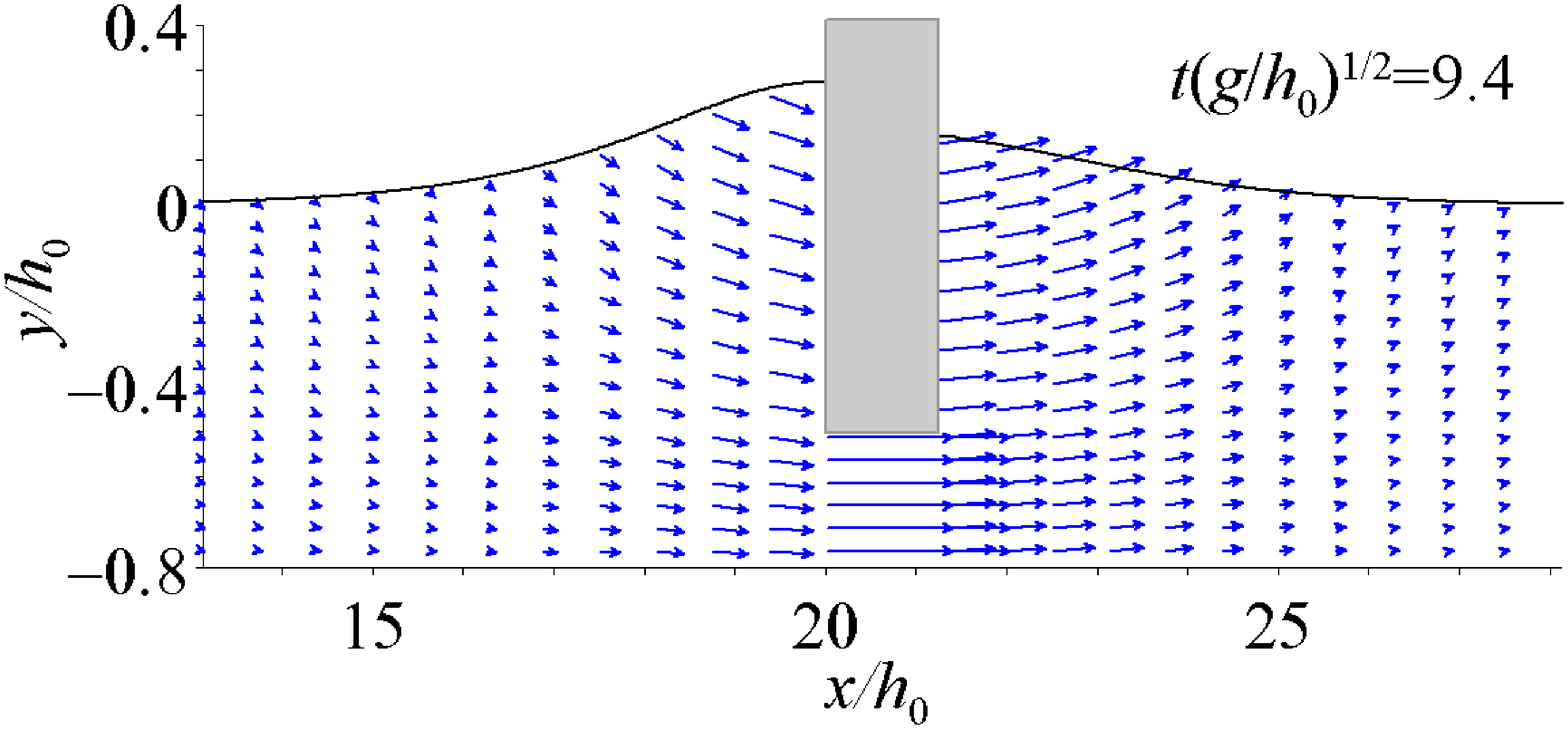}\hfill\includegraphics[width=0.49\textwidth]{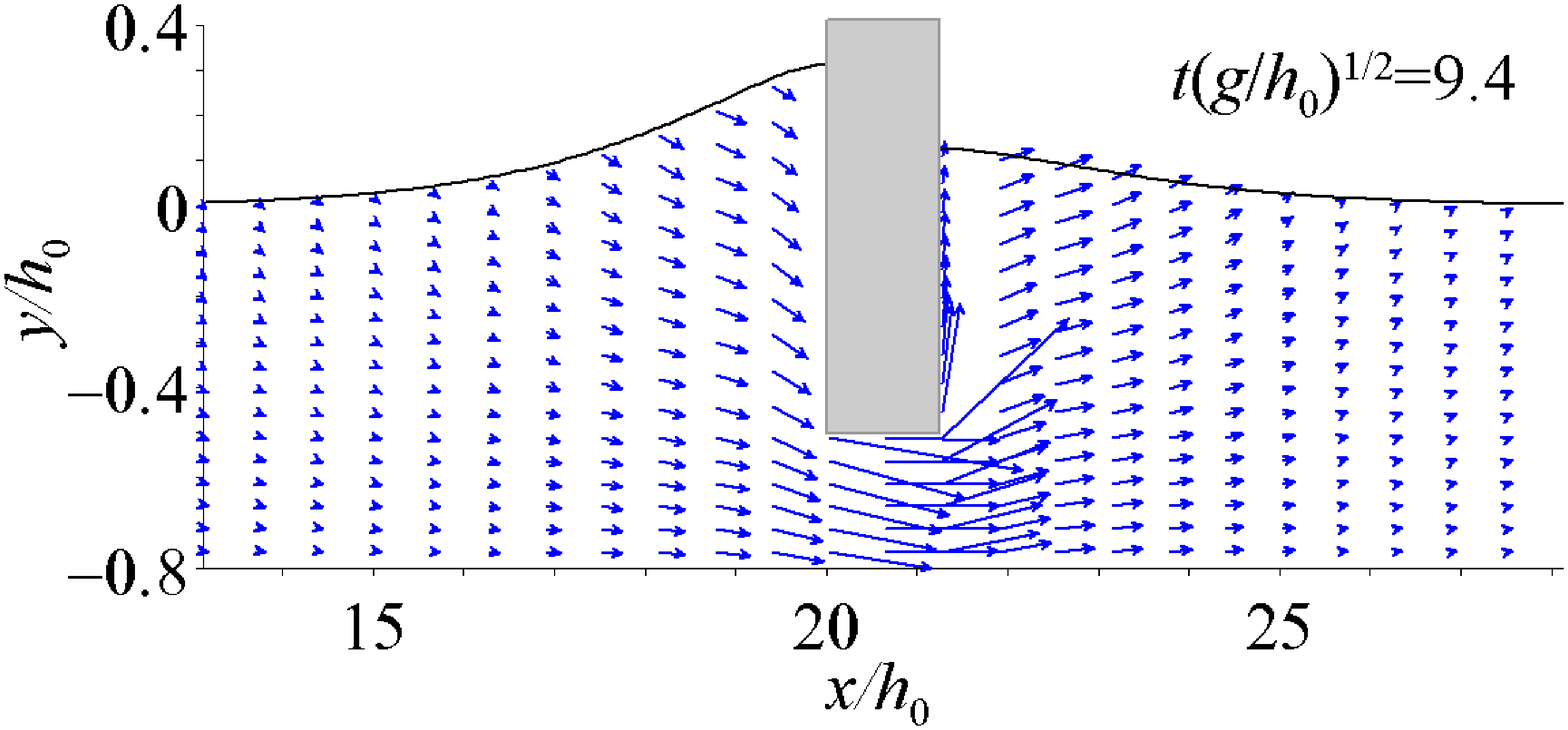}\\
\includegraphics[width=0.49\textwidth]{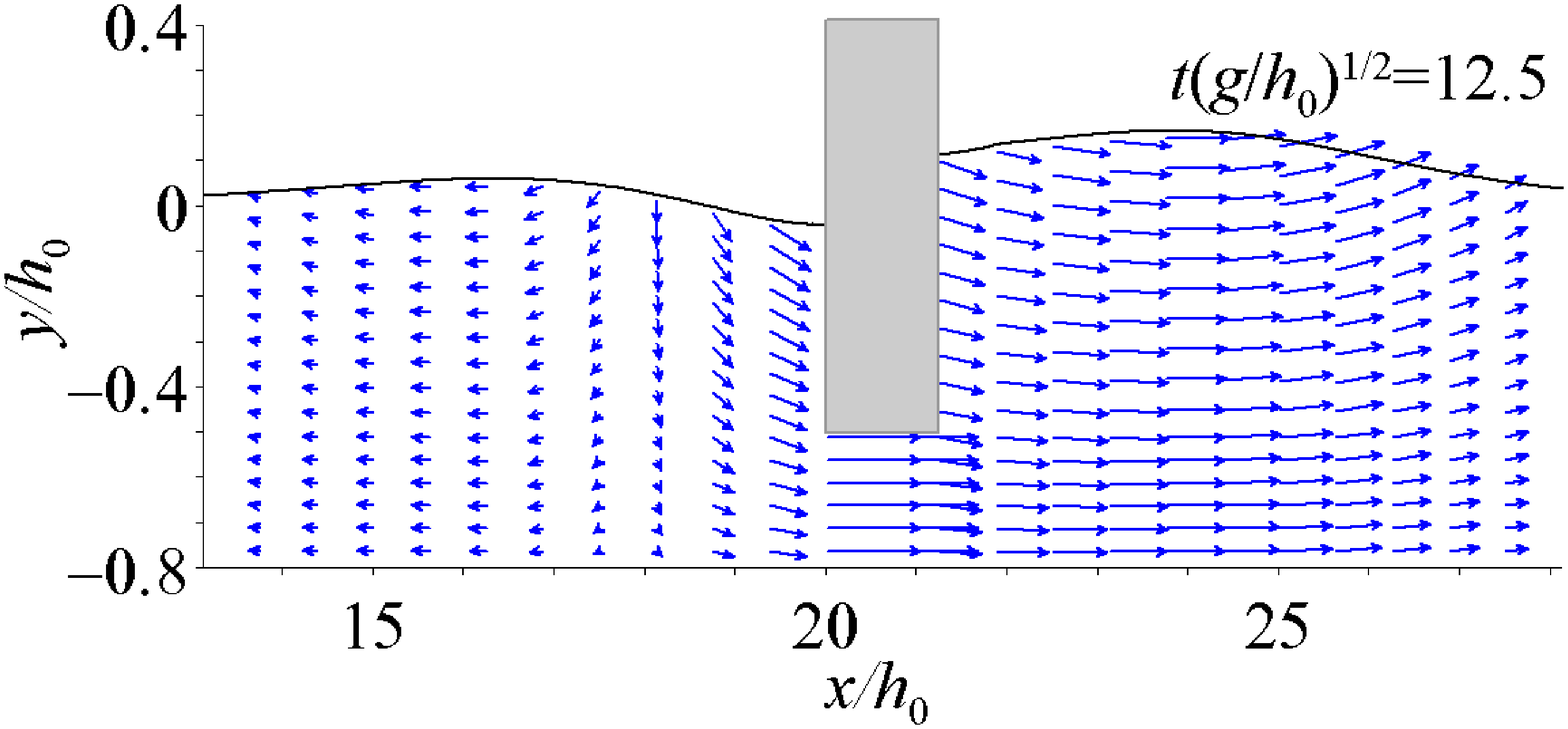}\hfill\includegraphics[width=0.49\textwidth]{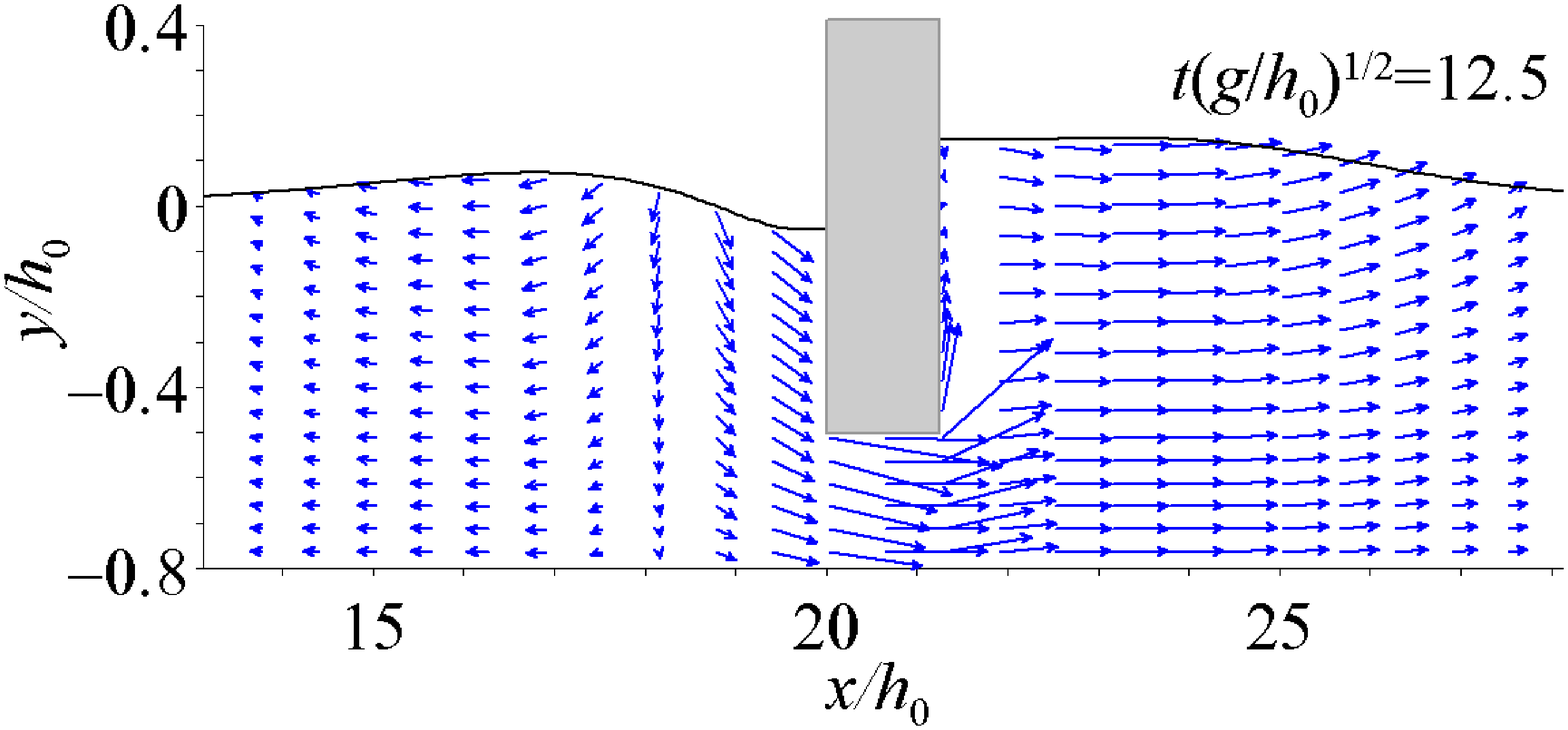}\\
\includegraphics[width=0.49\textwidth]{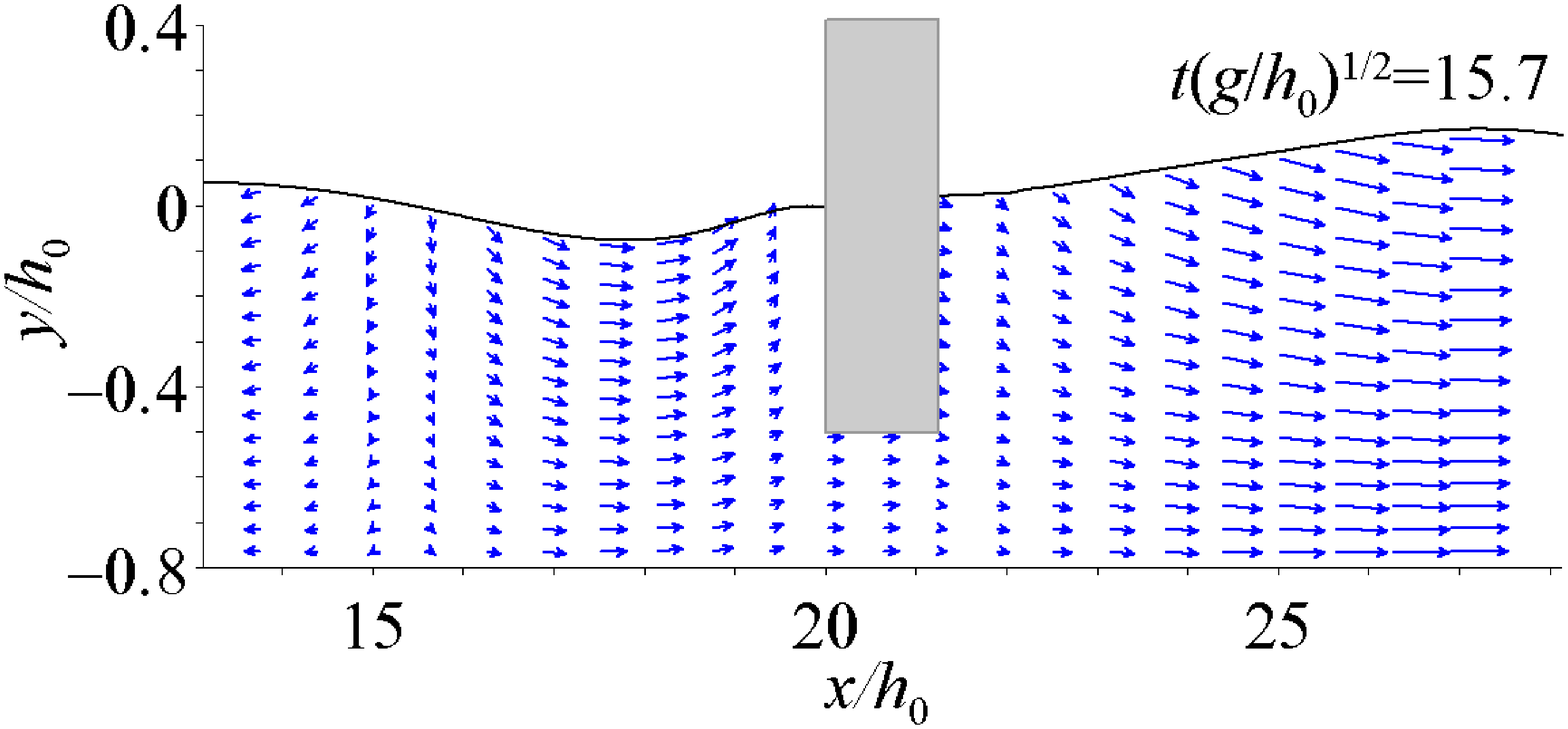}\hfill\includegraphics[width=0.49\textwidth]{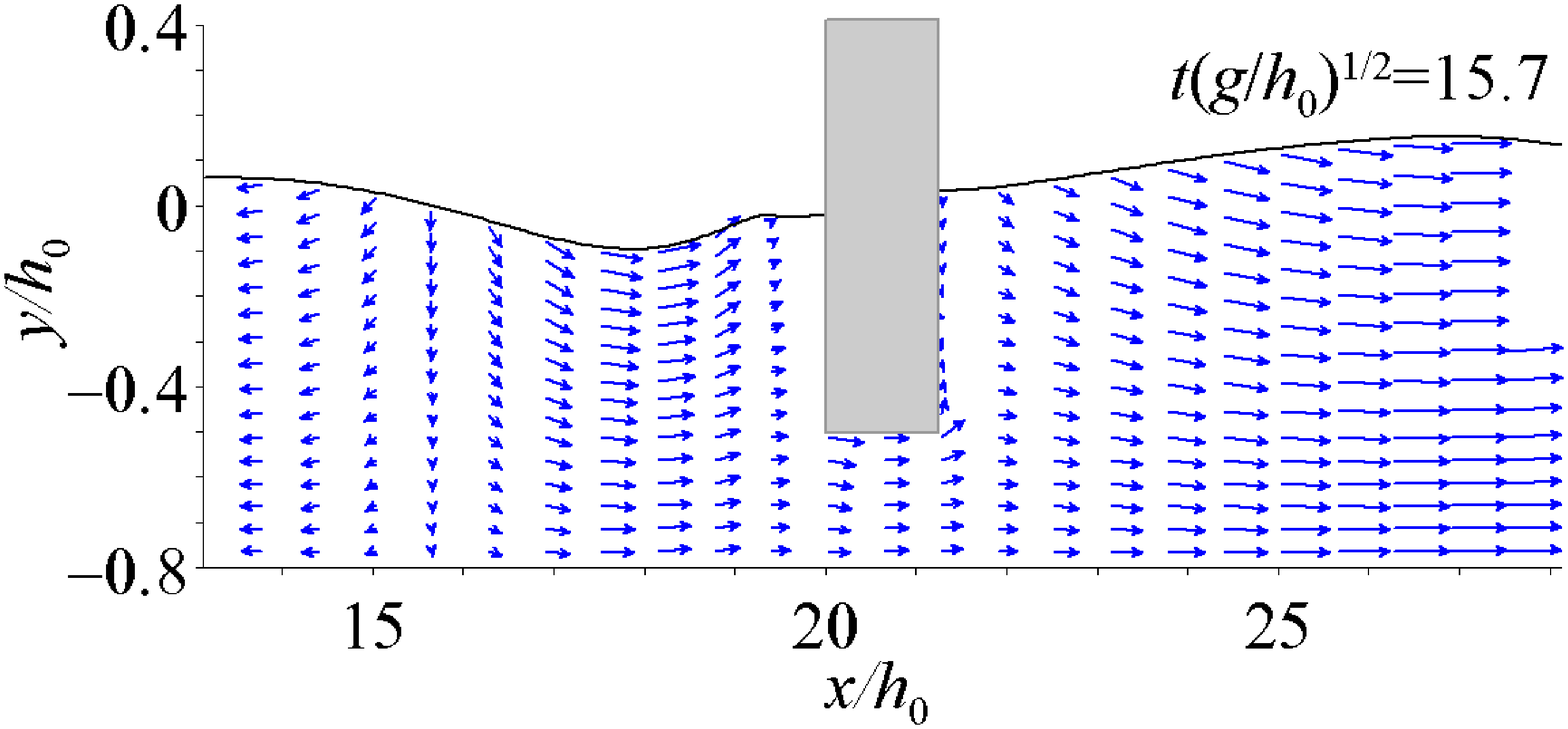}\\
\includegraphics[width=0.49\textwidth]{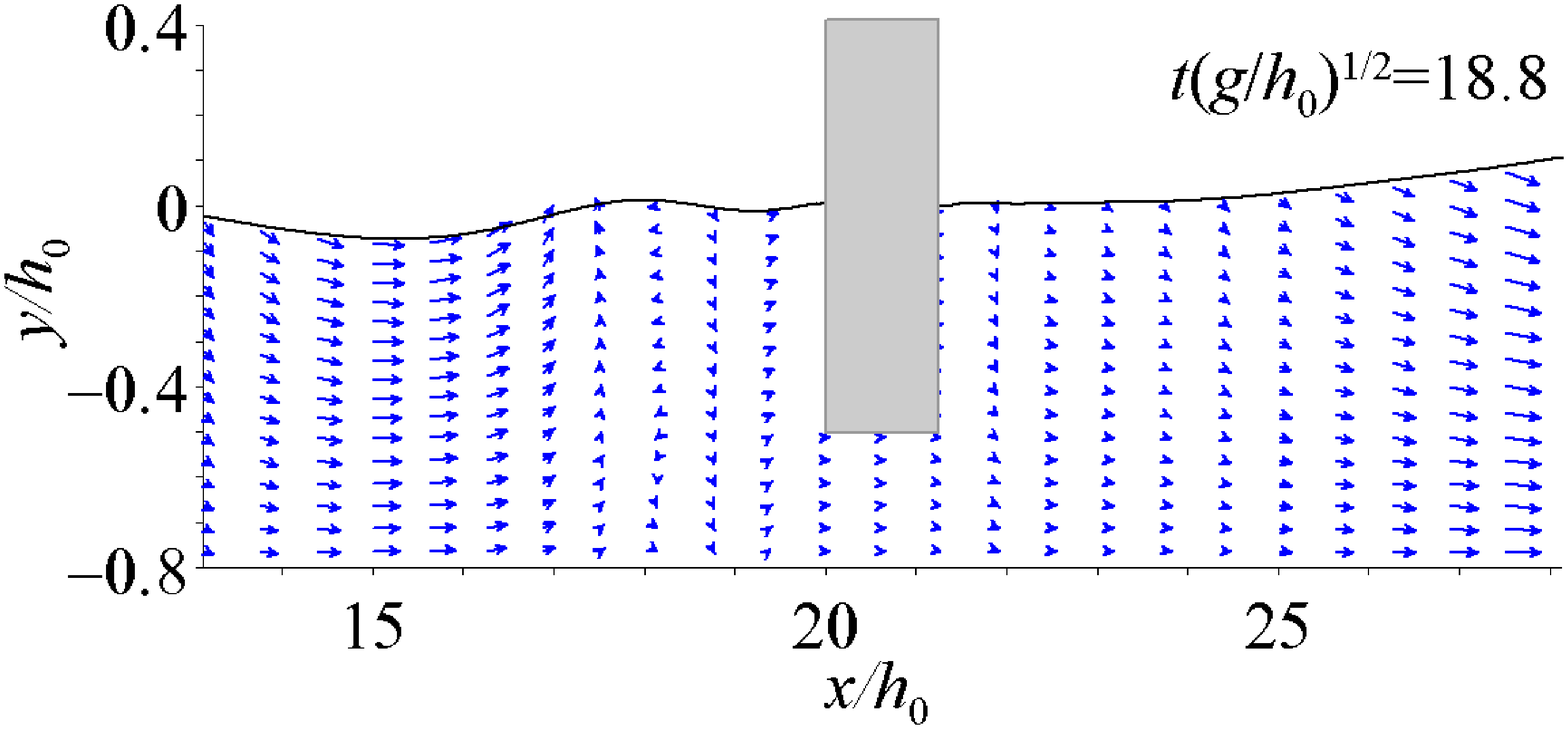}\hfill\includegraphics[width=0.49\textwidth]{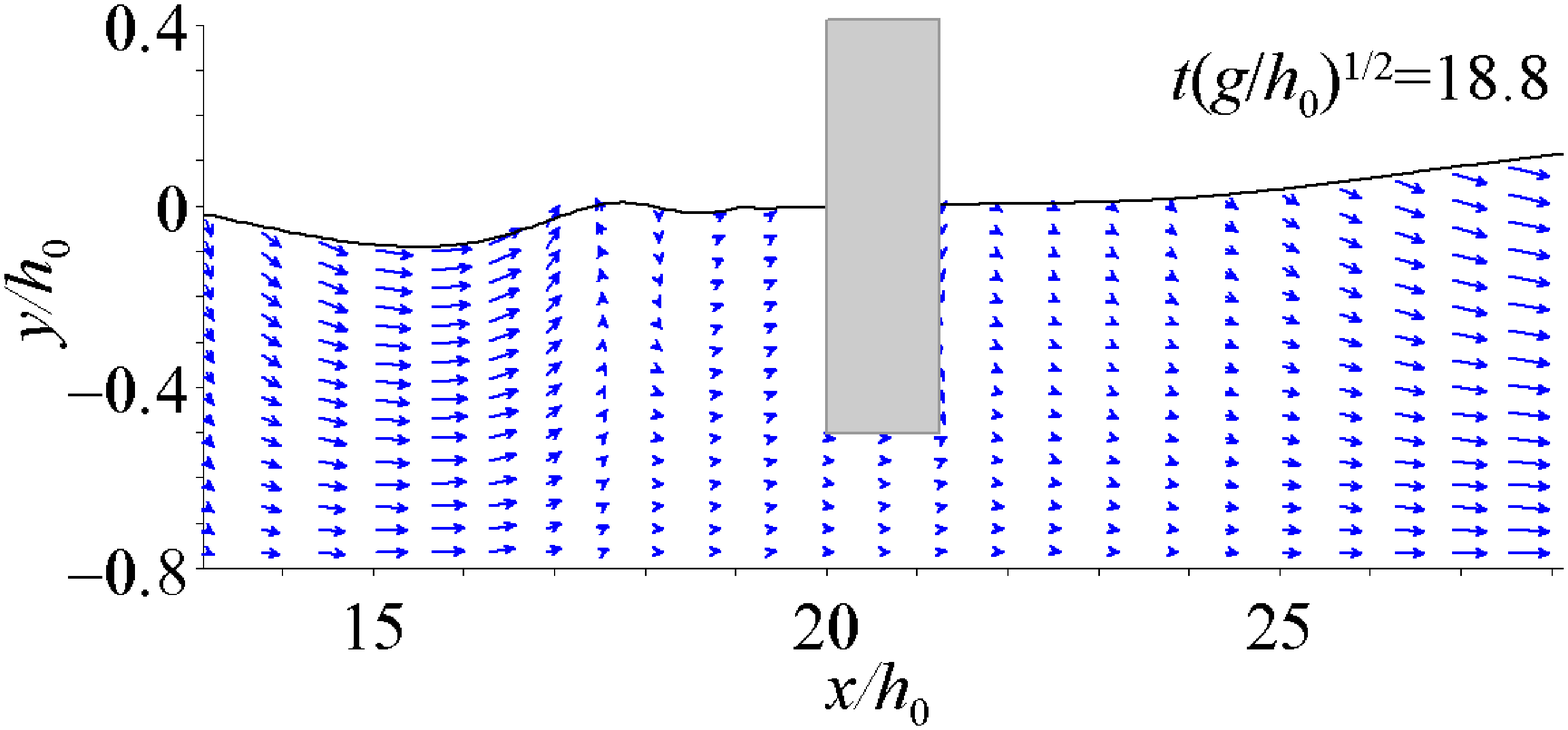}\\
{\caption{Runup of the solitary wave on the semi-immersed short body. Velocity vector fields in the vicinity of the body at different moments of time calculated by the $\SGN$ model (left) and the $\Po$ model (right). $a_0/h_0=0.2$, $L/h_0=1.25$, $d_0/h_0=-0.5$}
\label{Vel_Fields_a02_NLD+Pot}}
\end{figure}

It is interesting to compare the velocity vector fields calculated within the $\Po$ and $\SGN$ models. Generally speaking, such comparison is meaningless, since the velocity of the $\Po$ model is two-dimensional vector ${\vect U}=(U, V){}^{\top}$ (in this paper, and three-dimensional in the general case \cite{Khakimzyanov2018a}), whose horizontal $U$ and vertical $V$ components are related to the potential $\Phi$ by formulae (\ref{2D_11_1.1.71}), and the velocity in one-dimensional shallow water models is the scalar function (${\overline{u}}(x,t)$ in the outer subdomain ${\cal D}_e$ and ${\underline{u}}(x,t)$ in the inner subdomain ${\cal D}_i$) that approximates $U$. Nevertheless, the fully nonlinear $\SGN$ model allows reproducing the vertical structure of the flow with a certain accuracy by means of the so-called reconstruction formulas. For the general case, the formulas for the reconstruction of the velocity and pressure vector in the interaction problems of waves with semi-immersed bodies were given in \cite{Khakimzyanov2018a}. In \cite{Khakimzyanov2016} they were used to study the vertical structure of the flow in the problems of the wave generation by the underwater landslide and the problems of the interaction of surface waves with the underwater step. For the particular case considered in this paper (horizontal and fixed basin and body bottoms), the reconstruction formulas are greatly simplified and for the velocity the reconstruction result will be the vector ${\vect U}_{\textrm {SGN}}=(U_{\textrm {SGN}}, V_{\textrm {SGN}}){}^{\top}$ with the following components:
\begin{equation}
U_{\textrm {SGN}}(x,y,t)=\left\{\begin{array}{cl}
\displaystyle
{\overline{u}}(x,t)+\left(\frac{\H(x,t)^2}{6}-\frac{(y+h_0)^2}{2}\right){\overline{u}}_{\;xx}(x,t)\;,  & (x,y)\in \Omega_e(t)\;,\\[4mm]
\displaystyle
{\underline{u}}(x,t)+\left(\frac{S_0^2}{6}-\frac{(y+h_0)^2}{2}\right){\underline{u}}_{\;xx}(x,t)\;,  & (x,y)\in \Omega_i\;,
\end{array}\right.
\label{U_reconstr}
\end{equation}
\begin{equation}
V_{\textrm {SGN}}(x,y,t)=\left\{\begin{array}{cl}
\displaystyle
-(y+h_0){\overline{u}}_{\;x}(x,t)\;,  & (x,y)\in \Omega_e(t)\;,\\[3mm]
\displaystyle
-(y+h_0){\underline{u}}_{\;x}(x,t)\;,  & (x,y)\in \Omega_i\;.
\end{array}\right.
\label{V_reconstr}
\end{equation}
Using the first of equations (\ref{1D_NLD_to_NSWE_a}), we obtain that in the space between the bottom of the semi-immersed body and the bottom of the basin the reconstructed velocity vector is determined by the following formula:
\begin{equation*}
{\vect U}_{\textrm {SGN}} (x,y,t)= \left(\frac{Q(t)}{{\rho}S_0}\;, \; 0\right)^{\top}, \quad (x,y)\in \Omega_i\;,
\end{equation*}
i.e. the reconstructed velocity vector is parallel to the planes of the basin bottom and the body bottom and depends only on time there.

Figures \ref{Vel_Fields_a04_NLD+Pot},~\ref{Vel_Fields_a02_NLD+Pot} show the fields of velocity vectors ${\vect U}$ and ${\vect U}_{\textrm {SGN}}$ in the vicinity of a semi-immersed body at different times. To avoid cluttering the figures, velocity vectors are not drawn in all nodes of the grid: every fourth node in the horizontal direction and every second one in the vertical direction. The greatest differences in the values and directions of the velocity vectors are observed in the vicinity of the angular edges of the bottom. This is explained by the fact that exactly in the vicinity of these angular edges there are fast vertical movements of fluid in the moments of interaction of a solitary wave with a semi-immersed body. Nevertheless, a qualitative correspondence takes place, although the values of input parameters considered here are close to the limits of applicability of the $\SGN$ model and are ``unfavorable'' for it: in the first case (Fig. ~\ref{Vel_Fields_a04_NLD+Pot}), the amplitude of the incoming wave is large, which leads in the interaction to strong vertical displacements of water particles near the front and back faces of the body, while in the second (Fig. ~\ref{Vel_Fields_a02_NLD+Pot}) the body is short, which causes rapid flow restructuring  in its immediate vicinity.
Note that in more ``favorable'' cases (extended body, small submergence, small relative amplitude of the incoming wave), the velocity vector fields ${\vect U}_{\textrm {SGN}}$ restored by the reconstruction formulae are not only qualitatively, but also quantitatively close to the velocity vector fields ${\vect U}$ calculated in the $\Po$ model.

\subsection{Validation of the models}\label{sect_validation}

\begin{figure}[h!]
\centering
\includegraphics[width=0.47\textwidth]{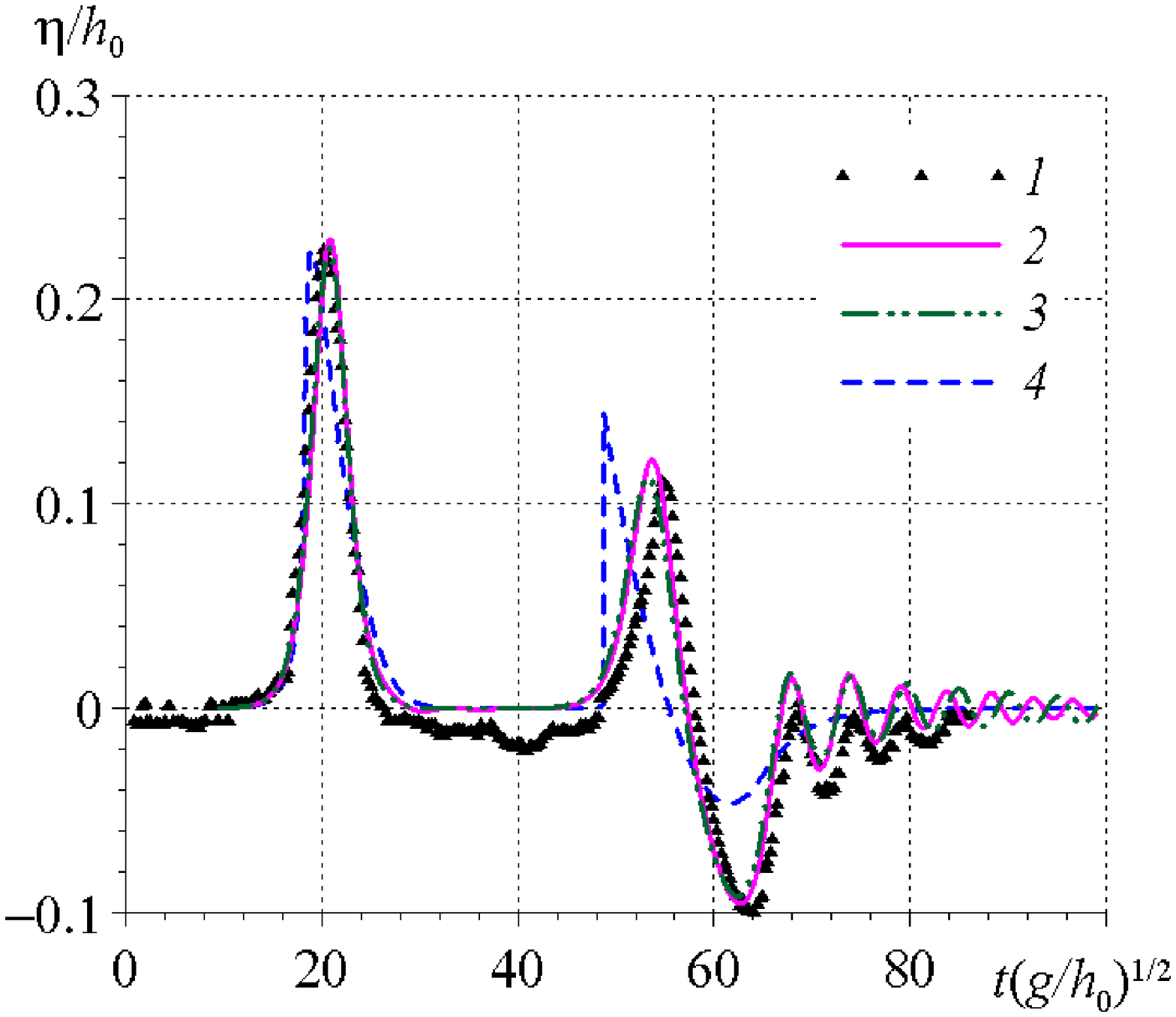}\hfill
\includegraphics[width=0.47\textwidth]{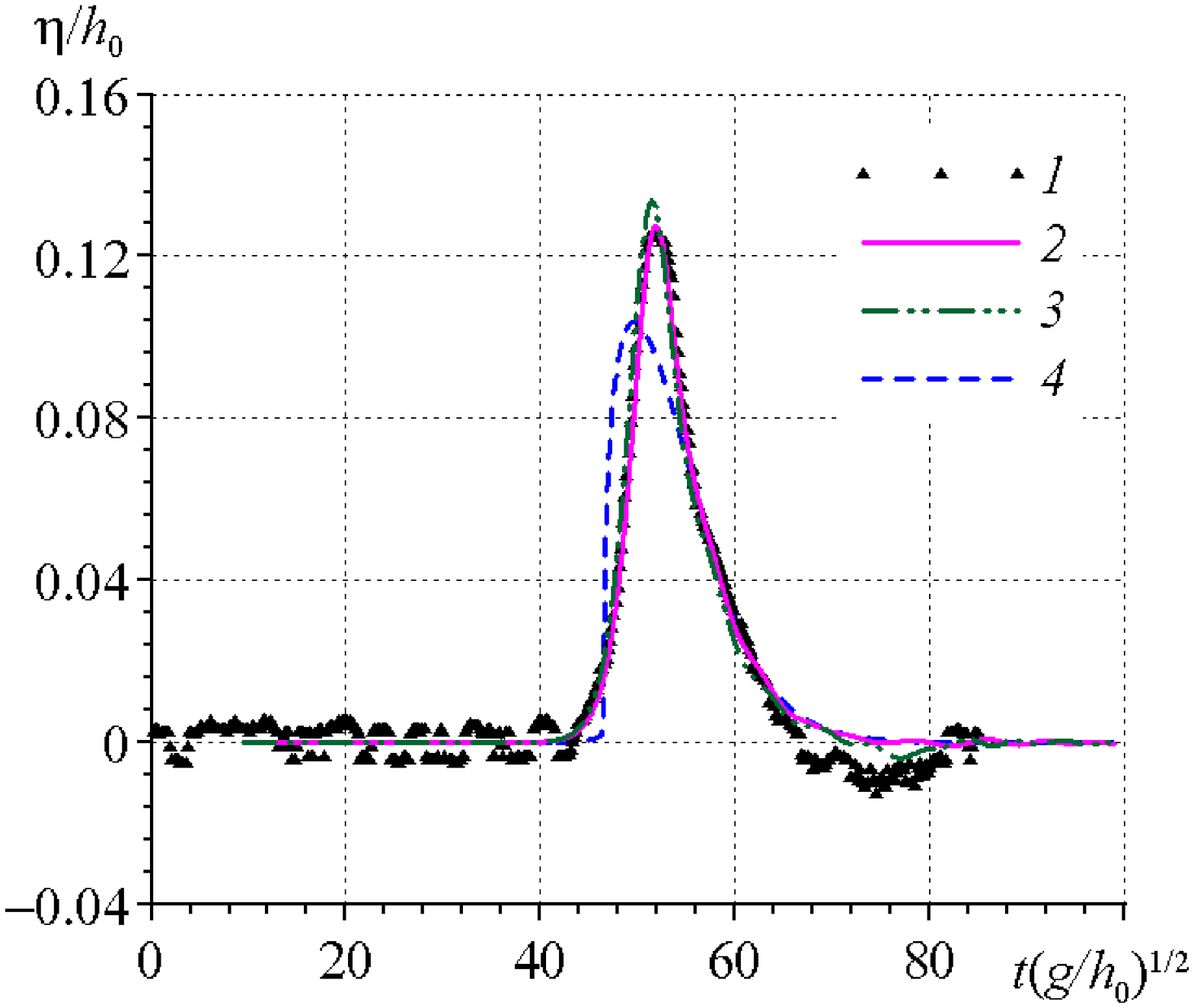}
\parbox[t]{0.49\textwidth}{\centering  ({\it a})}
\hfill
\parbox[t]{0.49\textwidth}{\centering ({\it b})}

{\caption{Time histories of free surface at gauges  $G_2$~({\it a}) and  $G_3$~({\it b}) obtained in the laboratory experiments \cite{Lu_Wang_2015}~({\sl 1}) and in the calculations by the $\Po$ model~({\sl 2}), the $\SGN$ model with compatibility conditions (C1)~({\sl 3}) and the $\SW$ model with compatibility conditions (C1)~({\sl 4}). $a_0/h_0=0.23$, $d_0/h_0=-0.5$}
\label{exp_d05}}
\end{figure}

\begin{figure}[h!]
\centering
\includegraphics[width=0.47\textwidth]{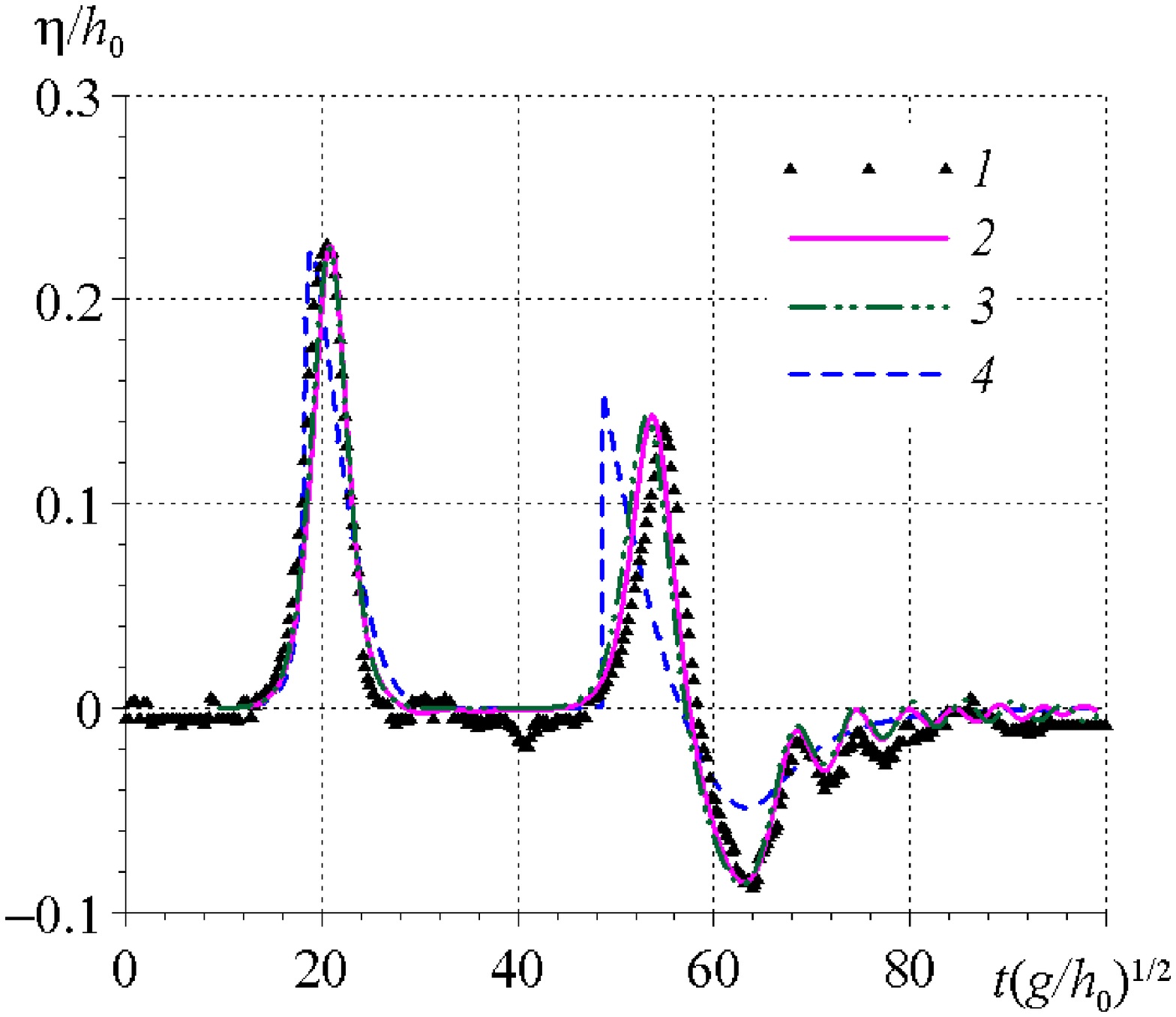}\hfill
\includegraphics[width=0.47\textwidth]{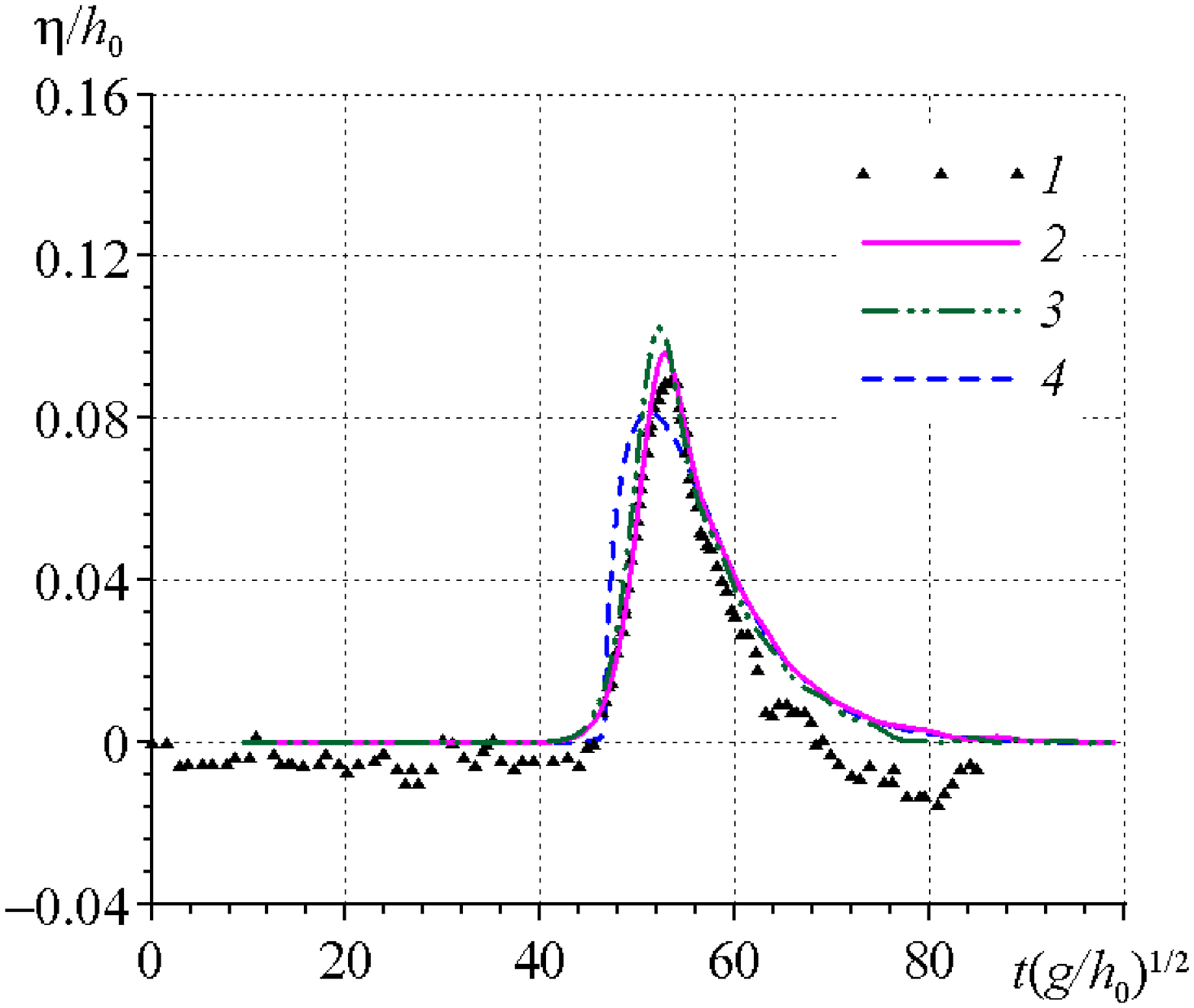}
\parbox[t]{0.49\textwidth}{\centering  ({\it a})}
\hfill
\parbox[t]{0.49\textwidth}{\centering ({\it b})}

{\caption{Time histories of free surface at gauges  $G_2$~({\it a}) and  $G_3$~({\it b}),  obtained in the laboratory experiments \cite{Lu_Wang_2015}~({\sl 1}) and in the calculations by the $\Po$ model~({\sl 2}), the $\SGN$ model with compatibility conditions (C1)~({\sl 3}) and the $\SW$ model with compatibility conditions (C1)~({\sl 4}). $a_0/h_0=0.23$, $d_0/h_0=-0.7$}
\label{exp_d07}}
\end{figure}

\begin{figure}[h!]
\centering
\includegraphics[width=0.47\textwidth]{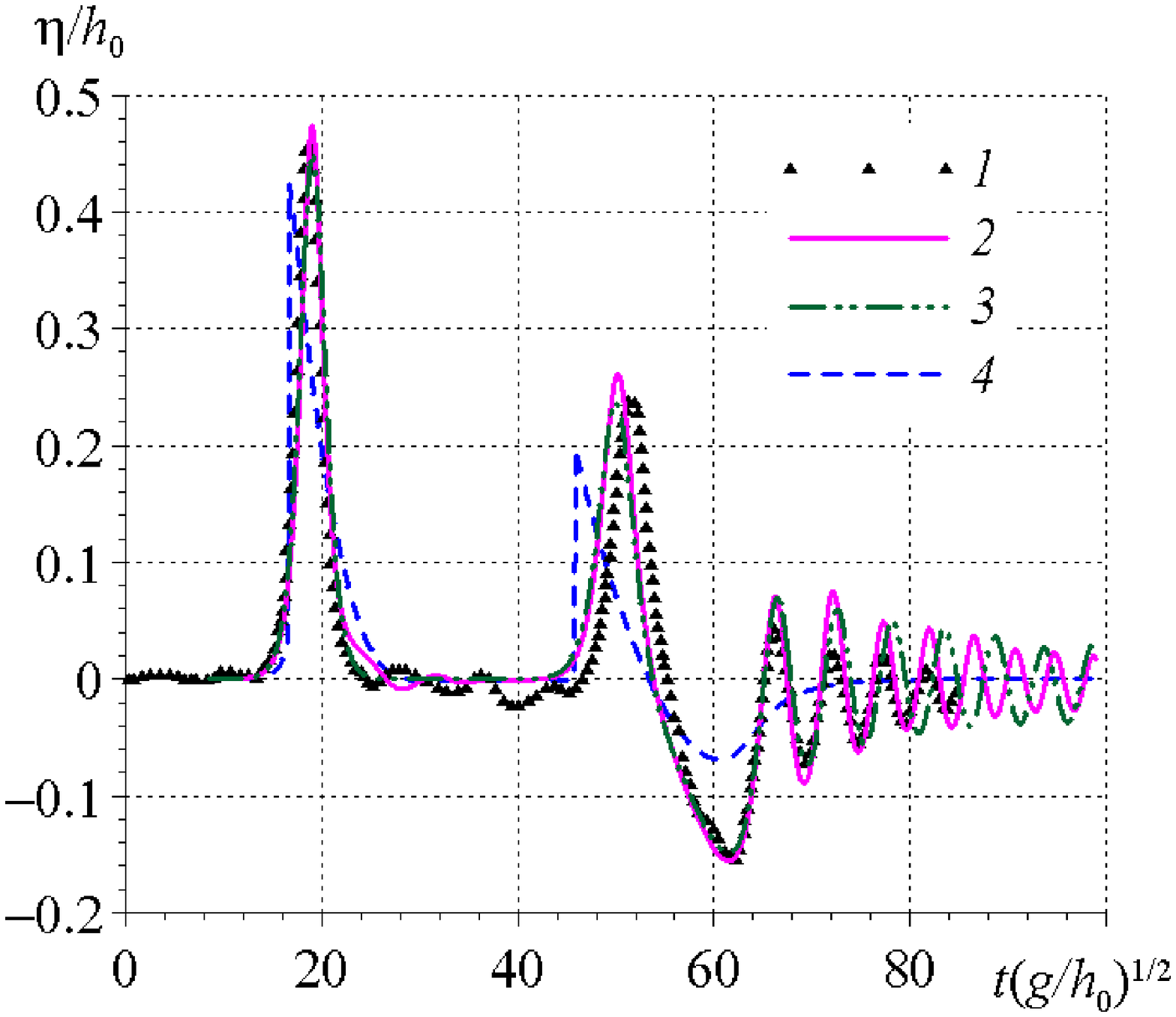}\hfill
\includegraphics[width=0.47\textwidth]{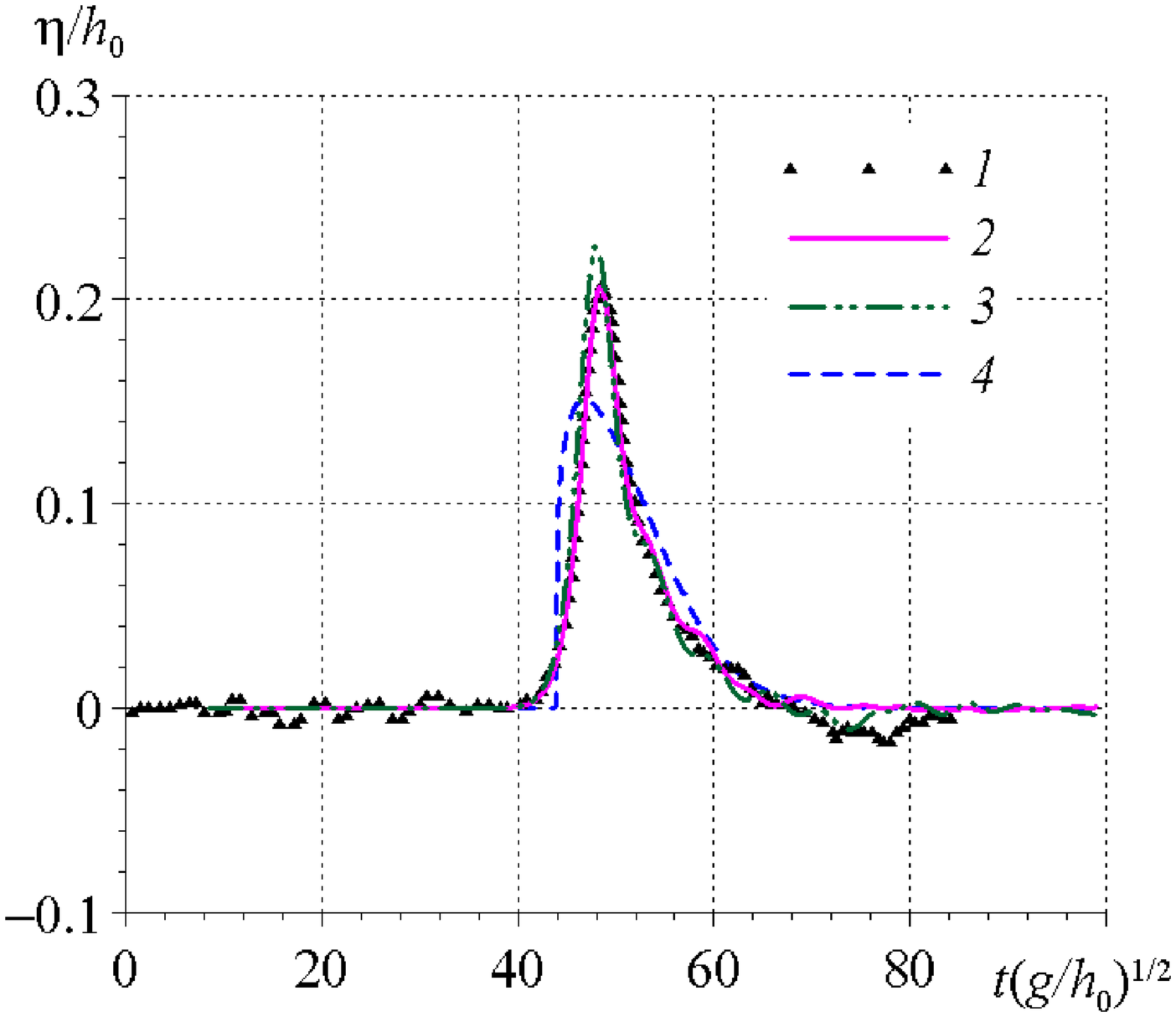}
\parbox[t]{0.49\textwidth}{\centering  ({\it a})}
\hfill
\parbox[t]{0.49\textwidth}{\centering ({\it b})}

{\caption{Time histories of free surface at gauges  $G_2$~({\it a}) and  $G_3$~({\it b}), obtained in the laboratory experiments \cite{Lu_Wang_2015}~({\sl 1}) and in the calculations by the $\Po$ model~({\sl 2}), the $\SGN$ model with compatibility conditions (C1)~({\sl 3}) and the $\SW$ model with compatibility conditions (C1)~({\sl 4}). $a_0/h_0=0.45$, $d_0/h_0=-0.5$}
\label{exp_d05_a045}}
\end{figure}

We present here the comparison of the numerical solutions with the data of laboratory experiments \cite{Lu_Wang_2015}. Figures~\ref{exp_d05}---\ref{exp_d05_a045} show the results of such comparisons from the records of two gauges located to the left and right of the partially immersed rectangular body.  Laboratory experiments were carried out in a tray with a horizontal bottom at the depth $h_0=7.63$~cm of water at rest, varying the wave amplitude, the body length and its submergence. The scheme of the hydro-wave flume~\cite{Lu_Wang_2015} is identical to that of the calculation domain shown in Fig.~\ref{scheme_of_task}. Fig.~\ref{exp_d05} shows the comparison of the experimental data with the calculations within the $\Po$, $\SGN$, and $\SW$ models. The input data for the calculations were the following parameter values written in the coordinate system of Fig.~\ref{scheme_of_task}:
\begin{equation}
\frac{a_0}{h_0}=0.23, \quad  \frac{L}{h_0}=4,\quad \frac{d_0}{h_0}=-0.5, \quad \frac{x_l}{h_0}=40, \quad \frac{x_0}{h_0}=10,\quad x_r=x_l+L,\quad \frac{l}{h_0}=70\;.
\label{input_Valid}
\end{equation}
The graphs of the other two figures were obtained at same values (\ref{input_Valid}) of the input parameters, except for the changed depth ${d_0}/{h_0}=-0.7$ (Fig.~\ref{exp_d07}) or the changed wave amplitude $a_0/h_0=0. 45$ (Fig.~\ref{exp_d05_a045}), which in calculations was set at $t=0$ by formulas (\ref{Full_an_sol_eta}), (\ref{init_cond_6_new}) and satisfied the initial data consistent conditions.

The experimental data \cite{Lu_Wang_2015} were given for the gauges $G_2$ and $G_3$ located in front and behind the body at points with the following abscissa values (in the coordinate system of the figure~\ref{scheme_of_task}):
\begin{equation*}
\frac{x_{G_2}}{h_0}=22.5, \quad \frac{x_{G_3}}{h_0}=57.5\;.
\end{equation*}
Note that the supplementary material to the article \cite{Lu_Wang_2015} was not present, so the graphs with the experimental data were obtained by digitizing. The plots when drawing the calculated data are shifted in time so that the time moments of the first maxima of elevations measured in the experiment and calculated in the $\Po$ model coincide at the gauge $G_2$.

The presented graphs show that the $\Po$ and $\SGN$ models reproduce the experiment very well. In the calculations of these models, the reflected wave is slightly ahead of the laboratory wave, and this can be explained by the influence of friction on the flow in the tray of small dimensions. $\SGN$ model slightly overestimates the amplitude of the transmitted wave in the wave gauge $G_3$ in comparison with the experiment and the $\Po$ model, which can be seen in the results of other numerical experiments. The $\SW$ model, on the contrary, underestimates the amplitude of the transmitted wave in cases with $a_0/h_0=0.23$ (see Fig.~\ref{exp_d05} and \ref{exp_d07}), but overestimates the amplitude of the reflected wave, simplifying the flow pattern and not reproducing the dispersion ``tail'', which is reproduced by other models and observed in the experiment. In the case of larger amplitude (Fig.~\ref{exp_d05_a045}), the $\SW$ model leads to wave breaking and underestimation of the amplitude of not only the passed wave, but also the reflected wave.
Comparing Fig.~\ref{exp_d05} with Fig.~\ref{exp_d07}, we see that with decreasing the submergence of the body the wave reflected from the body decreases, while the passed wave increases. This effect is observed in the experiment and in the calculations by all the models.

\subsection{On anomalous runups on a vertical wall in the presence of a semi-immersed body near it}

After the interaction of a solitary wave with a semi-immersed body, a passed wave is formed behind it, moving away from the body. In this section, we briefly consider the situation when there is a vertical impenetrable wall behind the body at a short distance from it. In this case, there are actually two obstacles for the propagating wave and the pattern of the wave-body interaction significantly changes. 

Analysis of the calculation results shows that if there is a body near the vertical wall, the pattern of the interaction becomes more complicated than in the case of normal wave runup on the wall (in the absence of a body), as well as more complicated than in the case of wave-body interaction far from the wall (i.e. actually as in the absence of the wall). For example, if the gap between the body and the wall is small, then the passed wave does not occur, but instead long-lasting vertical oscillations with a large amplitude occur in the gap. At that, maximum runup on the back face exceeds maximum runup on the front face, which was never observed when the body was placed far from the wall (see Fig. ~\ref{G123_vs_a_var_dL}---\ref{G123_vs_L_var_ad}, \ref{G123_vs_adL_PT_SW_NLD}). It is known that after reflection of a solitary wave of high relative amplitude from a vertical wall, a ``dispersion tail'' appears behind the reflected wave, and water level fluctuations with small amplitude are observed on the wall. Significantly greater amplitude oscillations occur if a semi-immersed body is placed near the wall, and these oscillations occur even at small amplitudes of the incoming wave. The value of the maximum decrease of the water level on the wall may exceed the amplitude of the incoming wave.

If there is a wall close behind the body, the wave pattern of interaction also changes in front of the body. Thus, on the front face of the body, water level oscillations also increase, although these oscillations are smaller than on the back face. Finally, note that when the body is close to the wall, the wave reflected from the body has a different shape than when the wave is reflected just from a vertical wall or from a body that is far away from the wall. 

\begin{figure}[h!]
\centering
\includegraphics[width=0.49\textwidth]{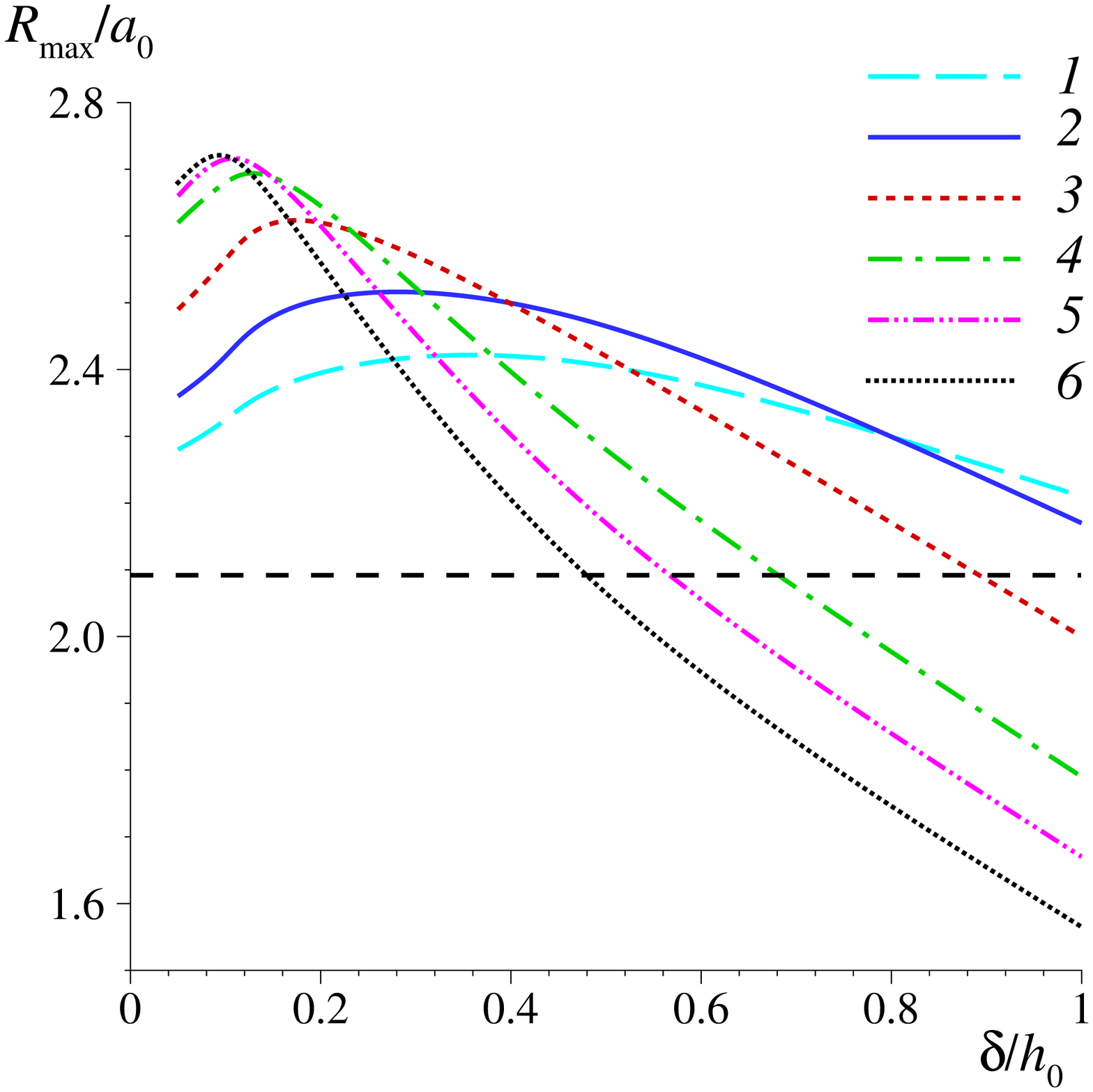}
\hfill
\includegraphics[width=0.49\textwidth]{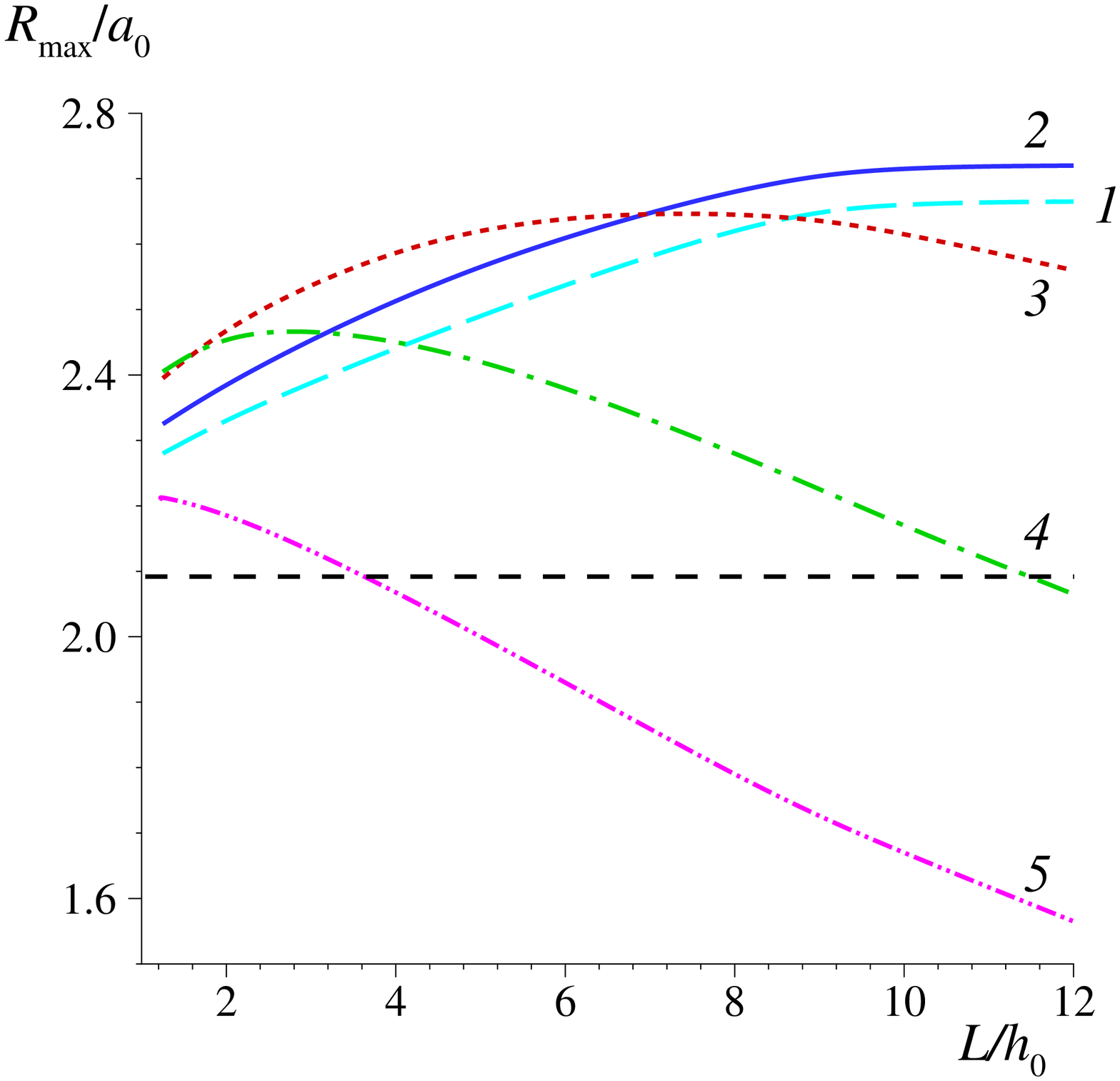}\\
\parbox[t]{0.49\textwidth}{\centering ({\it a})}
\hfill
\parbox[t]{0.49\textwidth}{\centering ({\it b})}\\

\vspace*{-1mm}

{\caption{Maximum runup $R_{\textrm {max}}/a_0$ on the vertical wall mounted behind the body in computations within the $\Po$ model: ({\it a}) the dependence of $R_{\textrm {max}}/a_0$ on the gap $\delta/h_0$ between the wall and the body fixing various values of the body length $L/h_0=1. 25$~({\sl 1}), $2.5$~({\sl 2}), $5.0$~({\sl 3}), $8.0$~({\sl 4}), $10.0$~({\sl 5}), $12. 0$~({\sl 6}); ({\it b}) the dependence of $R_{\textrm {max}}/a_0$ on the body length $L/h_0$ fixing various values of wall-body gap $\delta/h_0=0.05$~({\sl 1}), $0.1$~({\sl 2}), $0.2$~({\sl 3}), $0.5$~({\sl 4}), $1.0$~({\sl 5}). The dashed horizontal line corresponds to the maximum runup $R_{\textrm {max}}/a_0=2.09$ \cite{Maciej_Paprota_2017} computed in the problem without body. $a_0/h_0=0.2$, $d_0/h_0=-0.5$
}
\label{Rup_vs_delta_L_a02_d05}}
\end{figure}

Thus, if the gap between the body and the wall is small, strong vertical displacements arise in the close vicinity of the body, and under such conditions the $\SGN$ and $\SW$ models do not work so well (see section \ref{num_results_1}) compared to the $\Po$ model. If in this section we cover the results of calculations only within the $\Po$ model, this section will be out of the general scheme of our paper, in which we wanted to present the results of investigations within the hierarchy of mathematical models. In order not to deviate from the central line of this paper, we will present  in detail these results in the future publication, and here we will provide only one result in the form of graphs (Fig. ~\ref{Rup_vs_delta_L_a02_d05}), illustrating only the maximum runup height on the vertical wall, and only at one value of the wave amplitude  and one value of the body submergence.

Fig.~\ref{Rup_vs_delta_L_a02_d05} shows that by varying the length of the body $L$ and the gap $\delta$ between the body and the vertical wall, wave regimes with runups significantly higher than the usual runup on the vertical wall in the absence of the body may occur. This result seems paradoxical due to the fact that the semi-immersed body placed in front of the wall, partially reflecting the incoming wave and thus partially ``protecting'' the vertical wall, may seem to reduce the runup on the vertical wall. But this does not happen: the runup increases. At the same time, for some other parameter values, the runup with the body will be less than in the absence of the body.

Fig.~\ref{Rup_vs_delta_L_a02_d05} also shows that with a smaller gap the runup is greater than with a larger gap (when the wall is located further from the body), but this dependence on the gap value $\delta$ is not monotonic. The dependence of the maximum runup $R_{\textrm {max}}/a_0$ on the parameter $L$ is also not monotonic. And for each body length $L$ there is a different gap $\delta$ at which the runup on the vertical wall will be the largest. It is clear that this relationship between $L$ and $\delta$ essentially depends on both the amplitude of the incoming wave and the submergence of the body.

We can conclude that when a solitary wave interacts with a semi-immersed body placed with some gap in front of a vertical wall, light resonance regimes may occur, in which the runup on the vertical wall exceeds the runup value in the absence of the body. This result, obtained in our computational experiments for a long (solitary) wave, is analogous to the well-known fact of strong resonance amplification of the water level height in a narrow gap between rectangular shaped bodies when short waves of a certain length run into them. The latter phenomenon was investigated numerically within the models of ideal \cite{Miao_2001} and viscous \cite{Lu_Cheng_2010} fluids at different values of the gap between the bodies and their submergence \cite{Chen_He_Bingham_Shao_2019}, and by laboratory experiments in a hydro-wave tray \cite{Iwata_2007}.

\subsection{Wave-body interaction in the case of nonuniform gap between the body and the basin bottom}\label{sect_irregbott}

In contrast to the paper \cite{Lu_Wang_2015} which describes an integrated analytical-numerical approach for the dispersive shallow water model, the algorithms for $\SGN$ and $\SW$ models proposed here can be generalized to the case of a nonuniform gap between the body and the basin bottom.  Because of the limitations on the size of this article, we will not give a detailed description of the changes in the algorithms associated with an uneven gap, but will devote a separate study  to this case. However, here we provide an example of the calculation of one of such cases.

\begin{figure}[h!]
\centering
\includegraphics[width=0.49\textwidth]{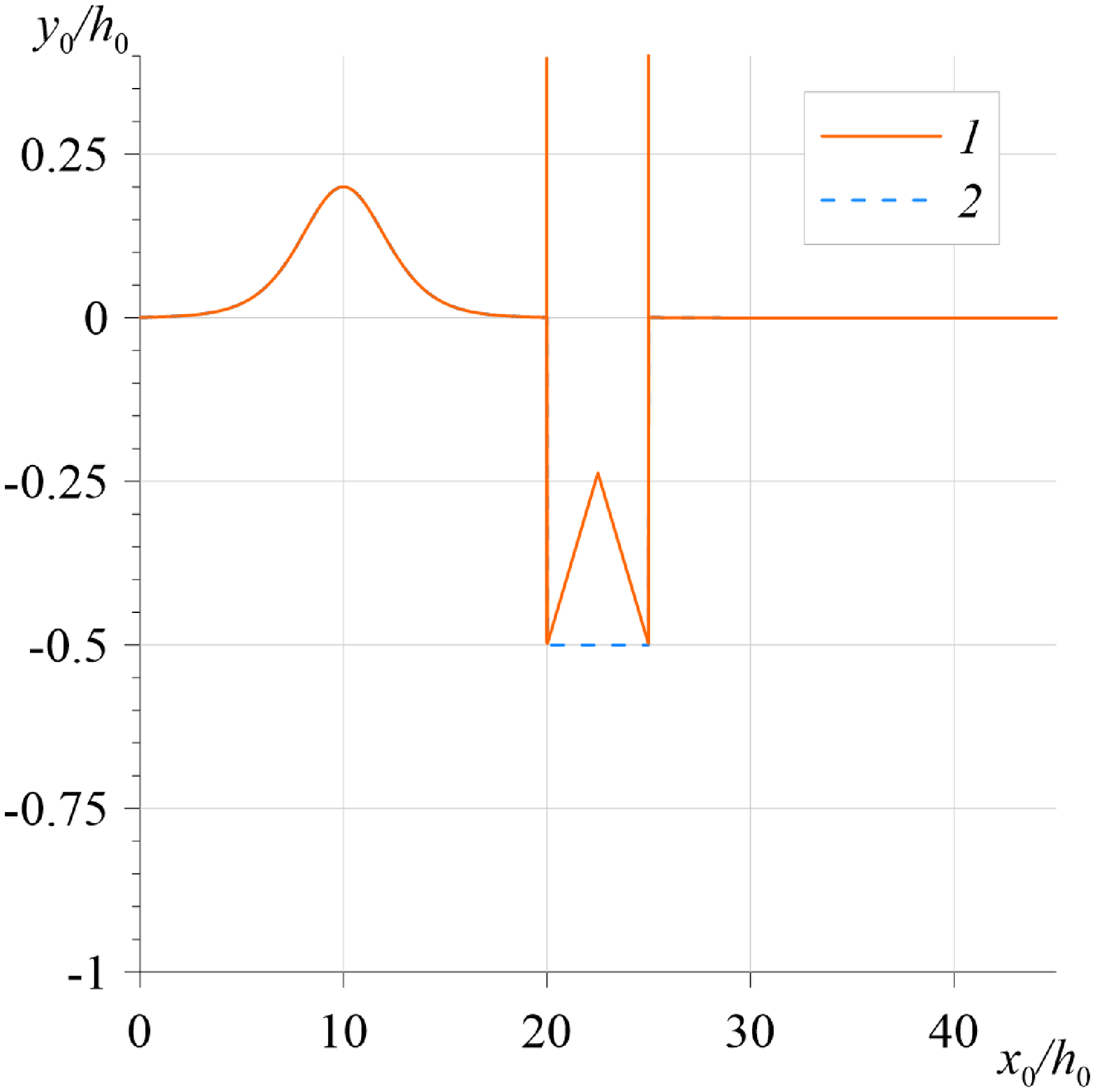}\\
\parbox[t]{0.49\textwidth}{\centering ({\it a})}\\
\includegraphics[width=0.49\textwidth]{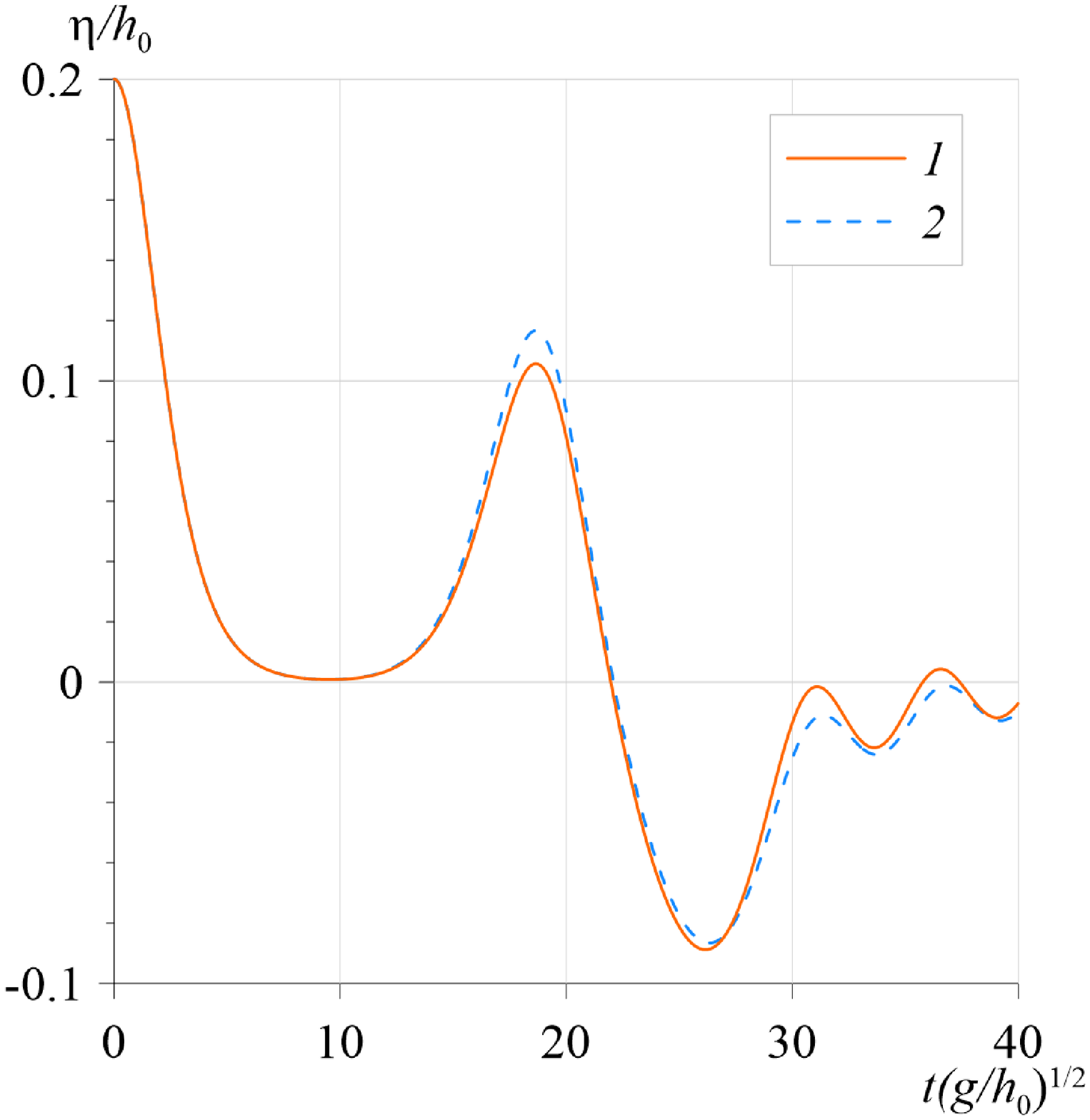}
\includegraphics[width=0.49\textwidth]{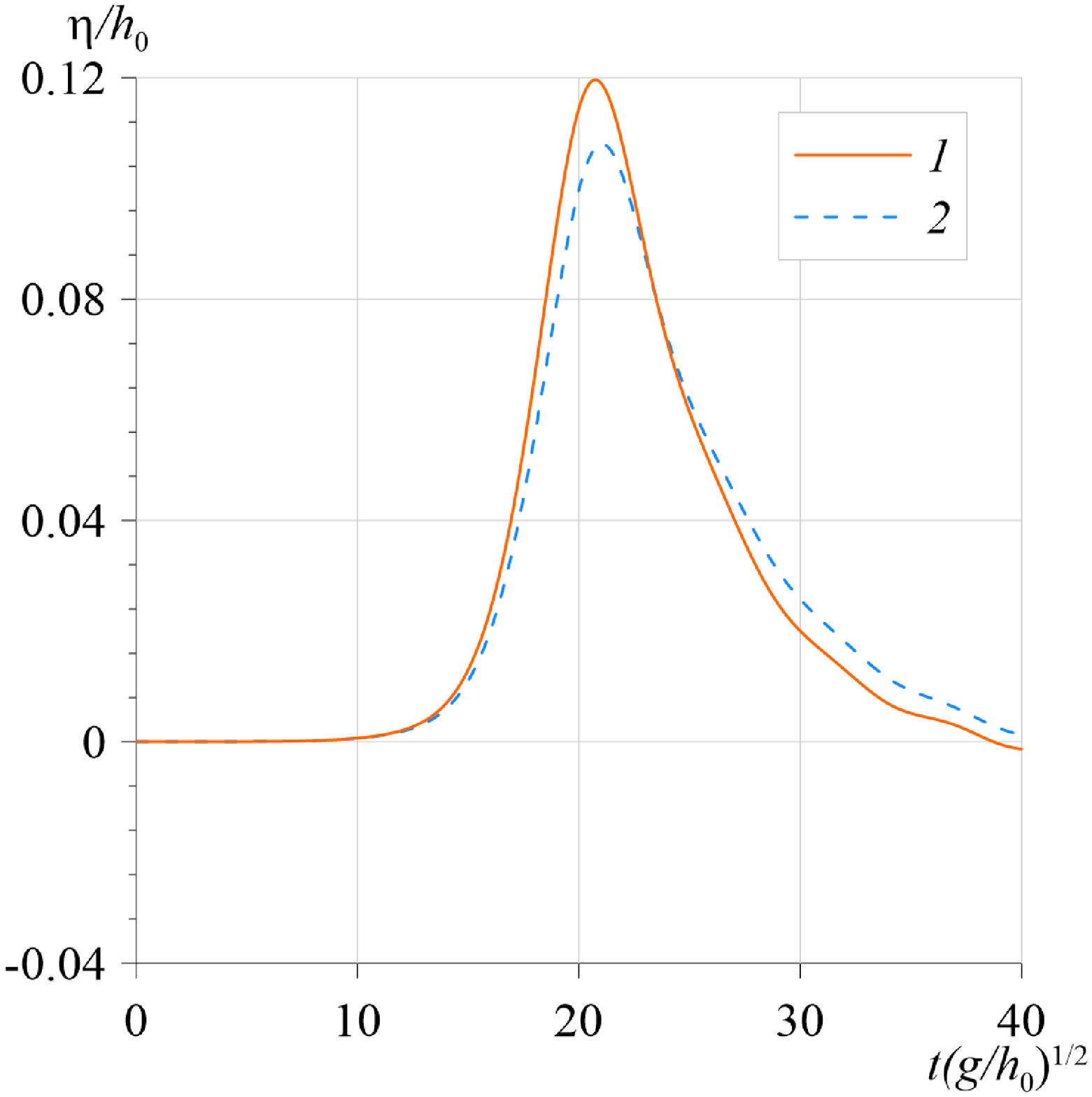}
\parbox[t]{0.49\textwidth}{\centering ({\it b})} \hfill \parbox[t]{0.49\textwidth}{\centering ({\it c})}\\
{\caption{Scheme of the problem ({\it a}) and the records of gauges $G_1$ ({\it b}) and $G_4$ ({\it c}) computed using the $\SGN$ model for the body bottom with ({\it 1}) and without ({\it 2}) triangular cutout}
\label{Irreg}}
\end{figure}

 Similarly to the above calculations, solitary wave is placed to the left from the body, and we use parameters from Eq.~\eqref{input_Pot} with one change only: $a_0/h_0=0.2$. Consider a simple case with a triangular-shaped cutout in the bottom of the body. Fig.~\ref{Irreg} shows the shape of the body and the records of gauges $G_1$ and $G_4$ (see Eq.~\eqref{x_mareogrs}) computed for the cases with and without cutout in the body bottom. The computations were performed in the framework of the $\SGN$ model using a sufficiently fine grid resolution $\Delta x/h_0=0.02$.

The graphs on Fig.~\ref{Irreg}~({\it b, c}) show that the cutout in the body bottom decreases the amplitude of the reflected from the body wave and increases the amplitude of the transmitted one. This can be explained by the fact that the volume of the body decreases and it becomes less ``noticeable'' to the wave. However, this result goes against the conclusions of the study \cite{Chang_Wang_Hseih_2017} and requires a more detailed study.

\section{Conclusions and perspectives}
\label{sec:concl}

In the manuscript text above this Section we presented the main results regarding the wave/floating body interaction problem. Below we outline the main conclusions and perspectives of this study.

\subsection{Conclusions}
In this study, we investigated the solitary wave interaction problem with a fixed floating and partially immersed obstacle. The starting point was the hierarchy of mathematical models presented in the \acrshort{3d} case in Part~I \cite{Khakimzyanov2018a} of the present series of manuscripts. In this Part~II we proposed and tested the numerical algorithms which allowed to study the wave/body interaction problem \emph{in silico}. The comparisons of numerical predictions, obtained in the framework of the hierarchy of mathematical models, allow us to draw the following preliminary conclusions regarding the solitary wave/fixed floating body interaction with the rectangular cross-section:
\begin{itemize}
  \item For the incident solitary wave amplitudes $\dfrac{a_{\,0}}{h_{\,0}}\ \lesssim\ 0.45$ (and in a certain range of other problem parameters) the best trade-off between the complexity and accuracy is offered by the $\SGN$ model.
  \item For the solitary wave amplitudes $\dfrac{a_{\,0}}{h_{\,0}}\ \lesssim\ 0.25$ it was found out that even the simple $\SW$ model gives accurate predictions for the maximal wave run-up on the fixed partially immersed body.
  \item In all other cases we recommend using the \acrshort{2d} $\Po$ formulation.
\end{itemize}
We would like to mention also that the proposed algorithms can be easily generalized to the case of the general uneven and, eventually, moving solid surfaces --- namely, fluid and floating body bottoms. This generalization can be easily done for the whole hierarchy of considered mathematical models ($\Po$, $\SGN$ and $\SW$). In this manuscript we made a choice of presenting the numerical algorithms in a slightly simplified situation. This choice allowed us to make the exposition clearer without focusing on unnecessary details and complications. Moreover, the flat bottom case turns out to be perfectly consistent with the incident solitary wave assumption. However, we do not exclude the possibility of presenting this generalization in one of our future publications.

\subsection{Perspectives}

In our future studies it would be desirable to develop a mathematical model (ideally, in the framework of \acrshort{fee}) with an associated numerical solver for the case when the body is truly floating (not being assumed to be fixed in the flow). This would allow us to study the influence of this simplifying assumption on the wave field before and after the obstacle. We would like to justify this assumption and understand its limits of applicability regarding the hierarchy of models considered in the present study. On the more technical side, one can think about the generalization of the proposed algorithms for the case of more general curvilinear bottoms as we mentioned earlier.

In the next Part~III of our series of articles devoted to the wave/floating body interaction problem, we shall investigate the \acrshort{3d} effects in the framework of the \acrshort{fee} model ($\Po$) which were neglected in the present Part~II. Moreover, we shall describe the properties of the proposed numerical method. In the same Part~III we shall present also our investigations on the wave forces acting on a \acrshort{3d} partially immersed body. Finally, this series of manuscripts will be finished by Part~IV where we shall describe the wave/body interaction in \acrshort{3d} using the long wave models ($\SGN$ and $\SW$). Of course, the validation of results in Part~IV will be done using the reference solution from Part~III. In these subsequent works, we shall try to highlight the quantitative and qualitative differences in generated wave fields as predicted by reduced (simplified) models.

\section*{Acknowledgments}
The work of D.~Dutykh has been supported by the French National Research Agency, through Investments for Future Program (contract \No ANR--$18$--EURE--$0016$ --- Solar Academy). The work of O. Gusev was partially supported by RSCF project \No 21-71-00127 (validation of the shallow water models in \ref{sect_validation}, the results of subsection \ref{sect_irregbott}).


\bigskip\bigskip
\invisiblesection{References}
\bibliographystyle{acm}
\bibliography{mybiblio}
\bigskip\bigskip


\printglossary[type=\acronymtype,title={}]


\section*{Appendix: finite-difference equations for calculating the dispersion component of pressure for the SGN model}

Let $x_j$ ($j=0,\ldots,N$) be coordinates of nodes of the uniform fixed grid with step $\Delta x=l/N$, covering the domain ${\cal D}$, $x_0=0$, $x_N=l$, $x_{l}=x_{j_l}$, $x_{r}=x_{j_r}$. The nodes $x_{j_l}$ and $x_{j_r}$ are common to the subdomains ${\cal D}_e$ and ${\cal D}_i$. In the one-dimensional approximation with a horizontal bottom, there is no calculation of the values in the grid nodes under the body, because instead of partial differential equations in the area under the body, we use one ordinary differential equation (\ref{1D_NLD_to_NSWE_b}). Velocity does not depend on $x$, and pressure is a linear function of $x$ given by formula (\ref{1D_p_in_Di}).

Assume that at the time layer with number $n$ all the sought functions were calculated. Thus, outside the body and at the common boundary $\Gamma$, the free surface $\eta^n_j$, velocity $u^n_j$ and dispersion component $\Pnh^n_j$ of the pressure  ($j=0,\ldots , j_l$, $j=j_r,\ldots , N$) are known. Under the body, the flow rate $Q^n$ satisfying the compatibility condition (\ref{1D_mass_conj_1}) is known, i.~e.
\begin{equation}
\rho (\H u)^n_{j_l}=Q^n=\rho (\H u)^n_{j_r}.
\label{Appen_B_FDE1}
\end{equation}
The rate $\dot{Q}^n$ of change in flow at the time layer $n$ is determined from the finite-difference analogue of relation (\ref{1D_NLD_to_NSWE_b}).  The derivatives $u_x$ included in the compatibility conditions (C2) are calculated using one-sided finite-differences:
\begin{equation}
u_x\big|_{x_{r}+0}  \sim \frac{u^n_{j_r+1}-u^n_{j_r}}{\Delta x}, \quad u_x\big|_{x_{l}-0}  \sim \frac{u^n_{j_l}-u^n_{j_l-1}}{\Delta x}.
\label{Appen_B_FDE4}
\end{equation}

Below we present formulas for calculating the dispersion component of the pressure in the predictor-corrector scheme \cite{Khakimzyanov2016} and describe the method of merging the numerical solutions from different sides of the semi-immersed body, that satisfies the finite-difference analogues of conditions (\ref{1D_NLD_to_NSWE_b}) and (\ref{1D_mass_conj_1}) at the predictor step and at the corrector step.

{\textit {\textbf {Predictor.}}} The predictor step \cite{Khakimzyanov2016,Khakimzyanov2019c} first calculates the total depth $\H^*_{j+1/2}$ and the velocity $u^*_{j+1/2}$ ($j=0,\ldots , j_l-1$, $j=j_r,\ldots , N-1$). These grid functions are defined on an intermediate time layer at the centers $x_{j+1/2}=x_j+\Delta x/2$ of the grid cells covering the outer region ${\cal D}_e$. Then the values of the dispersion component $\Pnh^*_{j+1/2}$ of the pressure are calculated. For this purpose, finite-difference equations approximating the differential equation (\ref{curve_phi_1}) are used. To do this, the integral form of equation (\ref{curve_phi_1}) is used:
\begin{equation}
\int\limits_{x_j}^{x_{j+1}} \left(k\Pnh_x\right)_x  dx-\int\limits_{x_j}^{x_{j+1}} k_0\Pnh  dx=\int\limits_{x_j}^{x_{j+1}}F  dx, \qquad \begin{array}{l}\displaystyle j=0,\ldots , j_l-1, \\\displaystyle j=j_r,\ldots , N-1.\end{array}
\label{Afd_phi_1}
\end{equation}

Consider first the cells $[x_j, x_{j+1}]$ that are not boundary cells, i.e. when $j=1,\ldots , j_l-2$, $j=j_r+1,\ldots , N-2$. In this case, we will use the following quadrature formulas to calculate integrals:
\begin{equation}
\int\limits_{x_j}^{x_{j+1}} \left(k\Pnh_x\right)_x  dx\sim \frac{k_{j+3/2}+k_{j+1/2}}{2}\cdot \Pnh^*_{x,j+1}-\frac{k_{j+1/2}+k_{j-1/2}}{2}\cdot \Pnh^*_{x,j},
\label{fd_phi_2}
\end{equation}
\begin{equation}
\int\limits_{x_j}^{x_{j+1}} k_0\Pnh  dx\sim  \Pnh^*_{j+1/2}\frac{3}{\left(\H^*_{j+1/2}\right)^3}\Delta x,
\label{fd_phi_3}
\end{equation}
\begin{equation}
\int\limits_{x_j}^{x_{j+1}}F dx \sim {\rho}g \left(\eta^*_{x,j+1}-\eta^*_{x,j}\right)+2{\rho}\Delta x\left(u^*_{x,j+1/2}\right)^2,
\label{fd_phi_4}
\end{equation}
where
\begin{equation}
k_{j+1/2}=\frac{1}{\H^*_{j+1/2}}, \quad \Pnh^*_{x,j}=\frac{\Pnh^*_{j+1/2}-\Pnh^*_{j-1/2}}{\Delta x},\quad \eta^*_{x,j}=\frac{\eta^*_{j+1/2}-\eta^*_{j-1/2}}{\Delta x},
\label{fd_phi_5}
\end{equation}
\begin{equation}
u^*_{x,j+1/2}=\frac{u^*_{j+1}-u^*_{j}}{\Delta x}, \quad u^*_{j}=\frac{u^*_{j+1/2}+u^*_{j-1/2}}{2}.
\label{fd_phi_5a}
\end{equation}

Thus, for the specified values of $j$ we obtain the three-point finite-difference equations
\begin{equation}
a_j\Pnh^*_{j-1/2}-c_j\Pnh^*_{j+1/2}+b_j\Pnh^*_{j+3/2}=d_j,
\label{3_point_eq_phi}
\end{equation}
with the coefficients
\begin{equation}
a_j=\frac{k_{j-1/2}+k_{j+1/2}}{2\Delta x}, \quad b_j=\frac{k_{j+3/2}+k_{j+1/2}}{2\Delta x}, \quad c_j=a_j+b_j+\frac{3\Delta x}{\left(\H^*_{j+1/2}\right)^3}
\label{koef_abc}
\end{equation}
and the right side
\begin{equation}
d_j={\rho}g \left(\eta^*_{x,j+1}-\eta^*_{x,j}\right)+2{\rho}\Delta x\left( u^*_{x,j+1/2}\right)^2.
\label{RS_d}
\end{equation}

Note that $a_j>0$, $b_j>0$, and the coefficients of equation (\ref{3_point_eq_phi}) satisfy the property of strict diagonal dominance:
\begin{equation}
c_j>a_j+b_j,
\label{3_point_eq_phi_1}
\end{equation}
which is important for numerical implementation.

Finite-difference equations (\ref{3_point_eq_phi}) together give a system of $(N-(j_r-j_l)-4)$ linear equations with respect to $(N-(j_r-j_l))$ unknowns $\Pnh^*_{j+1/2}$ ($j=0,\ldots , j_l-1$, $j=j_r,\ldots , N-1$). The missing four equations are obtained by approximating integral relations (\ref{Afd_phi_1}) in four boundary cells ($j=0, j_l-1, j_r, N-1$), one of whose boundaries corresponds to either the side wall of the basin ($x=0$ or $x=l$) or the side face of a semi-immersed body ($x=x_l$ or $x=x_r$).

Consider first the boundary cells corresponding to the walls of the basin, for example, the cell $[x_j, x_{j+1}]$ at $j=0$. We will use the following approximation of integral relation (\ref{Afd_phi_1}) for this cell:
\begin{equation*}
\frac{k_{j+3/2}+k_{j+1/2}}{2}\cdot \Pnh^*_{x,j+1}-\left.\left(\frac{1}{\H}\cdot \frac{\partial \Pnh}{\partial x}\right)\right|_{x=0}- \Pnh^*_{j+1/2}\frac{3\Delta x}{\left(\H^*_{j+1/2}\right)^3}=
\end{equation*}
\begin{equation}
={\rho}g \left(\eta^*_{x,j+1}-\left.\frac{\partial \eta}{\partial x}\right|_{x=0}\right)+2{\rho}\Delta x\left(u^*_{x,j+1/2}\right)^2, \qquad (j=0).
\label{dif_phi_x=0}
\end{equation}
In this equation the derivatives ${\partial \Pnh}/{\partial x}\\big|_{x=0}$ and ${\partial \eta}/{\partial x}\big|_{x=0}$ are zero due to boundary conditions (\ref{Pt_2_Gamma_0}), so formulas (\ref{fd_phi_5}) assume $\Pnh^*_{x,j}=0$, $\eta^*_{x,j}=0$ at $j=0$. When calculating the derivative $u^*_{x,j+1/2}$ using formula (\ref{fd_phi_5a}), it is taken into account that $u^*_0=0$. Thus, for the left boundary cell ($j=0$) we obtain a two-point finite-difference equation, which, assuming $a_j=0$, can formally be written as finite-difference equation (\ref{3_point_eq_phi}), with coefficients (\ref{koef_abc}) and right part (\ref{RS_d}), with $b_j>0$, $c_j>a_j+b_j$.

Similarly, the two-point finite-difference equation in the right boundary cell $[x_j, x_{j+1}]$ ($j=N-1$) is derived using boundary conditions (\ref{Pt_2_Gamma_0}). It also has form (\ref{3_point_eq_phi}), with $\eta^*_{x,j+1}=0$, $u_{j+1}^*=0$, $b_j=0$, $a_j>0$, $c_j>a_j+b_j$.

Let us now consider the boundary cells adjacent to the lateral faces of the body. The derivation of the finite-difference equations for these cells is also based on the approximation of integral relation (\ref{Afd_phi_1}), but in addition to the boundary conditions, the compatibility conditions are also used here. For this purpose, at first, knowing the rate of change of the flow $\dot{Q}^n$, the value $Q^*=Q^n+\tau \dot{Q}^n/2$ is found and it is required that the equality of type (\ref{Appen_B_FDE1}) is also satisfied at the intermediate time layer:
\begin{equation}
\rho (\H u)^*_{j_l}=Q^*=\rho (\H u)^*_{j_r}.
\label{Appen_B_FDE6}
\end{equation}
Given conditions (\ref{1D_NLD_Deta_dx}), we assume $\H^*_{j_l}=\H^*_{j_l-1/2}$, $\H^*_{j_r}=\H^*_{j_r+1/2}$. Then  we get from (\ref{Appen_B_FDE6}) the following expressions for the velocity to the left and right of the body:
\begin{equation}
u^*_{j_l}=\frac{Q^*}{\rho \H^*_{j_l}}, \quad u^*_{j_r}=\frac{Q^*}{\rho \H^*_{j_r}}.
\label{Appen_B_FDE7}
\end{equation}
The velocity values $u^*_{j_l}$ and $u^*_{j_r}$ are used in formula (\ref{fd_phi_5a}) to calculate the derivatives $u^*_{x,j_l-1/2}$ and $u^*_{x,j_r+1/2}$, respectively.

Let us take the cell $[x_j, x_{j+1}]$ for which $j=j_l-1$, and write for it the expression obtained by approximating relation (\ref{Afd_phi_1}):
\begin{equation*}
\left.\left(\frac{1}{\H}\cdot \frac{\partial \Pnh}{\partial x}\right)\right|_{x_{l}-0}-\frac{k_{j+1/2}+k_{j-1/2}}{2}\cdot \Pnh^*_{x,j}- \Pnh^*_{j+1/2}\frac{3\Delta x}{\left(\H^*_{j+1/2}\right)^3}=
\end{equation*}
\begin{equation}
={\rho}g \left(\left.\frac{\partial \eta}{\partial x}\right|_{x_{l}-0}-\eta^*_{x,j}\right)+2{\rho}\Delta x \left(u^*_{x,j+1/2}\right)^2, \qquad (j=j_l-1).
\label{dif_phi_x=x_l}
\end{equation}
Due to (\ref{1D_NLD_Deta_dx}), we obtain $\left.\eta_x\right|_{x_{l}-0}=0$. $\SGN$ equations (\ref{Pt_2_cont_eq1}), (\ref{Pt_2_mov_eq2}), formula (\ref{1D_pDe}) and condition (\ref{1D_NLD_Deta_dx}) are used in calculating $\left.\left(\Pnh_x/\H\right)\right|_{x_{l}-0}$. As a result, we obtain the expression
\begin{equation*}
\left.\left(\frac{1}{\H}\cdot \frac{\partial \Pnh}{\partial x}\right)\right|_{x_l-0}=\rho \left(u_t+uu_x\right)\big|_{x_l-0}.
\end{equation*}

We transform the right part of this equality using the continuity equation (\ref{Pt_2_cont_eq1}), the condition (\ref{1D_NLD_Deta_dx}), and the consequence of the compatibility condition (\ref{1D_mass_conj_1}):
\begin{equation}
\rho (\H u)_t\big|_{x_{l}-0}=\dot{Q}=\rho (\H u)_t\big|_{x_{r}+0}.
\label{dif_phi_x=x_l_1}
\end{equation}
Hence,
\begin{equation*}
{\rho} \left(u_t+uu_x\right)\Big|_{x_l-0}={\rho}\frac{(\H u)_t-u\H_t}{\H}+\frac{{\rho}}{2}(u^2)_x\Big|_{x_l-0} = \frac{\dot{Q}}{\H}+\frac{{\rho}u(\H u)_x}{\H}+\frac{{\rho}}{2}(u^2)_x\Big|_{x_l-0} =
\end{equation*}
\begin{equation*}
=\frac{\dot{Q}}{\H}+{\rho}(u^2)_x\Big|_{x_l-0}
\end{equation*}
as well as
\begin{equation*}
{\rho} \left(u_t+uu_x\right)\Big|_{x_r+0}=\frac{\dot{Q}}{\H}+{\rho}(u^2)_x\Big|_{x_r+0}.
\end{equation*}
The obtained equations provide a basis for using the following approximations:
\begin{equation}
\left.\left(\frac{1}{\H}\cdot \frac{\partial \Pnh}{\partial x}\right)\right|_{x_l-0}\sim \frac{\dot{Q}^*}{\H_{j_l-1/2}^*}+{\rho}(u^2)_{x,j_l-1/2}^*, \quad
\left.\left(\frac{1}{\H}\cdot \frac{\partial \Pnh}{\partial x}\right)\right|_{x_r+0}\sim \frac{\dot{Q}^*}{\H_{j_r+1/2}^*}+{\rho}(u^2)_{x,j_r+1/2}^*,
\label{Appen_B_main_fi_x}
\end{equation}
where, according to (\ref{1D_NLD_to_NSWE_b}), for compatibility conditions (C1) we obtain
\begin{equation}
\dot{Q}^*=-\frac{S_0}{L}\left[\left({\rho}g\eta+\frac{(S_0^2-3\H^2)}{2\H^3}\Pnh\right)^*_{j_r+1/2}\hspace*{-4mm} - \left({\rho}g\eta+\frac{(S_0^2-3\H^2)}{2\H^3}\Pnh\right)^*_{j_l-1/2}\right],
\label{Appen_B_FDE2}
\end{equation}
and for (C2):
\begin{equation}
\hspace*{-4mm}\dot{Q}^*=-\frac{{\rho}S_0}{L}\left[\left(g\eta+\frac{u^2}{2}+\frac{\H^2}{6}(u^2)_x-\frac{\Pnh}{{\rho}\H}\right)^*_{j_r+1/2}\hspace*{-4mm}  - \left(
g\eta+\frac{u^2}{2}+\frac{\H^2}{6}(u^2)_x-\frac{\Pnh}{{\rho}\H}\right)^*_{j_l-1/2}\right],
\label{Appen_B_FDE3}
\end{equation}
and $u^*_{j_l}$, $u^*_{j_r}$ from (\ref{Appen_B_FDE7}) are used to calculate the derivative $(u^2)_x$:
\begin{equation*}
\left(u^2\right)^*_{x,j_l-1/2} = \frac{(u^*_{j_l})^2-(u^*_{j_l-1})^2}{\Delta x}, \quad \left(u^2\right)^*_{x,j_r+1/2} = \frac{(u^*_{j_r+1})^2-(u^*_{j_r})^2}{\Delta x}.
\end{equation*}

Substituting these expressions into (\ref{dif_phi_x=x_l}), we obtain three-point equation (\ref{3_point_eq_phi}), in which $j=j_l-1$, $\Pnh^*_{j+3/2} = \Pnh^*_{j_r+1/2}$,
\begin{equation*}
a_j=\frac{k_{j-1/2}+k_{j+1/2}}{2\Delta x}.
\end{equation*}
In the case of  compatibility conditions (C1) we have
\begin{equation}
b_j=\frac{3S_0\big(\H^*_{j_r+1/2}\big)^2-S_0^3}{2L\H^*_{j+1/2}\big(\H^*_{j_r+1/2}\big)^3}, \quad c_j=a_j+\frac{3\Delta x}{\big(\H^*_{j+1/2}\big)^3}+\frac{3S_0\big(\H^*_{j+1/2}\big)^2-S_0^3}{2L\big(\H^*_{j+1/2}\big)^4},
\label{Appen_B_jlC1bcz}
\end{equation}
\begin{equation}
d_j=\frac{{\rho}S_0}{L\H^*_{j+1/2}}\Big[g\eta^*_{j_r+1/2}-g\eta^*_{j+1/2}\Big]-{\rho}g \eta^*_{x,j}+2{\rho}\Delta x\left( u^*_{x,j+1/2}\right)^2 - {\rho}(u^2)^*_{x,j+1/2}.
\label{Appen_B_jlC1dz}
\end{equation}
For  compatibility conditions (C2),  formulas (\ref{Appen_B_jlC1bcz}), (\ref{Appen_B_jlC1dz}) should be replaced by the following:
\begin{equation}
b_j=\frac{S_0}{L\H^*_{j+1/2}\H^*_{j_r+1/2}}, \quad c_j=a_j+\frac{3\Delta x}{\big(H^*_{j+1/2}\big)^3}+\frac{S_0}{L\big(\H^*_{j+1/2}\big)^2},
\label{Appen_B_jlC2bcz}
\end{equation}
\begin{equation*}
d_j=\frac{{\rho}S_0}{L\H^*_{j+1/2}}\left[\left(g\eta+\frac{u^2}{2}+\frac{\H^2}{6}(u^2)_x\right)^*_{j_r+1/2} - \left(g\eta+\frac{u^2}{2}+\frac{\H^2}{6}(u^2)_x\right)^*_{j+1/2}\right] -
\end{equation*}
\begin{equation}
-{\rho}g \eta^*_{x,j}+2{\rho}\Delta x\left( u^*_{x,j+1/2}\right)^2 - {\rho}(u^2)^*_{x,j+1/2}.
\label{Appen_B_jlC2dz}
\end{equation}

Similarly, using equalities (\ref{Appen_B_main_fi_x})---(\ref{Appen_B_FDE3}), we obtain three-point finite-difference equation (\ref{3_point_eq_phi}) at $j=j_r$, assuming $\Pnh^*_{j-1/2} = \Pnh^*_{j_l-1/2}$. The coefficients $a_j$, $c_j$ and the right-hand side $d_j$ of this equation depend on the chosen type of compatibility conditions, while the coefficient $b_j$ is calculated using the  formula
\begin{equation*}
b_j=\frac{k_{j+1/2}+k_{j+3/2}}{2\Delta x}, \quad (j=j_r).
\end{equation*}
In the case of  compatibility conditions (C1) we obtain
\begin{equation}
a_j=\frac{3S_0\big(\H^*_{j_l-1/2}\big)^2-S_0^3}{2L\H^*_{j+1/2}\big(\H^*_{j_l-1/2}\big)^3}, \quad c_j=b_j+\frac{3\Delta x}{\big(\H^*_{j+1/2}\big)^3}+\frac{3S_0\big(\H^*_{j+1/2}\big)^2-S_0^3}{2L\big(\H^*_{j+1/2}\big)^4},
\label{Appen_B_jrC1acz}
\end{equation}
\begin{equation}
d_j=-\frac{{\rho}S_0}{L\H^*_{j+1/2}}\Big[g\eta^*_{j+1/2}-g\eta^*_{j_l-1/2}\Big]+{\rho}g \eta^*_{x,j+1}+2{\rho}\Delta x\left( u^*_{x,j+1/2}\right)^2 + {\rho}(u^2)^*_{x,j+1/2}.
\label{Appen_B_jrC1dz}
\end{equation}
For compatibility conditions (C2) instead of formulas (\ref{Appen_B_jrC1acz}), (\ref{Appen_B_jrC1dz}) we have the following:
\begin{equation}
a_j=\frac{S_0}{L\H^*_{j+1/2}\H^*_{j_l-1/2}}, \quad c_j=b_j+\frac{3\Delta x}{\big(\H^*_{j+1/2}\big)^3}+\frac{S_0}{L\big(\H^*_{j+1/2}\big)^2},
\label{Appen_B_jrC2acz}
\end{equation}
\begin{equation*}
d_j=-\frac{{\rho}S_0}{L\H^*_{j+1/2}}\left[\left(g\eta+\frac{u^2}{2}+\frac{\H^2}{6}(u^2)_x\right)^*_{j+1/2} - \left(g\eta+\frac{u^2}{2}+\frac{\H^2}{6}(u^2)_x\right)^*_{j_l-1/2}\right] +
\end{equation*}
\begin{equation}
+ {\rho}g \eta^*_{x,j+1}+2{\rho}\Delta x\left( u^*_{x,j+1/2}\right)^2 + {\rho}(u^2)^*_{x,j+1/2}.
\label{Appen_B_jrC2dz}
\end{equation}

The resulting system of the finite-difference equations for the unknowns $\Pnh^*_{j+1/2}$ is solved by the Thomas algorithm.
The predictor step is completed by calculating the value of change of fluid flow under the body $\dot{Q}^*$. For this, depending on the type of compatibility conditions, either formula (\ref{Appen_B_FDE2}) or (\ref{Appen_B_FDE3}) is used.

{\textit {\textbf {Corrector.}}}  The corrector step calculates the total depth $\H^{n+1}_{j}$, velocity $u^{n+1}_{j}$, and pressure dispersion component $\Pnh^{n+1}_{j}$.  The values of the total depth and velocity at the inner nodes of this grid ($j=1,\ldots , j_l-1$, $j=j_r+1,\ldots , N_1$) are computed using the algorithm described in \cite{Khakimzyanov2016,Khakimzyanov2019c}. At the outer boundary $\Gamma_0$, condition (\ref{Pt_2_Gamma_0}) is used and $u_{0}^{n+1}=u_{N}^{n+1}=0$, $\eta^{n+1}_{0}=\eta^{n+1}_{1}$, $\ \eta^{n+1}_{N}=\eta^{n+1}_{N-1}$.
At the common boundary $\Gamma$ of the subregions ${\cal D}_e$ and ${\cal D}_i$ condition (\ref{1D_NLD_Deta_dx}) is used, which is implemented here in the following finite-difference form: $\eta^{n+1}_{j_l}=\eta^{n+1}_{j_l-1}$, $\eta^{n+1}_{j_r}=\eta^{n+1}_{j_r+1}$. Then, using the flow rate change  $\dot{Q}^*$ of the fluid under the body, which was found on the predictor step, we obtain the flow rate at the $(n+1)$ time step: $Q^{n+1}=Q^n+\tau \dot{Q}^*$. Then we can calculate the velocities at the time step $(n+1)$ using condition (\ref{Appen_B_FDE6}):
\begin{equation}
u^{n+1}_{j_l}=\frac{Q^{n+1}}{\rho \H^{n+1}_{j_l}}, \quad u^{n+1}_{j_r}=\frac{Q^{n+1}}{\rho \H^{n+1}_{j_r}}.
\label{Appen_B_FDE7_corr}
\end{equation}
Thus, the values $\H^{n+1}_{j}$, $u^{n+1}_{j}$ are now known at all the nodes $x_{j}$ ($j=0,\ldots , j_l$, $j=j_r,\ldots , N$).

Similar to the predictor step, the finite-difference equations for $\Pnh^{n+1}_{j}$ are derived based on the integral form of equation (\ref{curve_phi_1}), but the relations of  type (\ref{Afd_phi_1}) take other cells for integration. Thus, in the inner nodes $x_{j}$ instead of (\ref{Afd_phi_1}) the following integral relation is used:
\begin{equation}
\int\limits_{x_{j-1/2}}^{x_{j+1/2}} \left(k\Pnh_x\right)_x  dx-\int\limits_{x_{j-1/2}}^{x_{j+1/2}} k_0\Pnh  dx=\int\limits_{x_{j-1/2}}^{x_{j+1/2}}F  dx, \qquad \begin{array}{l}\displaystyle j=1,\ldots , j_l-1, \\\displaystyle j=j_r+1,\ldots , N-1,\end{array}
\label{Afd_phi_1_corr}
\end{equation}
where $x_{j+1/2}=x_j+h/2$. The quadrature formulas similar to (\ref{fd_phi_2})---(\ref{fd_phi_4}) are used to calculate integrals:
\begin{equation*}
\int\limits_{x_{j-1/2}}^{x_{j+1/2}} \left(k\Pnh_x\right)_x  dx\sim \frac{k_{j+1}+k_{j}}{2}\cdot \Pnh^{n+1}_{x,j+1/2}-\frac{k_{j}+k_{j-1}}{2}\cdot \Pnh^{n+1}_{x,j-1/2},
\end{equation*}
\begin{equation*}
\int\limits_{x_{j-1/2}}^{x_{j+1/2}} k_0\Pnh  dx\sim  \Pnh^{n+1}_{j}\frac{3}{\left(\H^{n+1}_{j}\right)^3}\Delta x,
\end{equation*}
\begin{equation*}
\int\limits_{x_{j-1/2}}^{x_{j+1/2}}F dx \sim {\rho}g \left(\eta^{n+1}_{x,j+1/2}-\eta^{n+1}_{x,j-1/2}\right)+2{\rho}\Delta x\left(u^{n+1}_{x,j}\right)^2,
\end{equation*}
where
\begin{equation*}
k_{j}=\frac{1}{\H^{n+1}_{j}}, \quad \Pnh^{n+1}_{x,j+1/2}=\frac{\Pnh^{n+1}_{j+1}-\Pnh^{n+1}_{j}}{\Delta x},\quad \eta^{n+1}_{x,j+1/2}=\frac{\eta^{n+1}_{j+1}-\eta^{n+1}_{j}}{\Delta x},
\end{equation*}
\begin{equation*}
u^{n+1}_{x,j}=\frac{1}{2}\left(u^{n+1}_{x,j+1/2}+u^{n+1}_{x,j-1/2}\right), \quad  u^{n+1}_{x,j+1/2}=\frac{u^{n+1}_{j+1}-u^{n+1}_{j}}{\Delta x}.
\end{equation*}

So, for the specified values of $j$ we obtain three-point difference equations
\begin{equation}
a_j\Pnh^{n+1}_{j-1}-c_j\Pnh^{n+1}_{j}+b_j\Pnh^{n+1}_{j+1}=d_j,
\label{3_point_eq_phi_corr}
\end{equation}
where
\begin{equation*}
a_j=\frac{k_{j-1}+k_{j}}{2\Delta x}, \quad b_j=\frac{k_{j+1}+k_{j}}{2\Delta x}, \quad c_j=a_j+b_j+\frac{3\Delta x}{\left(\H^{n+1}_{j}\right)^3},
\end{equation*}
\begin{equation*}
d_j={\rho}g \left(\eta^{n+1}_{x,j+1/2}-\eta^{n+1}_{x,j-1/2}\right)+2{\rho}\Delta x\left( u^{n+1}_{x,j}\right)^2.
\end{equation*}

At $j=0$ and $j=N$, we use the boundary cells $[x_j, x_{j+1/2}]$ and $[x_{j-1/2}, x_{j}]$, respectively, and consider boundary conditions (\ref{Pt_2_Gamma_0}). This results in two-point finite-difference equations, which formally can be written in form (\ref{3_point_eq_phi_corr}), assuming at $j=0$
\begin{equation*}
a_j=0, \quad b_j=\frac{k_{j+1}+k_{j}}{2\Delta x}, \quad c_j=a_j+b_j+\frac{3\Delta x}{2\left(\H^{n+1}_{j}\right)^3}, \quad d_j={\rho}g \eta^{n+1}_{x,j+1/2}+{\rho}\Delta x\left( u^{n+1}_{x,j+1/2}\right)^2,
\end{equation*}
and at $j=N$
\begin{equation*}
a_j=\frac{k_{j-1}+k_{j}}{2\Delta x}, \quad b_j=0, \quad c_j=a_j+b_j+\frac{3\Delta x}{2\left(\H^{n+1}_{j}\right)^3},\quad d_j=-{\rho}g \eta^{n+1}_{x,j-1/2}+{\rho}\Delta x\left( u^{n+1}_{x,j-1/2}\right)^2.
\end{equation*}

The approximation of the integral equations for $\Pnh$ at the boundary cells $[x_{j_l-1/2}, x_{j_l}]$ and $[x_{j_r}, x_{j_r+1/2}]$ is
\begin{equation*}
\left.\left(\frac{1}{\H}\cdot \frac{\partial \Pnh}{\partial x}\right)\right|_{x_{l}-0}-\frac{k_{j}+k_{j-1}}{2}\cdot \Pnh^{n+1}_{x,j-1/2}- \Pnh^{n+1}_{j}\frac{3\Delta x}{2\left(\H^{n+1}_{j}\right)^3}=F_j\frac{\Delta x}{2}, \qquad j=j_l,
\end{equation*}
\begin{equation*}
\frac{k_{j}+k_{j+1}}{2}\cdot \Pnh^{n+1}_{x,j+1/2}-\left.\left(\frac{1}{\H}\cdot \frac{\partial \Pnh}{\partial x}\right)\right|_{x_{r}+0}- \Pnh^{n+1}_{j}\frac{3\Delta x}{2\left(\H^{n+1}_{j}\right)^3}=F_j\frac{\Delta x}{2}, \qquad j=j_r.
\end{equation*}
In these equations, we use the relations of form (\ref{Appen_B_main_fi_x}) to approximate the limit values of the derivatives ${\partial \Pnh}/{\partial x}$ at the time layer $(n+1)$:
\begin{equation*}
\left.\left(\frac{1}{\H}\cdot \frac{\partial \Pnh}{\partial x}\right)\right|_{x_l-0}\sim \frac{\dot{Q}^{n+1}}{\H_{j_l}^{n+1}}+{\rho}(u^2)_{x,j_l}^{n+1}, \quad
\left.\left(\frac{1}{\H}\cdot \frac{\partial \Pnh}{\partial x}\right)\right|_{x_r+0}\sim \frac{\dot{Q}^{n+1}}{\H_{j_r}^{n+1}}+{\rho}(u^2)_{x,j_r}^{n+1},
\end{equation*}
and for compatibility conditions (C1) we have an analogue of (\ref{Appen_B_FDE2}):
\begin{equation}
\dot{Q}^{n+1}=-\frac{S_0}{L}\left[\left({\rho}g\eta+\frac{(S_0^2-3\H^2)}{2\H^3}\Pnh\right)^{n+1}_{j_r} - \left({\rho}g\eta+\frac{(S_0^2-3\H^2)}{2\H^3}\Pnh\right)^{n+1}_{j_l}\right],
\label{Appen_B_FDE2_corr}
\end{equation}
and for (C2) the analog of (\ref{Appen_B_FDE3}):
\begin{equation}
\hspace*{-4mm}\dot{Q}^{n+1}=-\frac{{\rho}S_0}{L}\left[\left(g\eta+\frac{u^2}{2}+\frac{H^2}{6}(u^2)_x-\frac{\Pnh}{{\rho}\H}\right)^{n+1}_{j_r}\hspace*{-2mm}  - \left(
g\eta+\frac{u^2}{2}+\frac{\H^2}{6}(u^2)_x-\frac{\Pnh}{{\rho}\H}\right)^{n+1}_{j_l}\right],
\label{Appen_B_FDE3_corr}
\end{equation}
where
\begin{equation*}
\left(u^2\right)^{n+1}_{x,j_l} = \frac{(u^{n+1}_{j_l})^2-(u^{n+1}_{j_l-1/2})^2}{\Delta x /2}, \quad \left(u^2\right)^{n+1}_{x,j_r} = \frac{(u^{n+1}_{j_r+1/2})^2-(u^{n+1}_{j_r})^2}{\Delta x /2}, \quad u^{n+1}_{j+1/2}=\frac{u^{n+1}_{j}+u^{n+1}_{j+1}}{2}.
\end{equation*}

Thus,  we have two additional three-point equations of type (\ref{3_point_eq_phi_corr}) at $x_{j_l}$ and $x_{j_r}$:
\begin{equation}
a_{j_l}\Pnh^{n+1}_{j_l-1}-c_{j_l}\Pnh^{n+1}_{j_l}+b_{j_l}\Pnh^{n+1}_{j_r}=d_{j_l},
\label{3_point_eq_phi_corr_j_l}
\end{equation}
\begin{equation}
a_{j_r}\Pnh^{n+1}_{j_l}-c_{j_r}\Pnh^{n+1}_{j_r}+b_{j_r}\Pnh^{n+1}_{j_r+1}=d_{j_r}.
\label{3_point_eq_phi_corr_j_r}
\end{equation}
In the case of compatibility conditions (C1), the coefficients of these equations are calculated by the following formulas:
\begin{equation*}
a_{j_l}=\frac{k_{j_l-1}+k_{j_l}}{2\Delta x},
\end{equation*}
\begin{equation}
b_{j_l}=\frac{3S_0\big(\H^{n+1}_{j_r}\big)^2-S_0^3}{2L\H^{n+1}_{j_l}\big(\H^{n+1}_{j_r}\big)^3}, \quad c_{j_l}=a_{j_l}+\frac{3\Delta x}{2\big(\H^{n+1}_{j_l}\big)^3}+\frac{3S_0\big(\H^{n+1}_{j_l}\big)^2-S_0^3}{2L\big(\H^{n+1}_{j_l}\big)^4},
\label{Appen_B_jlC1bc1}
\end{equation}
\begin{equation}
d_{j_l}=\frac{{\rho}S_0}{L\H^{n+1}_{j_l}}\Big[g\eta^{n+1}_{j_r}-g\eta^{n+1}_{j_l}\Big]-{\rho}g \eta^{n+1}_{x,j_l-1/2}+{\rho}\Delta x\left( u^{n+1}_{x,j_l-1/2}\right)^2 - {\rho}(u^2)^{n+1}_{x,j_l},
\label{Appen_B_jlC1d1}
\end{equation}
\begin{equation*}
b_{j_r}=\frac{k_{j_r}+k_{j_r+1}}{2\Delta x},
\end{equation*}
\begin{equation}
a_{j_r}=\frac{3S_0\big(\H^{n+1}_{j_l}\big)^2-S_0^3}{2L\H^{n+1}_{j_r}\big(\H^{n+1}_{j_l}\big)^3}, \quad c_{j_r}=b_{j_r}+\frac{3\Delta x}{2\big(\H^{n+1}_{j_r}\big)^3}+\frac{3S_0\big(\H^{n+1}_{j_r}\big)^2-S_0^3}{2L\big(\H^{n+1}_{j_r}\big)^4},
\label{Appen_B_jrC1ac1}
\end{equation}
\begin{equation}
d_{j_r}=-\frac{{\rho}S_0}{L\H^{n+1}_{j_r}}\Big[g\eta^{n+1}_{j_r}-g\eta^{n+1}_{j_l}\Big]+{\rho}g \eta^{n+1}_{x,j_r+1/2}+{\rho}\Delta x\left( u^{n+1}_{x,j_r+1/2}\right)^2 + {\rho}(u^2)^{n+1}_{x,j_r}.
\label{Appen_B_jrC1d1}
\end{equation}
In the case of compatibility conditions (C2), the coefficients (\ref{Appen_B_jlC1bc1})---(\ref{Appen_B_jrC1d1}) will be changed:
\begin{equation}
b_{j_l}=\frac{S_0}{L\H^{n+1}_{j_l}\H^{n+1}_{j_r}}, \quad c_{j_l}=a_{j_l}+\frac{3\Delta x}{2\big(\H^{n+1}_{j_l}\big)^3}+\frac{S_0}{L\big(\H^{n+1}_{j_l}\big)^2},
\label{Appen_B_jlC2bc1}
\end{equation}
\begin{equation*}
d_{j_l}=\frac{{\rho}S_0}{L\H^{n+1}_{j_l}}\left[\left(g\eta+\frac{u^2}{2}+\frac{\H^2}{6}(u^2)_x\right)^{n+1}_{j_r} - \left(g\eta+\frac{u^2}{2}+\frac{\H^2}{6}(u^2)_x\right)^{n+1}_{j_l}\right] -
\end{equation*}
\begin{equation}
-{\rho}g \eta^{n+1}_{x,j_l-1/2}+{\rho}\Delta x\left( u^{n+1}_{x,j_l-1/2}\right)^2 - {\rho}(u^2)^{n+1}_{x,j_l},
\label{Appen_B_jlC2d1}
\end{equation}
\begin{equation}
a_{j_r}=\frac{S_0}{L\H^{n+1}_{j_r}\H^{n+1}_{j_l}}, \quad c_{j_r}=b_{j_r}+\frac{3\Delta x}{2\big(\H^{n+1}_{j_r}\big)^3}+\frac{S_0}{L\big(\H^{n+1}_{j_r}\big)^2},
\label{Appen_B_jrC2ac1}
\end{equation}
\begin{equation*}
d_{j_r}=-\frac{{\rho}S_0}{L\H^{n+1}_{j_r}}\left[\left(g\eta+\frac{u^2}{2}+\frac{\H^2}{6}(u^2)_x\right)^{n+1}_{j_r} - \left(g\eta+\frac{u^2}{2}+\frac{\H^2}{6}(u^2)_x\right)^{n+1}_{j_l}\right] +
\end{equation*}
\begin{equation}
+ {\rho}g \eta^{n+1}_{x,j_r+1/2}+{\rho}\Delta x\left( u^{n+1}_{x,j_r+1/2}\right)^2 + {\rho}(u^2)^{n+1}_{x,j_r}.
\label{Appen_B_jrC2d1}
\end{equation}

The resulting system of finite-difference equations of type (\ref{3_point_eq_phi_corr}) for $\Pnh^{n+1}_{j}$ ($j=0,\ldots , j_l$, $\ j=j_r,\ldots , N$) is also solved by Thomas algorithm. When the values of $\Pnh^{n+1}_{j}$ are found, the rate of flow change $\dot{Q}^{n+1}$ is determined. To do this, one of formulas (\ref{Appen_B_FDE2_corr}) or (\ref{Appen_B_FDE3_corr}) is used. After that everything is ready for the next time step.

{\bf Remark}. For dispersionless shallow water ($\SW$) equations, the calculation of $\Pnh^*_{j+1/2}$ and $\Pnh^{n+1}_{j}$ values is not required and, according to
(\ref{Pt_2_conj_p_SW}), (\ref{Alter_SW_conj_5}), formulas (\ref{Appen_B_FDE2}), (\ref{Appen_B_FDE3}), (\ref{Appen_B_FDE2_corr}), (\ref{Appen_B_FDE3_corr}) must be modified by excluding expressions containing $\Pnh$ and $(u^2)_x$.

\end{document}